 \documentclass[final,1p,times]{elsarticle}
\usepackage{amssymb}
\usepackage{amsmath}
\usepackage{amsthm}

\usepackage{xspace}    % Leerzeichen oder nicht nach Kommando einfuegen
\usepackage{color}
\DeclareMathAlphabet\mathbfcal{OMS}{cmsy}{b}{n}
\usepackage{tikz}
\graphicspath{{figures/}}
\usepackage{standalone}
\usepackage{tabularx}
\usepackage{subcaption}
\usepackage{graphics}
\usepackage{gensymb}
\usepackage{textcomp}
\usepackage{floatrow}

\newcommand{\Flexi}{\textit{FLEXI}\xspace}
\newcommand{\Hopr}{\textit{HOPR}\xspace}
\newcommand{\Posti}{\textit{POSTI}\xspace}
\newcommand{\NSE}{Navier--Stokes equations\xspace}

\newcommand{\cunderline}[2]{{\color{#1}\underline{{\color{black}#2}}}}

\newcommand{\Ngeo}{N_{\mathrm{geo}}}
\newcommand{\nout}{\hat{n}}
\newcommand{\noutRef}{\mathcal{N}}
\newcommand{\J}{\mathcal{J}}
\newcommand{\meshVel}{\nu}
\newcommand{\pderivative}[2]{\frac{\partial #1}{\partial #2}}

\newcommand{\gradient}{\nabla}
\newcommand{\gradientXI}{\nabla_\xi}

\newcommand{\F}{\mathcal{F}}
\newcommand{\N}{\mathbb{N}} 
\newcommand{\R}{\mathbb{R}} 
\newcommand{\avg}[1]{\ensuremath{\{\!\!\{#1\}\!\!\}} }

\newcommand{\fracp}[2]{\frac{\partial #1}{\partial #2}}

\newcommand{\unit}[1]{\ensuremath{\, \mathrm{#1}}}

\newcommand{\tikzcircle}[2][red,fill=red]{\tikz[baseline=-0.5ex]\draw[#1,radius=#2] (0,0) circle ;}%
\usetikzlibrary{spy,arrows}
\usetikzlibrary{arrows.meta}
\usetikzlibrary{shapes.geometric}
\usetikzlibrary{intersections}
\usetikzlibrary{positioning}
\usetikzlibrary{backgrounds}
\usepackage{pgfplots}
\pgfplotsset{compat=1.13}
\usetikzlibrary{calc}
\tikzstyle{rounded}  = [rounded corners]     
\tikzstyle{function} = [rectangle, text centered, draw=black,minimum width=7.8cm,minimum height=1.0cm]
\tikzstyle{mpi}      = [anchor=south west, xshift=0.2cm] 

\usepackage{tikz}
\usepackage{xifthen}
\usetikzlibrary{spy,arrows}
\usetikzlibrary{arrows.meta}
\usetikzlibrary{shapes.geometric}
\usetikzlibrary{intersections}
\usetikzlibrary{positioning}
\usetikzlibrary{backgrounds}
\usepackage{pgfplots}
\pgfplotsset{compat=1.13}
\pgfplotsset{/pgfplots/colormap={hot2}{[1cm]rgb255(0cm)=(0,0,0)rgb255(3cm)=(255,0,0)rgb255(6cm)=(255,255,0)rgb255(8cm)=(255,255,255)}}
\usetikzlibrary{calc}
\tikzstyle{rounded}  = [rounded corners]     
\tikzstyle{function} = [rectangle, text centered, draw=black,minimum width=6.0cm,minimum height=1.0cm]
\tikzstyle{mpi}      = [anchor=south west, xshift=0.2cm] 

\newcommand{\xGP}{-0.866, -0.339,  0.339, 0.866}

\newcommand{\innerborders}{-0.652,0,0.652}
\newcommand{\xGPeq}{-0.75, -0.25,  0.25, 0.75}

\newcommand{\innerborderseq}{-0.5,0,0.5}

\usetikzlibrary{3d}

\usetikzlibrary{fillbetween}
\usetikzlibrary{decorations,decorations.text,backgrounds}

\tikzset{boximg/.style={remember picture,black,thick,draw,inner sep=0pt,outer sep=0pt}}
\DeclareRobustCommand\full  {\tikz[baseline=-0.6ex]\draw[thick] (0,0)--(0.5,0);}
\DeclareRobustCommand\dotted{\tikz[baseline=-0.6ex]\draw[thick,dotted] (0,0)--(0.54,0);}
\DeclareRobustCommand\dashed{\tikz[baseline=-0.6ex]\draw[thick,dashed] (0,0)--(0.54,0);}

\input{coordimg}
\pgfplotsset{
    colormap={plasma}{
      rgb=(0.050383, 0.029803, 0.527975)
      rgb=(0.063536, 0.028426, 0.533124)
      rgb=(0.075353, 0.027206, 0.538007)
      rgb=(0.086222, 0.026125, 0.542658)
      rgb=(0.096379, 0.025165, 0.547103)
      rgb=(0.105980, 0.024309, 0.551368)
      rgb=(0.115124, 0.023556, 0.555468)
      rgb=(0.123903, 0.022878, 0.559423)
      rgb=(0.132381, 0.022258, 0.563250)
      rgb=(0.140603, 0.021687, 0.566959)
      rgb=(0.148607, 0.021154, 0.570562)
      rgb=(0.156421, 0.020651, 0.574065)
      rgb=(0.164070, 0.020171, 0.577478)
      rgb=(0.171574, 0.019706, 0.580806)
      rgb=(0.178950, 0.019252, 0.584054)
      rgb=(0.186213, 0.018803, 0.587228)
      rgb=(0.193374, 0.018354, 0.590330)
      rgb=(0.200445, 0.017902, 0.593364)
      rgb=(0.207435, 0.017442, 0.596333)
      rgb=(0.214350, 0.016973, 0.599239)
      rgb=(0.221197, 0.016497, 0.602083)
      rgb=(0.227983, 0.016007, 0.604867)
      rgb=(0.234715, 0.015502, 0.607592)
      rgb=(0.241396, 0.014979, 0.610259)
      rgb=(0.248032, 0.014439, 0.612868)
      rgb=(0.254627, 0.013882, 0.615419)
      rgb=(0.261183, 0.013308, 0.617911)
      rgb=(0.267703, 0.012716, 0.620346)
      rgb=(0.274191, 0.012109, 0.622722)
      rgb=(0.280648, 0.011488, 0.625038)
      rgb=(0.287076, 0.010855, 0.627295)
      rgb=(0.293478, 0.010213, 0.629490)
      rgb=(0.299855, 0.009561, 0.631624)
      rgb=(0.306210, 0.008902, 0.633694)
      rgb=(0.312543, 0.008239, 0.635700)
      rgb=(0.318856, 0.007576, 0.637640)
      rgb=(0.325150, 0.006915, 0.639512)
      rgb=(0.331426, 0.006261, 0.641316)
      rgb=(0.337683, 0.005618, 0.643049)
      rgb=(0.343925, 0.004991, 0.644710)
      rgb=(0.350150, 0.004382, 0.646298)
      rgb=(0.356359, 0.003798, 0.647810)
      rgb=(0.362553, 0.003243, 0.649245)
      rgb=(0.368733, 0.002724, 0.650601)
      rgb=(0.374897, 0.002245, 0.651876)
      rgb=(0.381047, 0.001814, 0.653068)
      rgb=(0.387183, 0.001434, 0.654177)
      rgb=(0.393304, 0.001114, 0.655199)
      rgb=(0.399411, 0.000859, 0.656133)
      rgb=(0.405503, 0.000678, 0.656977)
      rgb=(0.411580, 0.000577, 0.657730)
      rgb=(0.417642, 0.000564, 0.658390)
      rgb=(0.423689, 0.000646, 0.658956)
      rgb=(0.429719, 0.000831, 0.659425)
      rgb=(0.435734, 0.001127, 0.659797)
      rgb=(0.441732, 0.001540, 0.660069)
      rgb=(0.447714, 0.002080, 0.660240)
      rgb=(0.453677, 0.002755, 0.660310)
      rgb=(0.459623, 0.003574, 0.660277)
      rgb=(0.465550, 0.004545, 0.660139)
      rgb=(0.471457, 0.005678, 0.659897)
      rgb=(0.477344, 0.006980, 0.659549)
      rgb=(0.483210, 0.008460, 0.659095)
      rgb=(0.489055, 0.010127, 0.658534)
      rgb=(0.494877, 0.011990, 0.657865)
      rgb=(0.500678, 0.014055, 0.657088)
      rgb=(0.506454, 0.016333, 0.656202)
      rgb=(0.512206, 0.018833, 0.655209)
      rgb=(0.517933, 0.021563, 0.654109)
      rgb=(0.523633, 0.024532, 0.652901)
      rgb=(0.529306, 0.027747, 0.651586)
      rgb=(0.534952, 0.031217, 0.650165)
      rgb=(0.540570, 0.034950, 0.648640)
      rgb=(0.546157, 0.038954, 0.647010)
      rgb=(0.551715, 0.043136, 0.645277)
      rgb=(0.557243, 0.047331, 0.643443)
      rgb=(0.562738, 0.051545, 0.641509)
      rgb=(0.568201, 0.055778, 0.639477)
      rgb=(0.573632, 0.060028, 0.637349)
      rgb=(0.579029, 0.064296, 0.635126)
      rgb=(0.584391, 0.068579, 0.632812)
      rgb=(0.589719, 0.072878, 0.630408)
      rgb=(0.595011, 0.077190, 0.627917)
      rgb=(0.600266, 0.081516, 0.625342)
      rgb=(0.605485, 0.085854, 0.622686)
      rgb=(0.610667, 0.090204, 0.619951)
      rgb=(0.615812, 0.094564, 0.617140)
      rgb=(0.620919, 0.098934, 0.614257)
      rgb=(0.625987, 0.103312, 0.611305)
      rgb=(0.631017, 0.107699, 0.608287)
      rgb=(0.636008, 0.112092, 0.605205)
      rgb=(0.640959, 0.116492, 0.602065)
      rgb=(0.645872, 0.120898, 0.598867)
      rgb=(0.650746, 0.125309, 0.595617)
      rgb=(0.655580, 0.129725, 0.592317)
      rgb=(0.660374, 0.134144, 0.588971)
      rgb=(0.665129, 0.138566, 0.585582)
      rgb=(0.669845, 0.142992, 0.582154)
      rgb=(0.674522, 0.147419, 0.578688)
      rgb=(0.679160, 0.151848, 0.575189)
      rgb=(0.683758, 0.156278, 0.571660)
      rgb=(0.688318, 0.160709, 0.568103)
      rgb=(0.692840, 0.165141, 0.564522)
      rgb=(0.697324, 0.169573, 0.560919)
      rgb=(0.701769, 0.174005, 0.557296)
      rgb=(0.706178, 0.178437, 0.553657)
      rgb=(0.710549, 0.182868, 0.550004)
      rgb=(0.714883, 0.187299, 0.546338)
      rgb=(0.719181, 0.191729, 0.542663)
      rgb=(0.723444, 0.196158, 0.538981)
      rgb=(0.727670, 0.200586, 0.535293)
      rgb=(0.731862, 0.205013, 0.531601)
      rgb=(0.736019, 0.209439, 0.527908)
      rgb=(0.740143, 0.213864, 0.524216)
      rgb=(0.744232, 0.218288, 0.520524)
      rgb=(0.748289, 0.222711, 0.516834)
      rgb=(0.752312, 0.227133, 0.513149)
      rgb=(0.756304, 0.231555, 0.509468)
      rgb=(0.760264, 0.235976, 0.505794)
      rgb=(0.764193, 0.240396, 0.502126)
      rgb=(0.768090, 0.244817, 0.498465)
      rgb=(0.771958, 0.249237, 0.494813)
      rgb=(0.775796, 0.253658, 0.491171)
      rgb=(0.779604, 0.258078, 0.487539)
      rgb=(0.783383, 0.262500, 0.483918)
      rgb=(0.787133, 0.266922, 0.480307)
      rgb=(0.790855, 0.271345, 0.476706)
      rgb=(0.794549, 0.275770, 0.473117)
      rgb=(0.798216, 0.280197, 0.469538)
      rgb=(0.801855, 0.284626, 0.465971)
      rgb=(0.805467, 0.289057, 0.462415)
      rgb=(0.809052, 0.293491, 0.458870)
      rgb=(0.812612, 0.297928, 0.455338)
      rgb=(0.816144, 0.302368, 0.451816)
      rgb=(0.819651, 0.306812, 0.448306)
      rgb=(0.823132, 0.311261, 0.444806)
      rgb=(0.826588, 0.315714, 0.441316)
      rgb=(0.830018, 0.320172, 0.437836)
      rgb=(0.833422, 0.324635, 0.434366)
      rgb=(0.836801, 0.329105, 0.430905)
      rgb=(0.840155, 0.333580, 0.427455)
      rgb=(0.843484, 0.338062, 0.424013)
      rgb=(0.846788, 0.342551, 0.420579)
      rgb=(0.850066, 0.347048, 0.417153)
      rgb=(0.853319, 0.351553, 0.413734)
      rgb=(0.856547, 0.356066, 0.410322)
      rgb=(0.859750, 0.360588, 0.406917)
      rgb=(0.862927, 0.365119, 0.403519)
      rgb=(0.866078, 0.369660, 0.400126)
      rgb=(0.869203, 0.374212, 0.396738)
      rgb=(0.872303, 0.378774, 0.393355)
      rgb=(0.875376, 0.383347, 0.389976)
      rgb=(0.878423, 0.387932, 0.386600)
      rgb=(0.881443, 0.392529, 0.383229)
      rgb=(0.884436, 0.397139, 0.379860)
      rgb=(0.887402, 0.401762, 0.376494)
      rgb=(0.890340, 0.406398, 0.373130)
      rgb=(0.893250, 0.411048, 0.369768)
      rgb=(0.896131, 0.415712, 0.366407)
      rgb=(0.898984, 0.420392, 0.363047)
      rgb=(0.901807, 0.425087, 0.359688)
      rgb=(0.904601, 0.429797, 0.356329)
      rgb=(0.907365, 0.434524, 0.352970)
      rgb=(0.910098, 0.439268, 0.349610)
      rgb=(0.912800, 0.444029, 0.346251)
      rgb=(0.915471, 0.448807, 0.342890)
      rgb=(0.918109, 0.453603, 0.339529)
      rgb=(0.920714, 0.458417, 0.336166)
      rgb=(0.923287, 0.463251, 0.332801)
      rgb=(0.925825, 0.468103, 0.329435)
      rgb=(0.928329, 0.472975, 0.326067)
      rgb=(0.930798, 0.477867, 0.322697)
      rgb=(0.933232, 0.482780, 0.319325)
      rgb=(0.935630, 0.487712, 0.315952)
      rgb=(0.937990, 0.492667, 0.312575)
      rgb=(0.940313, 0.497642, 0.309197)
      rgb=(0.942598, 0.502639, 0.305816)
      rgb=(0.944844, 0.507658, 0.302433)
      rgb=(0.947051, 0.512699, 0.299049)
      rgb=(0.949217, 0.517763, 0.295662)
      rgb=(0.951344, 0.522850, 0.292275)
      rgb=(0.953428, 0.527960, 0.288883)
      rgb=(0.955470, 0.533093, 0.285490)
      rgb=(0.957469, 0.538250, 0.282096)
      rgb=(0.959424, 0.543431, 0.278701)
      rgb=(0.961336, 0.548636, 0.275305)
      rgb=(0.963203, 0.553865, 0.271909)
      rgb=(0.965024, 0.559118, 0.268513)
      rgb=(0.966798, 0.564396, 0.265118)
      rgb=(0.968526, 0.569700, 0.261721)
      rgb=(0.970205, 0.575028, 0.258325)
      rgb=(0.971835, 0.580382, 0.254931)
      rgb=(0.973416, 0.585761, 0.251540)
      rgb=(0.974947, 0.591165, 0.248151)
      rgb=(0.976428, 0.596595, 0.244767)
      rgb=(0.977856, 0.602051, 0.241387)
      rgb=(0.979233, 0.607532, 0.238013)
      rgb=(0.980556, 0.613039, 0.234646)
      rgb=(0.981826, 0.618572, 0.231287)
      rgb=(0.983041, 0.624131, 0.227937)
      rgb=(0.984199, 0.629718, 0.224595)
      rgb=(0.985301, 0.635330, 0.221265)
      rgb=(0.986345, 0.640969, 0.217948)
      rgb=(0.987332, 0.646633, 0.214648)
      rgb=(0.988260, 0.652325, 0.211364)
      rgb=(0.989128, 0.658043, 0.208100)
      rgb=(0.989935, 0.663787, 0.204859)
      rgb=(0.990681, 0.669558, 0.201642)
      rgb=(0.991365, 0.675355, 0.198453)
      rgb=(0.991985, 0.681179, 0.195295)
      rgb=(0.992541, 0.687030, 0.192170)
      rgb=(0.993032, 0.692907, 0.189084)
      rgb=(0.993456, 0.698810, 0.186041)
      rgb=(0.993814, 0.704741, 0.183043)
      rgb=(0.994103, 0.710698, 0.180097)
      rgb=(0.994324, 0.716681, 0.177208)
      rgb=(0.994474, 0.722691, 0.174381)
      rgb=(0.994553, 0.728728, 0.171622)
      rgb=(0.994561, 0.734791, 0.168938)
      rgb=(0.994495, 0.740880, 0.166335)
      rgb=(0.994355, 0.746995, 0.163821)
      rgb=(0.994141, 0.753137, 0.161404)
      rgb=(0.993851, 0.759304, 0.159092)
      rgb=(0.993482, 0.765499, 0.156891)
      rgb=(0.993033, 0.771720, 0.154808)
      rgb=(0.992505, 0.777967, 0.152855)
      rgb=(0.991897, 0.784239, 0.151042)
      rgb=(0.991209, 0.790537, 0.149377)
      rgb=(0.990439, 0.796859, 0.147870)
      rgb=(0.989587, 0.803205, 0.146529)
      rgb=(0.988648, 0.809579, 0.145357)
      rgb=(0.987621, 0.815978, 0.144363)
      rgb=(0.986509, 0.822401, 0.143557)
      rgb=(0.985314, 0.828846, 0.142945)
      rgb=(0.984031, 0.835315, 0.142528)
      rgb=(0.982653, 0.841812, 0.142303)
      rgb=(0.981190, 0.848329, 0.142279)
      rgb=(0.979644, 0.854866, 0.142453)
      rgb=(0.977995, 0.861432, 0.142808)
      rgb=(0.976265, 0.868016, 0.143351)
      rgb=(0.974443, 0.874622, 0.144061)
      rgb=(0.972530, 0.881250, 0.144923)
      rgb=(0.970533, 0.887896, 0.145919)
      rgb=(0.968443, 0.894564, 0.147014)
      rgb=(0.966271, 0.901249, 0.148180)
      rgb=(0.964021, 0.907950, 0.149370)
      rgb=(0.961681, 0.914672, 0.150520)
      rgb=(0.959276, 0.921407, 0.151566)
      rgb=(0.956808, 0.928152, 0.152409)
      rgb=(0.954287, 0.934908, 0.152921)
      rgb=(0.951726, 0.941671, 0.152925)
      rgb=(0.949151, 0.948435, 0.152178)
      rgb=(0.946602, 0.955190, 0.150328)
      rgb=(0.944152, 0.961916, 0.146861)
      rgb=(0.941896, 0.968590, 0.140956)
      rgb=(0.940015, 0.975158, 0.131326)
    }
}

\journal{Computers and Mathematics with Applications}

\begin{document}

{\begin{frontmatter}



\title{FLEXI: A high order discontinuous Galerkin framework for hyperbolic-parabolic conservation laws}


\cortext[cor1]{Corresponding author}

\address[stuttgart]{Institute of Aerodynamics and Gasdynamics, University of Stuttgart}
\address[koeln]{Mathematical Institute, Center for Data and Simulation Science (CDS), University of Cologne}
\address[muenchen]{Max Planck Institute for Plasma Physics}

\author[stuttgart]{Nico Krais\corref{cor1}}
\ead{krais@iag.uni-stuttgart.de}
\author[stuttgart]{Andrea Beck}
\ead{beck@iag.uni-stuttgart.de}
\author[stuttgart]{Thomas Bolemann}
\ead{thomas.bolemann@daimler.com}
\author[stuttgart]{Hannes Frank}
\ead{hannes.frank@stihl.de}
\author[stuttgart]{David Flad}
\ead{david.g.flad@nasa.gov}
\author[koeln]{Gregor Gassner}
\ead{ggassner@uni-koeln.de}
\author[muenchen]{Florian Hindenlang}
\ead{florian.hindenlang@ipp.mpg.de}
\author[stuttgart]{Malte Hoffmann}
\ead{hoffmann@iag.uni-stuttgart.de}
\author[stuttgart]{Thomas Kuhn}
\ead{thomas.kuhn@iag.uni-stuttgart.de}
\author[stuttgart]{Matthias Sonntag}
\ead{matthias.sonntag@zeiss.com}
\author[stuttgart]{Claus-Dieter Munz}
\ead{munz@iag.uni-stuttgart.de}

\begin{abstract}

High order (HO) schemes are attractive candidates for the numerical solution of multiscale problems occurring in fluid dynamics and
related disciplines. Among the HO discretization variants, discontinuous Galerkin schemes offer a collection of advantageous
features which have lead to a strong increase in interest in them and related formulations in the last decade. The methods have
matured sufficiently to be of practical use for a range of problems, for example in direct numerical and large eddy
simulation of turbulence. However, in order to take full advantage of the potential benefits of these methods, all steps in
the simulation chain must be designed and executed with HO in mind. Especially in this area, many commercially available closed-source solutions fall short. In this work, we therefor present the \Flexi framework, a HO consistent, open-source
simulation tool chain for solving the compressible Navier-Stokes equations in a high performance computing setting. We
describe the numerical algorithms and implementation details and give an overview of the features and capabilities of all
parts of the framework. Beyond these technical details, we also discuss the important, but often overlooked issues of code
stability, reproducibility and user-friendliness. The benefits gained by developing an open-source framework are
discussed, with a particular focus on usability for the open-source community. We close with sample applications that
demonstrate the wide range of use cases and the expandability of \Flexi and an overview of current and future developments. 

\end{abstract}



\begin{keyword}

discontinuous Galerkin \sep high order \sep large eddy simulation \sep computational fluid dynamics \sep open-source software \sep
conservation laws \sep shock capturing \sep finite volume



\end{keyword}

\end{frontmatter}}



%%%%%%%%%%%%%%%%%%%%%%%%%%%%%% ------ INTRO ----- %%%%%%%%%%%%%%%%%%%%%%%%%%%%%%
\section{Introduction}
\label{sec:intro}
The favorable properties of high order (HO) methods for smooth flow problems have long been recognized. Initially, \emph{global}
spectral methods, where the solution in the computational domain is discretized by a single, global set of basis functions were
favored. One of the earliest applications of a spectral Galerkin method to flow problems was published by Silberman in
1954~\cite{silberman1954planetary}. Many others followed, in particular in the area of turbulence using Fourier (pseudo)-spectral
methods, see e.g.~\cite{orszag1969numerical,dahlburg1990pseudospectral,yokokawa200216}. While these methods can offer unsurpassed
accuracy per invested degree of freedom, they are naturally limited by their restriction to the global, i.e. single element, domain
discretization. This makes them too inflexible for real word applications, and this shortcoming has sparked a lot of research
efforts into \emph{local} high order methods during the last decades. From these efforts, high order variants of essentially every
grid-based approximation method to partial differential equations (PDE) have sprung - there exist a plethora of finite difference,
finite volume, finite element and other high order formulations nowadays, see e.g.
~\cite{dumbser2007arbitrary,shu2003high,svard2007stable,maday1989spectral,bassi1997high,cockburn2012discontinuous} and many more. 
What all these methods have in common is that they
surrender the notion of a global solution representation in favor of a local but high order one. This locality entails that the
differential operators acting on the solution themselves now can become local, which in turn allows a multi-element
domain discretization. Since on a conceptual level, these HO methods differ in the way they define the approximate solution ansatz, the 
locality of their bases and the mathematical principle that recovers the global approximate solution through connection of the now local PDE
approximations also vary. These differences do not only influence the numerical properties of the resulting schemes, but have also
far-reaching consequences with regards to practical aspects: Which meshes are suitable for which method? How are the physical
properties and limitations reflected in the design of the method? How computationally efficient is the algorithm; can it exploit
modern vectorization in CPUs and what are its limiting components? How well can the method be run in parallel on tens or hundreds of
thousand of CPUs or on multiple GPUs? In particular the last two issues are of importance when considering practical applications of
HO methods beyond canonical test cases, most of which stem from the global method community. The multitude of questions and choices
to consider should also make it clear that no ``best'' HO method exists - instead, it depends on the specific requirements of the
targeted applications at hand.

In our case and for the framework presented here, these applications are found in the realm of the unsteady, compressible
Navier-Stokes equations and derived formulations like the Euler system. The particular focus of interest are non-linear multiscale
problems like turbulence and aeroacoustics in the compressible flow regime. For these types of problems, discontinuous Galerkin (DG)
methods offer a collection of advantageous features. They are based on the variational form of the equations realized by an $L_2$
Galerkin projection. The polynomial basis and test functions are both restricted to each element, and thus the projection becomes a
purely local operator. Connection to adjacent elements is achieved by the weak imposition of a unique surface flux. Thus, it is
convenient to interpret DG as a hybrid of an element-local finite element scheme and a finite volume (FV) discretization. This view
on DG methods has direct consequences for the approximation properties, the stability and the data connectivity of these methods.
Due to their local high order accuracy, the global method is also high order accurate and has very low numerical dispersion and
dissipation errors~\cite{gassner2011comparison}. The possibility to introduce upwinding through the surface fluxes provides
stability for linear hyperbolic problems, and naturally adds stabilizing dissipation in the non-linear case. The connection to
neighboring elements is only through weakly imposed numerical fluxes at the interfaces, which avoids strong global coupling and
reduces the communication pattern to that of first order finite volume schemes, while achieving arbitrary high order of accuracy
through the local ansatz degree. 

With a certain target application in mind, choosing the most suitable HO method to solve the governing equations is however only the
first choice along the way. Any meaningful application of such methods involves making certain that the whole simulation chain is up
to the task. In other words, choosing a HO method entails HO at all stages of the process, starting from the pre-processing steps,
in particular the generation of grids suitable for high order (both in terms of surface representations as well as sufficient grid
coarseness). Other issues to consider at pre-processing for example include the availability of high order accurate initial and
boundary conditions.  Post-processing of the solution also poses challenges. For the typical target application such as highly
resolved and time-accurate simulation of turbulence, large amounts of data are generated during the simulation, which must be dumped
to disk in a raw format to avoid efficiency losses. This raw data has to be post-processed, manipulated and visualized in a way that
is also consistent with the HO approach and is compatible with the usually low order data format used by third party software.

Since a successful HO strategy thus requires a concerted approach across all
links of the simulation chain and as suitable third party pre- and post-processing solutions are not available, we designed and
developed (and continue to do so) the \Flexi\footnote{www.flexi-project.org} framework, which will be discussed and presented in this work. The basic idea behind
\Flexi is not only to provide a HO DG solver for the compressible Navier-Stokes equations that is well-documented and easy to modify
in the hopes others may find it useful, but to create a full, HO consistent simulation tool chain with a focus on high performance
computing (HPC) in mind. Thus, all parts of \Flexi are designed to deal with HO data, and to  make efficient use of a large number of
cores. In order to make our work available to the scientific community and to facilitate peer-review and reproducibility, \Flexi has
been made open-source under the GNU GPL v3.0 license. We will present \Flexi, its features, parts, numerical schemes and algorithms as well as selected applications in the following sections. 

In this work, we focus on applications of \Flexi in the field of computational fluid dynamics (CFD). In particular, scale-resolving and time-accurate
simulation methodologies such as large eddy simulation (LES) and direct numerical simulation (DNS) will be considered. But at its heart, \Flexi is a solver
for general hyperbolic-parabolic conservation laws. For instance, the solver \textit{Fluxo}\footnote{https://github.com/project-fluxo/} is a spin-off from \Flexi and is used for 
applications in astro- and plasmaphysics, solving the equations of magnetohydrodynamics including non-conservative terms. Its main
focus is on the split-form DG formulation (see Sec.~\ref{sec:splitDG}) and entropy stability~\cite{fluxo_esmhd}.

%%%%%%%%%%%%%%%%%%%%%%%%%%%%%% ------ FRAMEWORK ----- %%%%%%%%%%%%%%%%%%%%%%%%%%%%%%
\section{Framework overview}
\begin{figure}
\begin{center}
	\includegraphics[width=\textwidth]{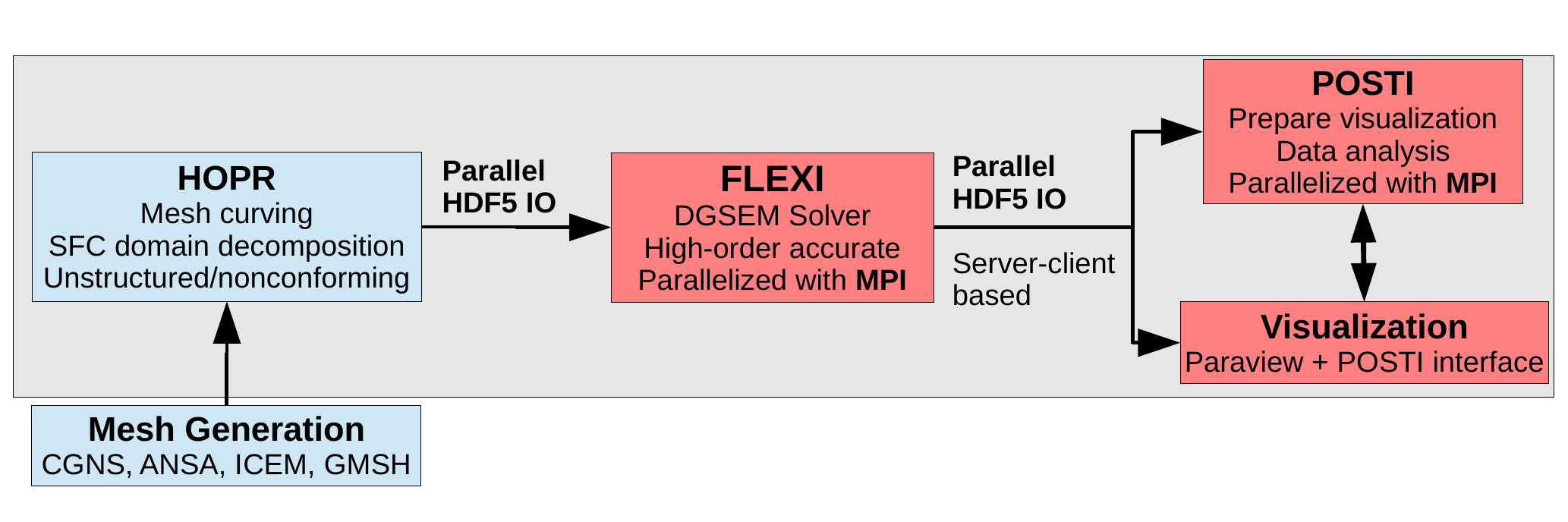}
	\caption{\label{fig:framework} Overview of the high order framework. 
	}
\end{center}
\end{figure}

The overall \Flexi framework consists of open-source tools for pre-processing and generation of high order meshes, the CFD solver
itself and a post-processing and visualization suite, see Fig.~\ref{fig:framework}. In the following sections, we will discuss each
of the parts and give an overview over their methods and features.

% !!!!!!!!! Hopr
\subsection{HOPR: Mesh generation tool}
\label{sec:hopr}

\newcommand\ANSA{$\text{ANSA}^{\!\text{\tiny\copyright}}$}

In this section, we give an overview of the high order mesh preprocessor \Hopr, which generates the mesh input file for \Flexi. Documentation and tutorials for \Hopr can be found under \texttt{www.hopr-project.org} and details of the mesh curving techniques are given in \cite{diss_hindenlang,hindenlang2015mesh}.

When using high order methods, wall boundary conditions at curved geometries need a high order representation to maintain the high order
accuracy of the method \cite{BRcurveds}. Especially CFD applications have complex geometries and pose high requirements for the quality of 
geometry representation. While the open-source software GMSH \cite{gmsh2009} is constantly improving the generation of high order meshes, 
it is still a topic of ongoing research \cite{toulorge_untangle2013,fortunato_persson_JCP2016,marcon_kopriva_JCP2019}. 
In \Hopr, the main approach  is to rely on  linear meshes from standard grid generators and give the user different techniques 
to curve the mesh. The curved mesh geometry is represented using polynomials of degree $N_{geo}$, which can be chosen depending
on the considered case.

\begin{figure}[!htp]
 	\centering
 	\includegraphics[trim= 0 0 0 0,clip,width=0.6\linewidth]{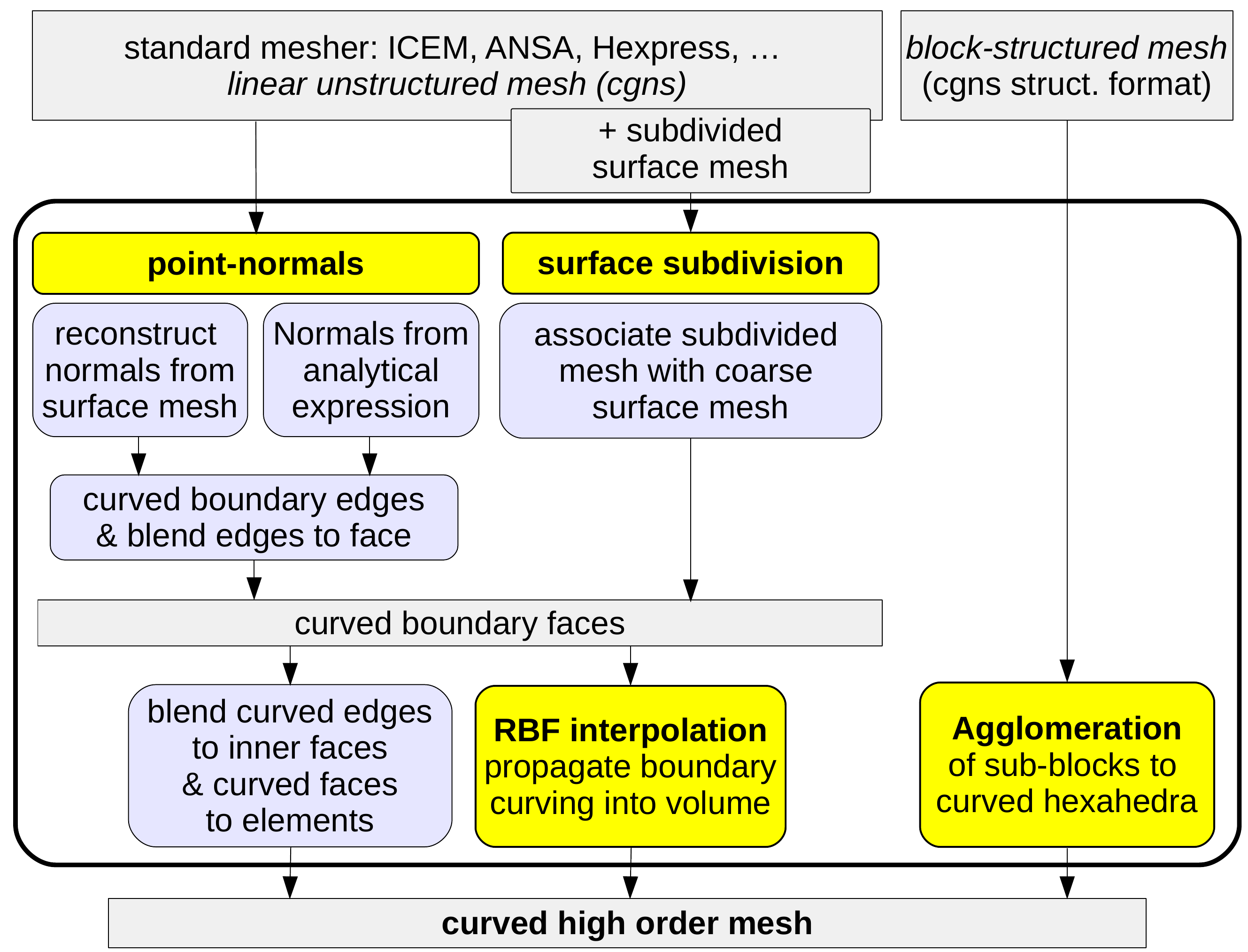}
	\caption{\label{fig:flow_mesh_gen} Flowchart of the curved mesh generation process in \Hopr.}
 \end{figure}

First, we present the high order mesh format that supports direct parallel read-in and meshes with curved non-conforming element interfaces.
Further, we discuss the main curving techniques that are summarized in the flowchart in Fig.~\ref{fig:flow_mesh_gen}. 

In \Hopr, simple multi-block meshes can also be generated internally, without the need to rely on external grid generators. 
Such meshes can including the use of periodic boundary conditions and non-conforming element interfaces. In addition, the mesh can be curved a-posteriori via user-defined mappings. 
A detailed explanation of these two features will not be given here, but can be found in the tutorials on the webpage.

\subsubsection{Mesh format}\label{sec:meshFormat}

\begin{figure}[ht]
\centering
\includegraphics[trim = 4 4 4 4,clip, width=0.48\textwidth] {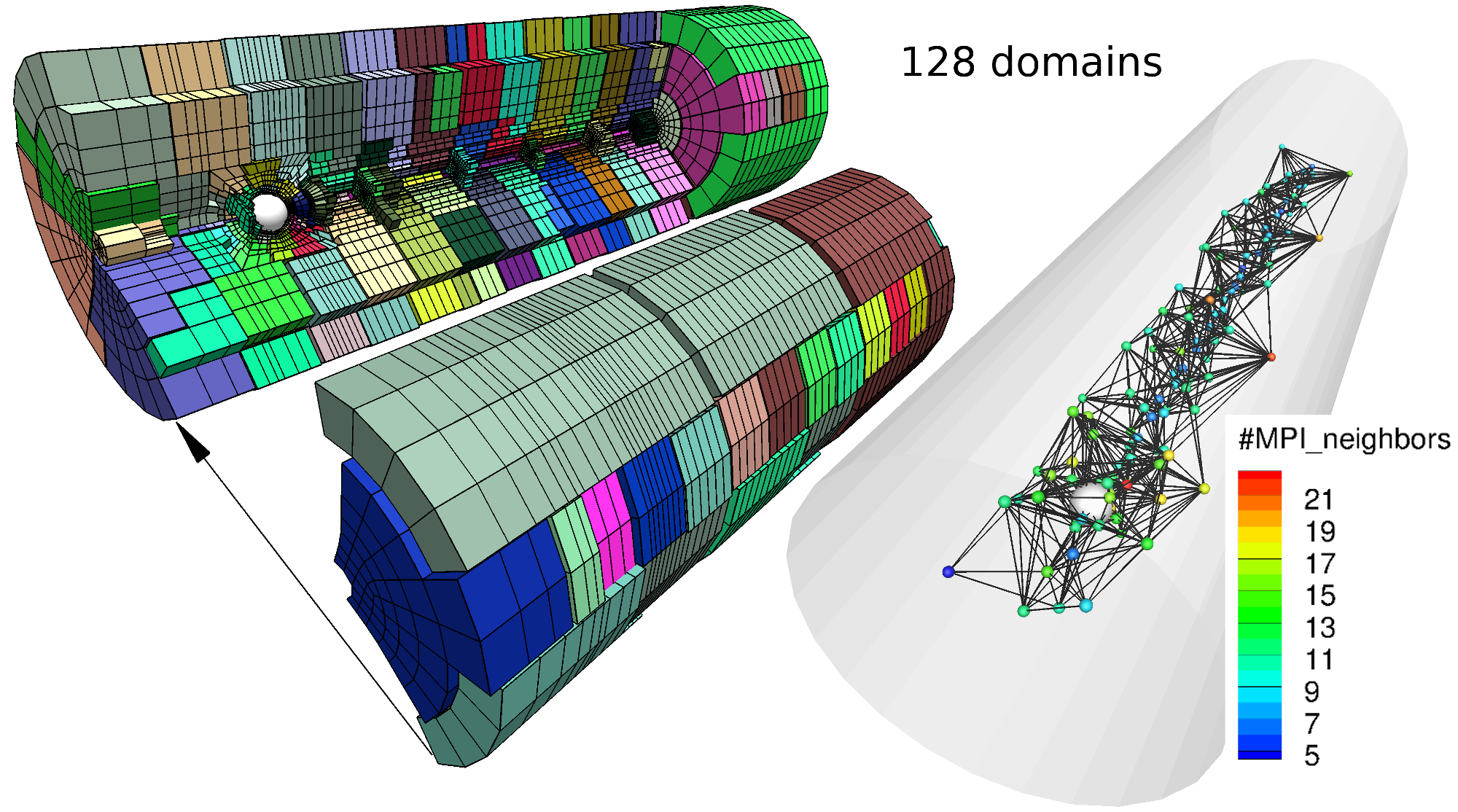}
\includegraphics[trim = 4 4 4 4,clip, width=0.48\textwidth] {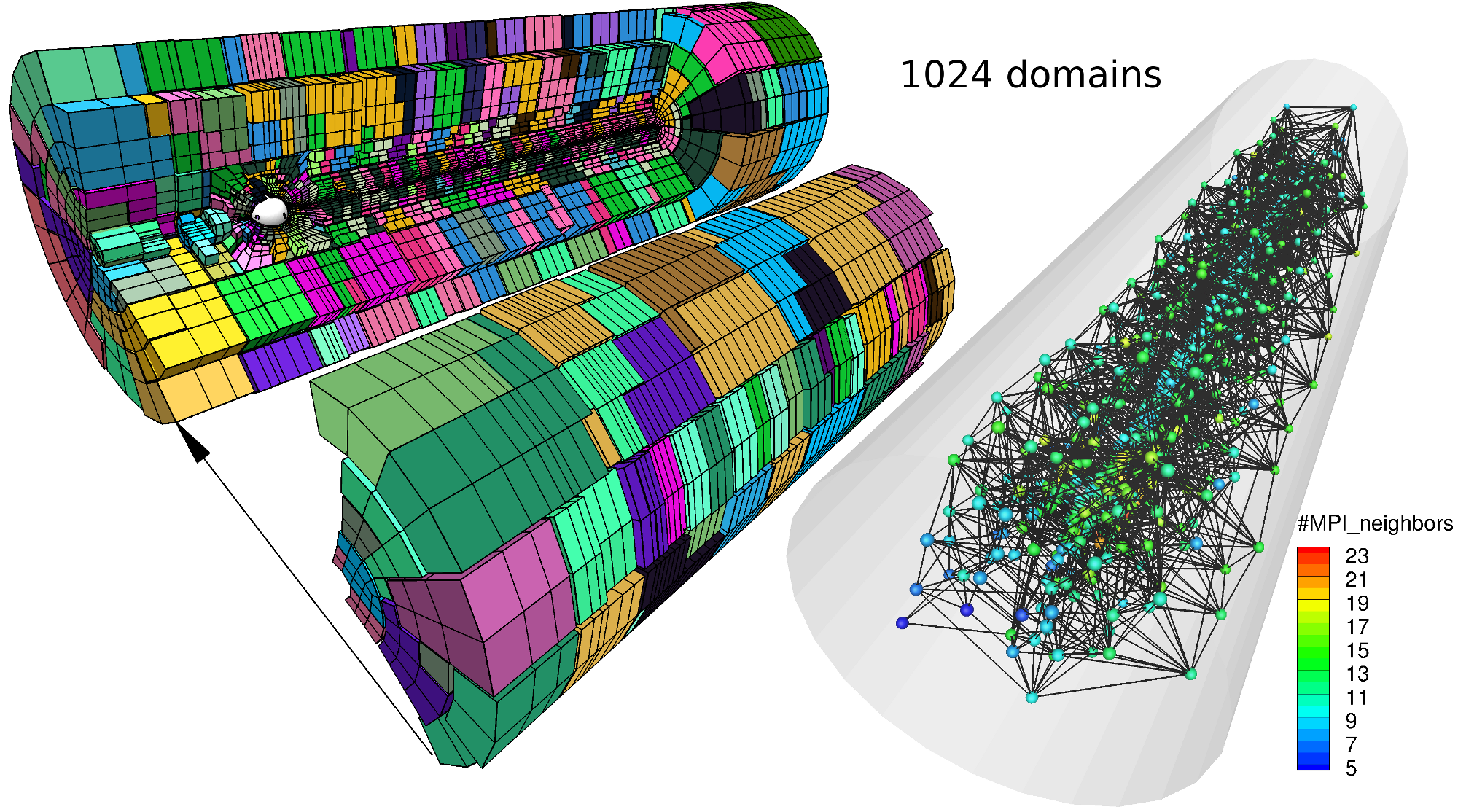}
\caption{Space-filling curve based domain decomposition of an unstructured mesh around a sphere, with communication graph between the domains, for $128$ domains of $165-166$ elements (left) and $1024$ domains of $20-21$ elements (right), from \cite{diss_hindenlang}.}
\label{fig:space-filling-curve}
\end{figure}

The mesh format in \Hopr is designed for fast parallel read-in in \Flexi. Therefore, the mesh file is written in binary HDF5 format \cite{hdf5}.
An important feature is that the elements are ordered along a space-filling curve (SFC), providing a direct domain decomposition 
during parallel read-in. That means one can directly start the parallel computation with an arbitrary 
number of domains ($\geq$ number of elements) and always read the same  mesh file. The element list is simply 
divided by the number of domains, and each domain is associated with a contiguous range of elements. 
The SFC curve guarantees that these elements are, on average, in proximity of each other. 
An example is shown in Fig.~\ref{fig:space-filling-curve} for an unstructured mesh, with two different domain decompositions 
based on the same \Hopr mesh file. In \Flexi, the same element ordering is used to write the restart files.
 
In contrast to standard mesh formats, which provide the element node connectivity list and the node list separately, 
the \Hopr mesh format provides direct neighbor connectivity information of the element sides (conforming and non-conforming) 
and the element node information (index and $(x,y,z)$ position) as a package per element. Hence, for a given range of elements, 
non-overlapping contiguous parts of the arrays can be read from file in parallel. As a consequence, the coordinates of the 
same physical nodes are stored several times, but remain associated by a unique global node index.

\subsubsection{Surface curving}\label{sec:surfCurv}
\begin{figure}[ht]
\centering
\includegraphics[trim = 204 125 4 174,clip, width=0.43\textwidth] {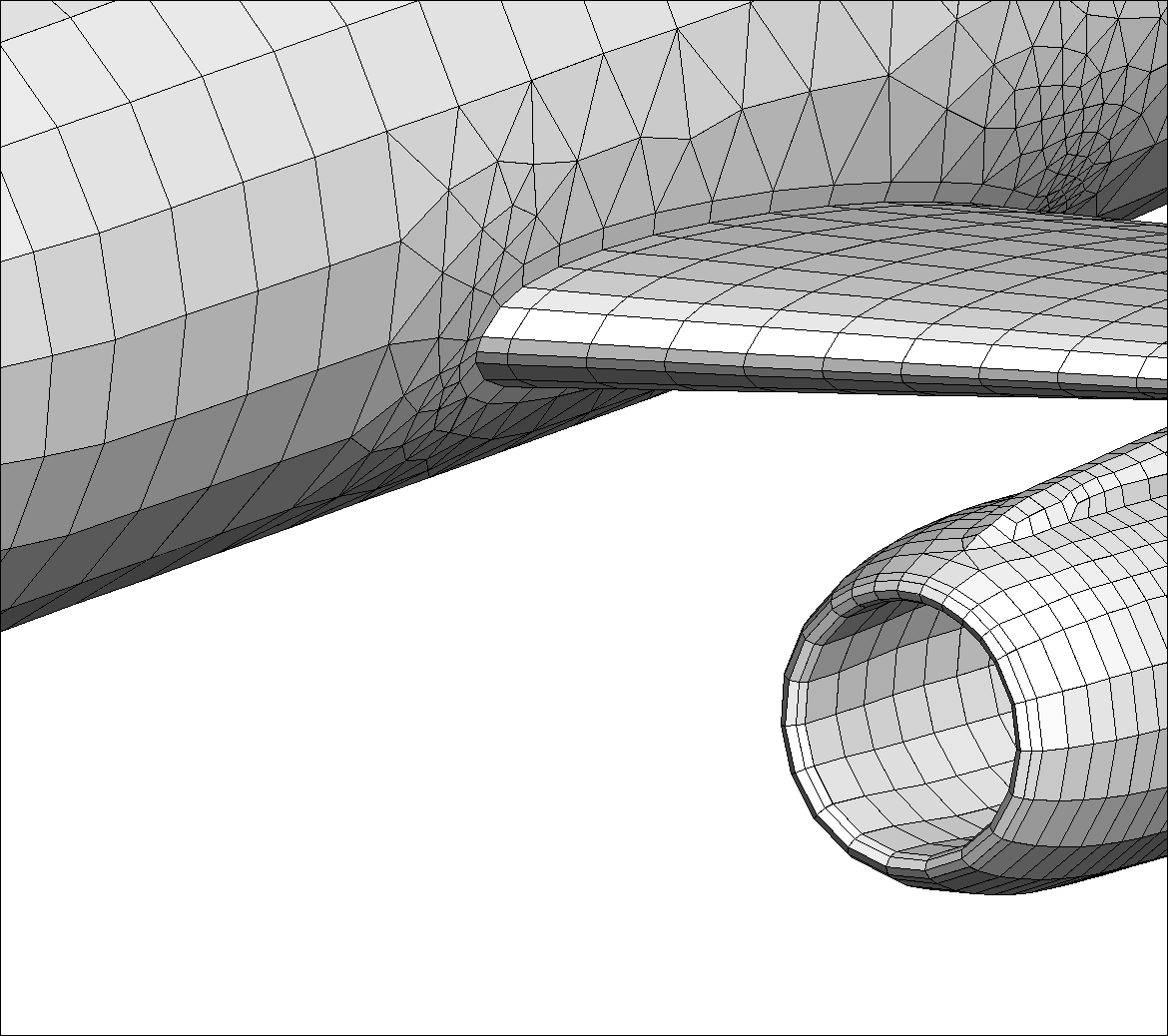}\hspace{0.02\textwidth}
\includegraphics[trim = 204 125 4 174,clip, width=0.43\textwidth] {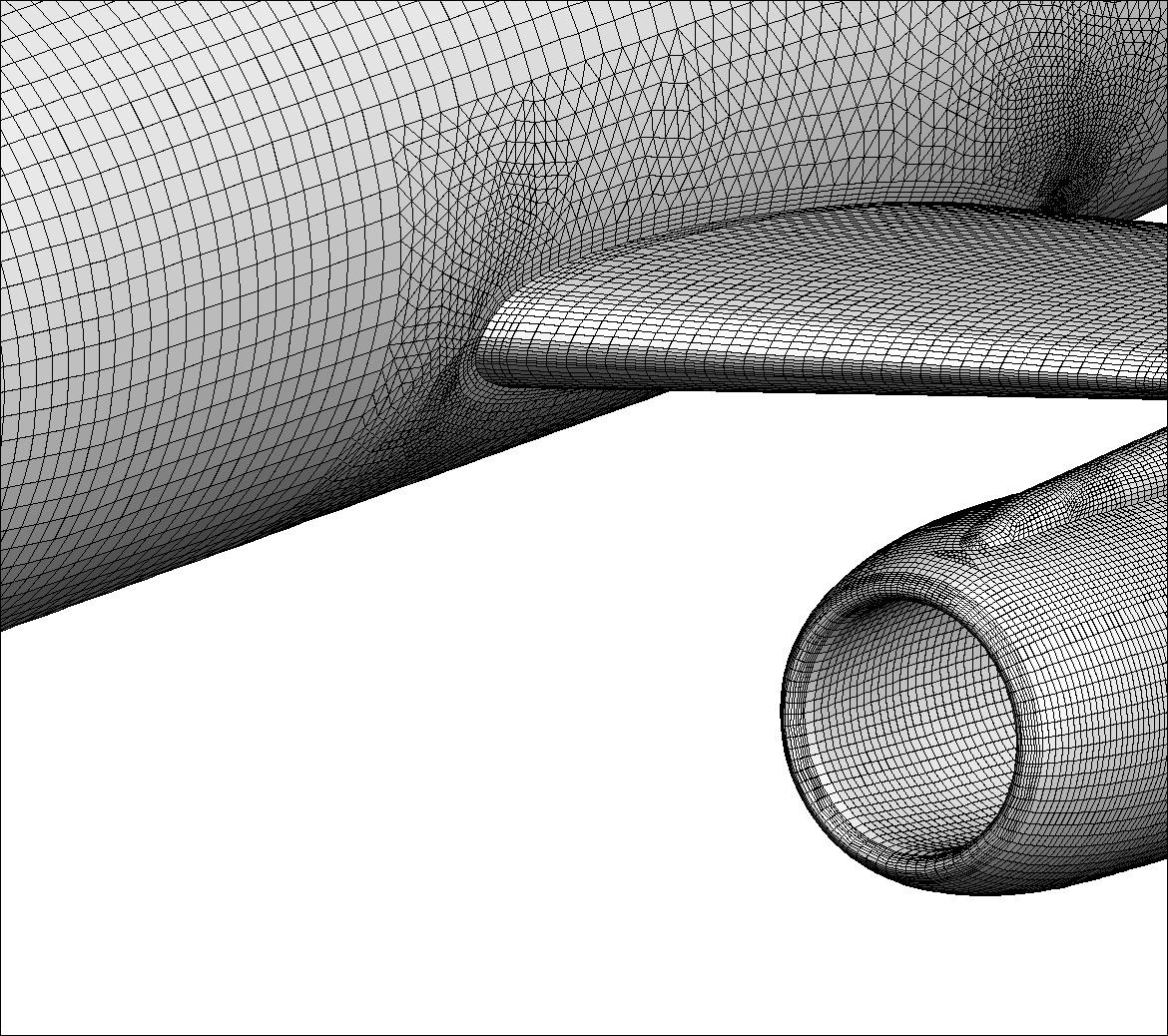}
\caption{Surface mesh of the DLR-F6 body-wing intersection, after two steps of isotropic refinement, from \cite{hindenlang2015mesh}.}
\label{fig:ansa_split}
\end{figure}

In \Hopr, there are two main strategies to curve the boundary surfaces, the 'surface-subdivision' and the 'point-normal' approach
\cite{hindenlang2015mesh}. The surface-subdivision approach is the most convenient and accurate, but it needs an additional refined surface mesh.
As shown in Fig.~\ref{fig:ansa_split}, a subdivided mesh was created with the mesh generator \ANSA, with the new points lying 
on the exact boundary surface, and the coarse mesh points being unchanged. 
Given the polynomial degree of the surface mapping ($\Ngeo=2,4,8$) that corresponds to the number of refinement steps, 
\Hopr finds automatically the connection between each boundary face and its refined surface elements by making use of 
the connectivity information of the subdivided surface mesh. The curved face is then defined by an interpolation polynomial 
through the new points.

The point-normal approach, proposed in \cite{FarinBook}, relies on the reconstruction of the curved surface using only normal vectors 
at the grid points of the boundary of the linear volume mesh. The surface elements are then $G^1$ continuous at the corner nodes. 
Clearly, the most difficult part is to provide the normal vector. In \Hopr, analytical expressions for simple geometric shapes are included,
and it is easy to add other user-defined functions. If no analytical expression is available, the normal vector is reconstructed 
from the linear surface mesh, with a grid-size dependent approximation error. We also need to consider sharp edges at the 
intersections of surfaces, where two point-normals exist. To find these edges, curved boundary patches can be associated with an unique index,
and a sharp edge is assumed between patches of different indices. 

\begin{figure}[ht]
\centering
\includegraphics[width=0.34\textwidth]{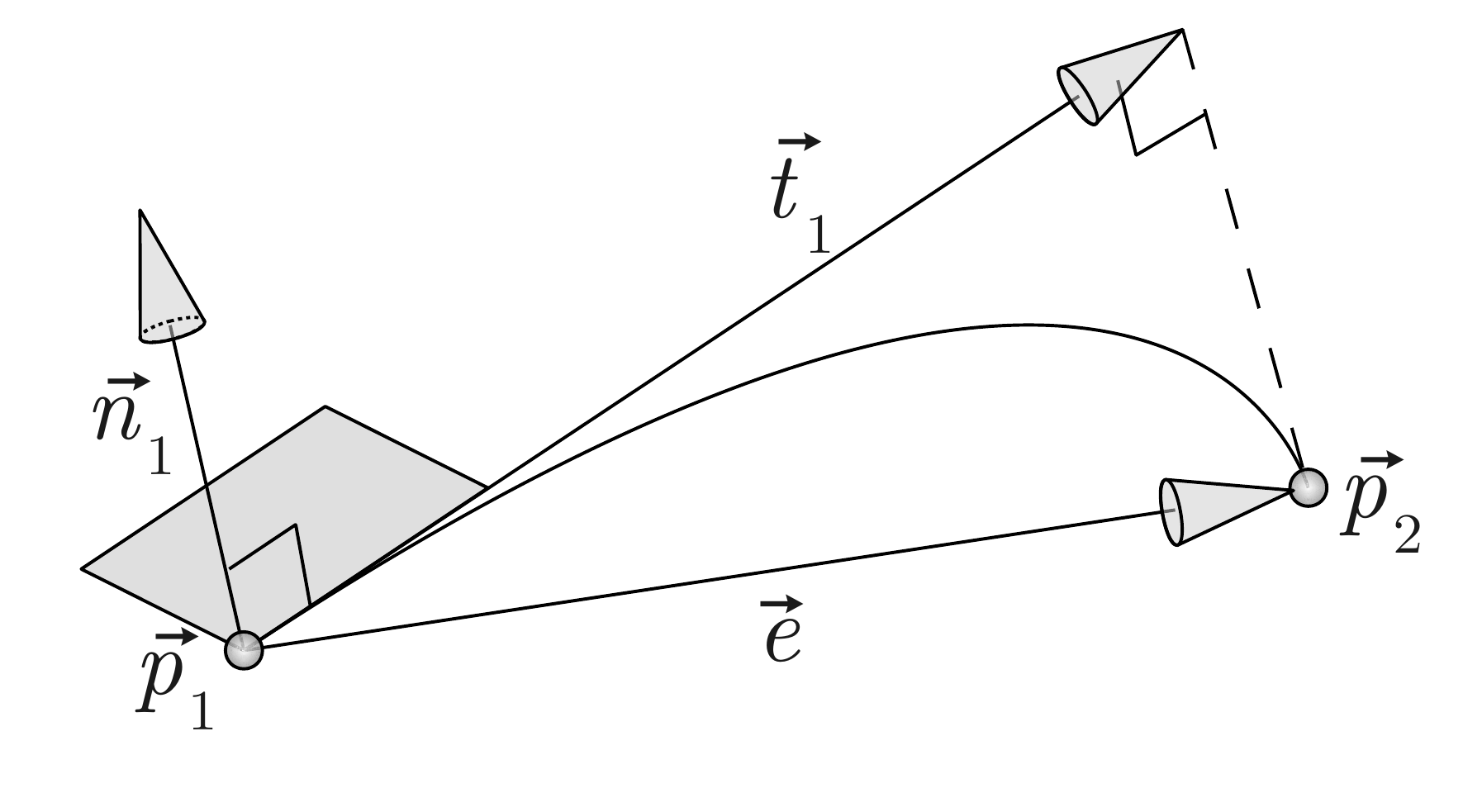} %\hspace{0.02\textwidth}
\includegraphics[width=0.64\textwidth]{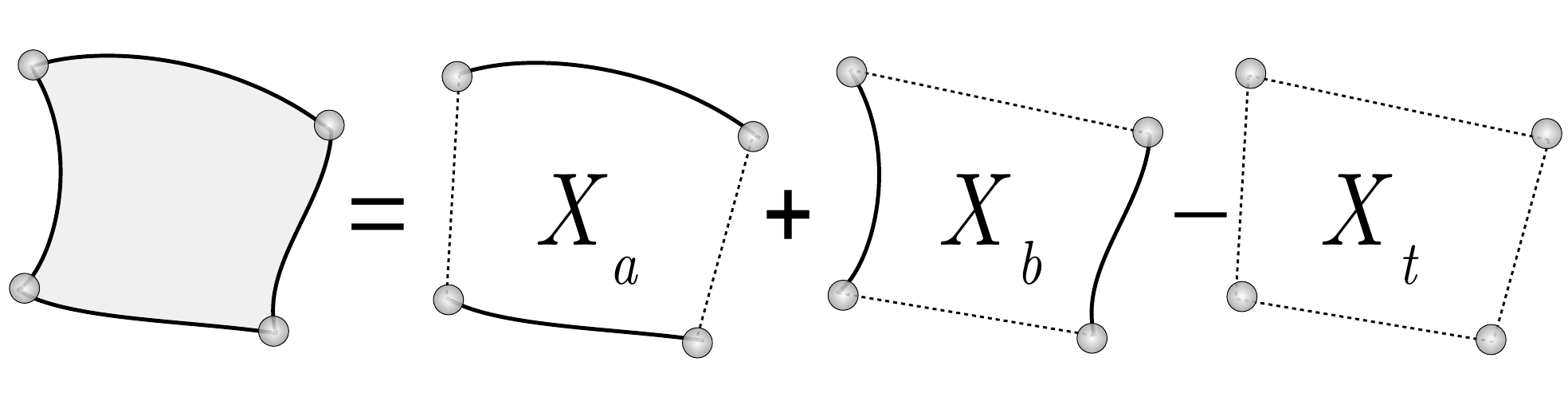}
\caption{Construction of curved edge tangential from point-normal vector (left) and blending of curved edges to curved surface (right), from \cite{hindenlang2015mesh}.}
\label{fig:normal_rec}
\end{figure}

The process to generate the curved faces from the normals is  depicted in Fig.~\ref{fig:normal_rec}. The tangential vectors of the face edges
are constructed first, by projecting the straight edge into the tangential plane. If two point-normals are given (at sharp edges), 
the direction of the tangential is found by a cross-product and its length again by projection onto the straight edge. 
The curved edge is a cubic polynomial, computed from the edge end points and the tangential vectors. In a following step, 
the curved faces are computed from a blending of the curved edges \cite{FarinCoons}.  

Once all boundary faces are curved, the 3D element mapping is computed using blending functions of curved edges and faces, 
for all elements that share at least one edge with a curved boundary face ('local curving'). If the elements are highly stretched,
additional curving of inner elements would be required %\cite{persson09,gmsh} 
('global curving') to avoid invalid cells, which is addressed in the next section.

\subsubsection{Volume curving using RBF interpolation}\label{sec:RBF}

\begin{figure}[!htp]
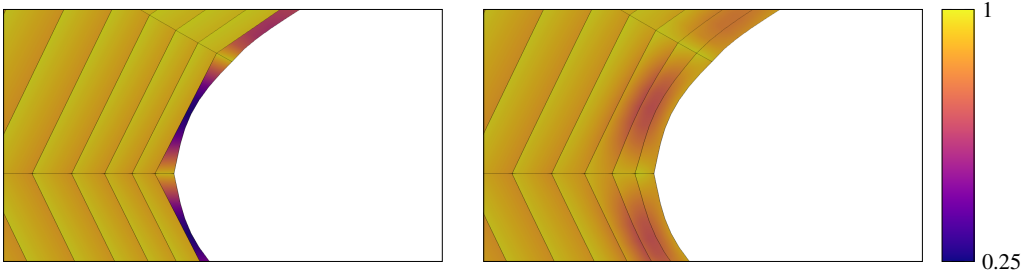

	\centering
	\includestandalone[width=1.0\textwidth]{./figures/RBFcurving}
	\caption{Visualization of mesh and scaled Jacobian around the leading edge of a NACA0012 airfoil. Left: Curved
	surface without volume curving, right: curved surface and volume curving based on RBF interpolation.}
	\label{fig:RBFcurving}
\end{figure}
The stategies presented in the last section allow to create curved boundary faces of
the domain and curve only the elements adjacent to the boundary. No information about volume curving is present in those approaches. 
This can lead to serious problems, as
demonstrated in the left plot in Fig.~\ref{fig:RBFcurving}. It shows the mesh around the leading edge of a NACA0012 airfoil, where
the surface of the airfoil has been curved, but the volume mesh is still linear. This leads to strongly deformed cells in the first
layer around the airfoil, especially since the cells are highly anisotropic as it is typical for boundary layer meshes. 
As a quantitative quality criterion, the mesh is colored by the scaled Jacobian, defined as the ratio between the Jacobian of the high
order mapping and its maximum value in each cell (see Sec.~\ref{subsec:dgsem} for details on that quantity). Lower values indicate larger
deformation, lowering the approximation quality in these regions. Since the physical boundaries are typically of high
interest and their representation affects the solution quality globally, a way to remedy this loss is necessary. In \Hopr, a possible way to
achieve this is based on the interpolation of the surface curving to the volume based on radial basis function (RBF) interpolation.
RBFs can in general be used to interpolate scattered data, and are for instance used to deform meshes in response to a moving
boundary~\cite{de2009radial}. In our application, we use them to interpolate the displacement between the linear and the curved
representation of the airfoil surface to the volume. This spreads the distortion across the first few layers of cells, dramatically
improving the mesh quality in the critical first layer. The effect of this can be seen by comparing the two plots in
Fig.~\ref{fig:RBFcurving}, where the mesh on the right was created with the RBF interpolation 
approach.

\subsubsection{Curved mesh by agglomeration}
The most simple way to generate a fully curved hexahedral mesh is by using block-structured meshes as input. 
The main requirement is that all blocks can be coarsened by a unique factor of $n_b$ elements in all directions. 
Then, the high order element mapping is simply found via interpolation of an agglomerated block of $(n_b\times n_b\times n_b)$ cells.
Note that the polynomial degree of the mapping can be chosen independently from $n_b$, with $\Ngeo\leq n_b$. 

 \begin{figure}[!htp]
 	\centering
 	\includegraphics[trim= 0 0 0 0,clip,width=0.5\linewidth]{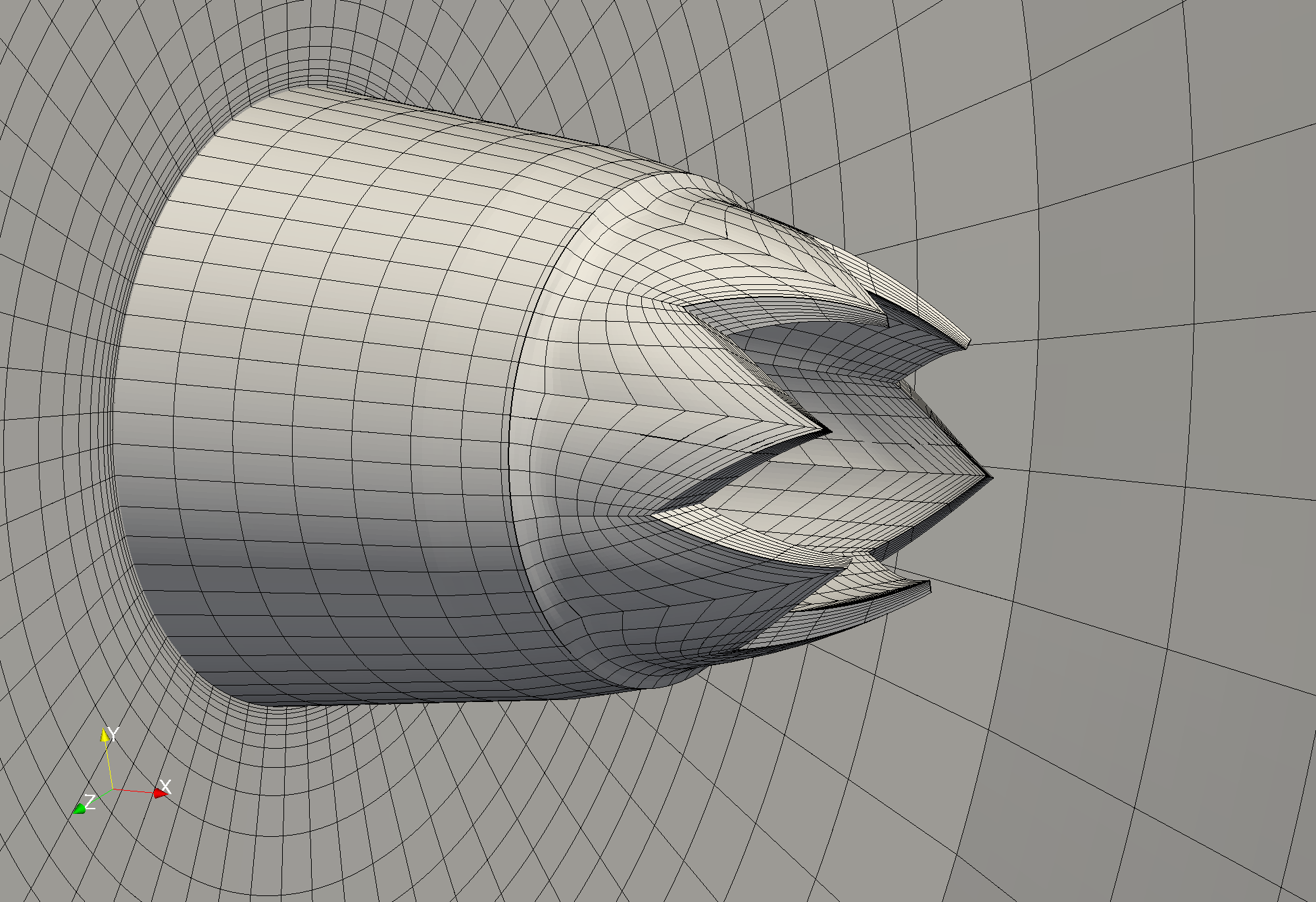}
 	\includegraphics[trim= 0 0 0 0,clip,width=0.36\linewidth]{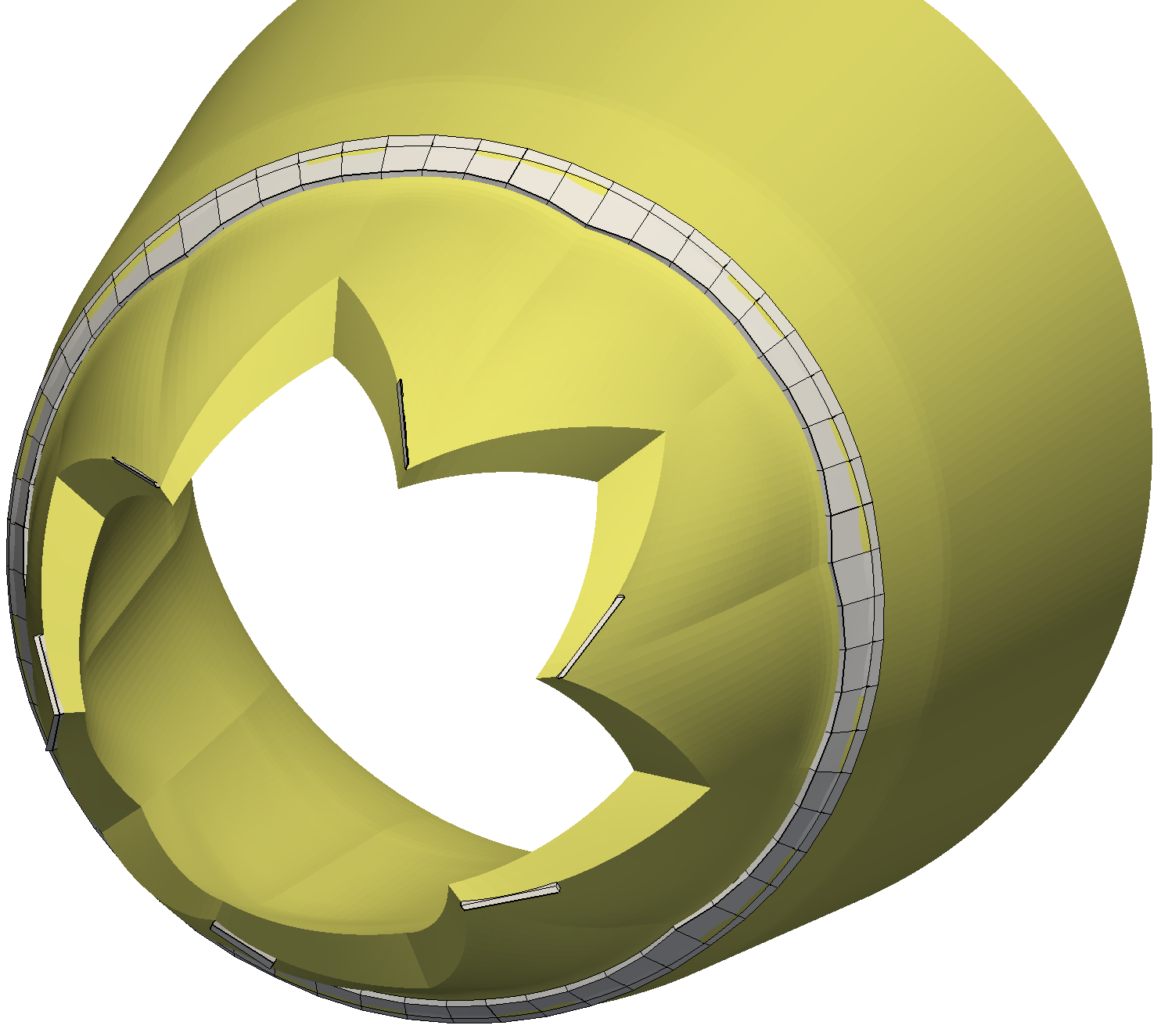}\\[0.5ex]
  	\includegraphics[trim= 0 0 0 0,clip,width=0.5\linewidth]{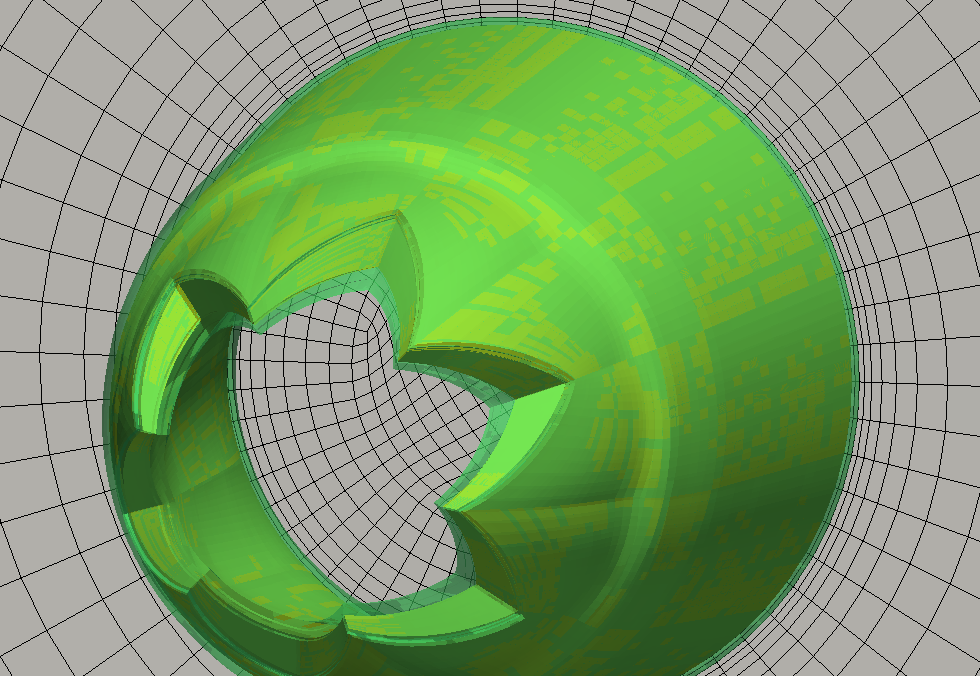} 
	\includegraphics[trim= 0 0 0 0,clip,width=0.36\linewidth]{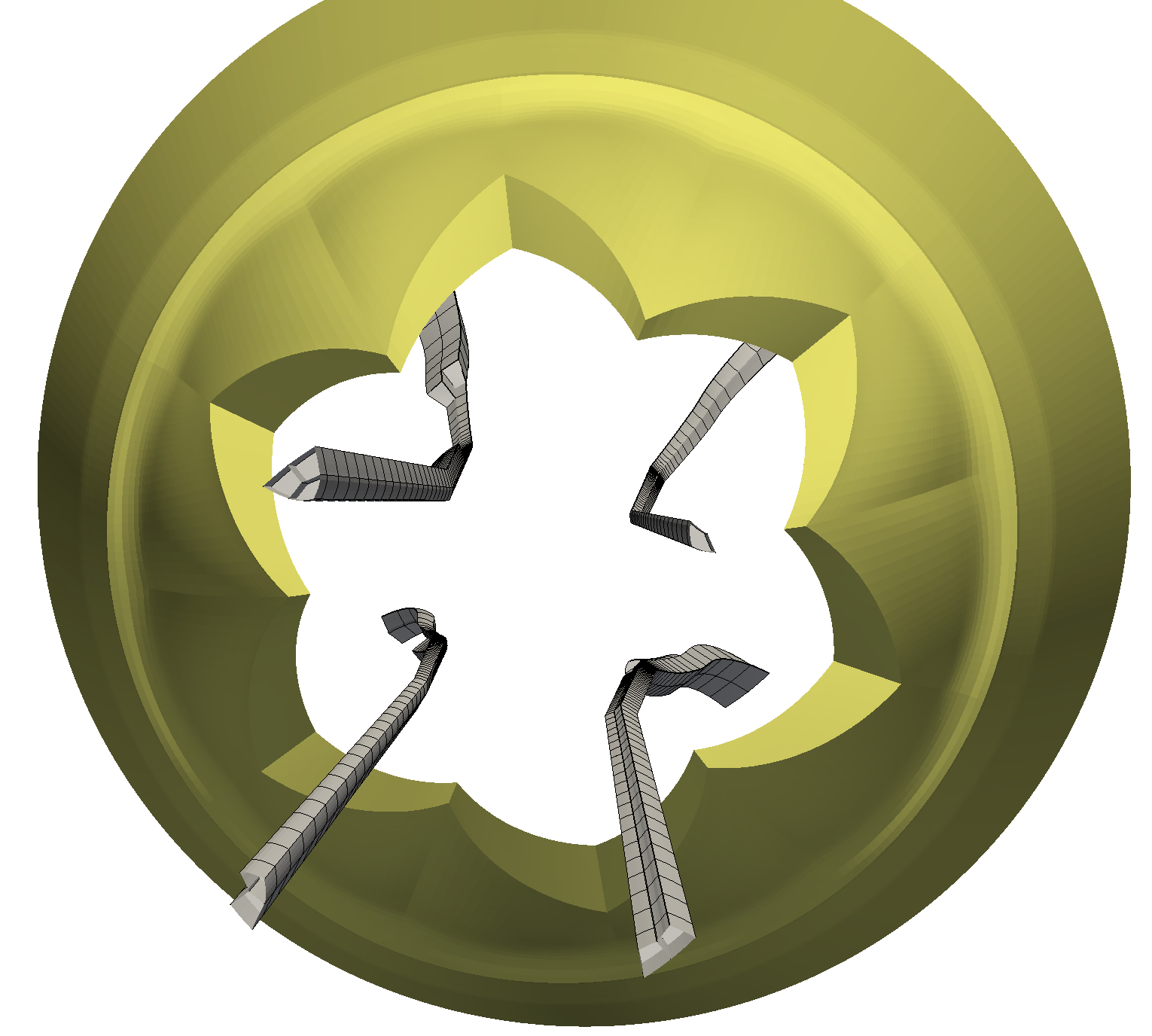}
	\caption{\label{fig:nozzle} Agglomeration of block-structured mesh of a serrated nozzle (courtesy of A. Cimpoero), with $\Ngeo=4$. Upper left: curved surface mesh, upper right: elements with $J_\text{scaled}<0.1$, if only boundary is curved, lower left: valid mesh, with $2$ layers of curved elements, lower right: elements with $J_\text{scaled}<0.1$, if all elements curved.     }
\end{figure}

A complex application example is shown in the upper left of Fig.\ref{fig:nozzle}. The original CGNS block-structured mesh has 85
blocks and a total of $20.6$ million linear elements. After agglomeration with $\Ngeo=n_b=4$, the high order mesh consisted of $322,110$ elements. 
 
There is no guarantee that the agglomerated mesh is a valid high order mesh. In \Hopr, we always provide the mesh statistics of the
scaled Jacobian, to inform the user about the mesh quality, and if 
invalid elements with a negative scaled Jacobian exist. 
In Tbl.~\ref{tab:agglo}, we show that the linear agglomerated mesh is valid ($n_b\!=\!4$, $\Ngeo\!=\!1$), but if all elements are curved, 
elements with small scaled Jacobian and even invalid elements exist. Therefore, an additional parameter controls to keep the high order mapping 
only in a certain number of element layers adjacent to the curved boundary, and all other elements become linear. If only the boundary faces 
or the first layer of elements is curved, the mesh is still invalid, but increasing the number of layers to $2-4$ leads to a valid mesh and 
an increased quality. In Fig.~\ref{fig:nozzle}, we show the  failed cases of only boundary curving or curving of all elements and the valid mesh 
with $2$ layers of curved elements.

\begin{table}[!htp]
\centering\small
\begin{tabular}{|l|c|c|c|c|c|}\hline
\#elements with $J_\text{scaled}$ & $<0$ & $0-0.1$ & $0.1-0.2$& $0.2-0.3$ & $\geq 0.3$ \\\hline \hline 
no curving        &        0   &     0 &    65 &   204 &  321,841 \\\hline
all curved        &\textbf{10} &   737 &   896 & 1,684 &  318,783 \\\hline
boundary only        &\textbf{61} &     6 &    70 &   216 & 321,757  \\\hline
$1$ layer         &\textbf{10} &    50 &   104 &   376 & 321,570  \\\hline  
$2$ layers        &        0   &     0 &   197 &   520 & 321,393  \\\hline 
$3$ layers        &  0         &     0 &   178 &   667 & 321,265  \\\hline 
$4$ layers        &  0         &     0 &   160 &   746 & 321,204  \\\hline 
\end{tabular}
	\caption{Scaled Jacobian distribution between all elements of the agglomerated serrated nozzle mesh, for different settings. Invalid elements have $J_\text{scaled}<0$. }
\label{tab:agglo}
\end{table}

% !!!!!!!!! Flexi
\subsection{FLEXI: DG solver}
\Flexi is a general purpose solver for hyperbolic/parabolic systems following the method of lines, where the spatial operator is
treated with a discontinuous Galerkin spectral element method (DGSEM), optionally combined with a subcell finite volume method to
capture regions with shocks. High order explicit Runge-Kutta methods are used to integrate the semi-discrete DGSEM in time. Explicit
time-integration in combination with the spectral element version of DG leads to a very efficient scheme that is well suited for
massively parallel computing hardware. Details on the numerical schemes, methods and implementations will be given in
Sec.~\ref{sec:numericalmethods},~\ref{sec:shockcapturing} and ~\ref{sec:implementation}. In this section, we will give an overview
of its features.

% !!!!!!!!! Code design
\subsubsection{Code design}
In the context of this work, we present \Flexi with a focus on CFD, in particular for solving the unsteady, compressible \NSE. At
its core however, \Flexi is intended to be a general purpose solver for hyperbolic-parabolic systems of partial differential
equations. This goal requires a design with a focus on modularity, i.e. a collection of features or methods in functional units,
which can be used and modified largely independently from each other. \Flexi and the framework are predominantly written in modern,
procedural Fortran for its suitability and performance for scientific computing in HPC environments. In Fortran, code flexibility
can be achieved by arranging functional units in modules, each with their own encapsulated set of variables. Within this concept,
five major functional units exist in our solver: 
\begin{enumerate}
	\item The spatial operator, including the DG and FV discretization schemes 
	\item The temporal operator
	\item Mesh preparation and handling
	\item The equation system to be solved
	\item Analysis and output routines
\end{enumerate}
The inter-dependencies of these functional units are kept to a minimum. For example, the temporal integrator is essentially an ODE
solver completely independent from all other units. The spatial discretization operators are written for general hyperbolic and
parabolic terms, and the specific equation system to be solved can be added in a ``plug and play'' style, as all code parts specific
to the equation system (e.g. fluxes, time step restrictions, initial conditions, numerical flux functions) are organized in a single,
interchangeable functional unit. In order to provide custom features specific to a user-defined case, a test case environment can be
defined, which injects information at specific parts of the code. For example, within this test case, extra analysis routines or
source terms can be defined, which are then invoked automatically.

% !!!!!!!!! Parallelization
\subsubsection{Parallelization}
\Flexi and its associated tools use a distributed memory parallelization based on the Message Passing Interface (MPI) standard. In
this paradigm, each parallel process manages its own memory, and information interchange between processes occurs through messages
being sent over the network only. Like for all element-based discretizations, a single grid cell is a natural conceptual, granular
unit for parallelization, i.e. all domain decomposition is element-based. Due to the local approximation space in DG and the absence
of continuity requirements on the solution across element interfaces, the communication pattern of DG schemes is side-based. Thus,
for a computation in a domain of dimensionality $d$, only information of dimensionality $d-1$ has to be exchanged. Furthermore
information only has to be exchanged between neighbours that share a face. In a communication sense, DG is thus comparable to FV
schemes, with the important differences being the amount of data being sent (for high order DG) and the absence of volume work in
FV, where the evaluation of the numerical flux requires the overwhelming amount of computational effort. In DG, where additional
volume work exists, these local computations can be used for communication latency hiding. This high density of local operations and
the small communication footprint make DG operators naturally parallel and allow for a rather simple parallelization of
explicit-in-time DG schemes. 

\begin{figure}
 	\centering
 	\includegraphics{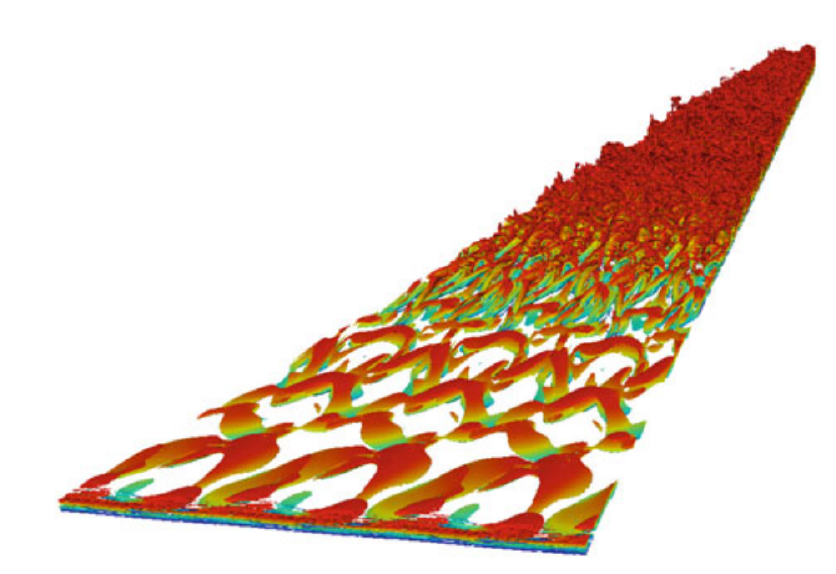}
	\caption{\label{fig:tbl}Spatially-developing supersonic turbulent boundary layer up to $Re_\theta=3878$ at $Ma=2.67$,
	 computed with 1.458 billion DOF per variable. $\lambda_2$-Visualization of the turbulent structures along the flat plate
	 colored by the streamwise velocity component $u$~\cite{10.1007/978-3-319-20340-9_14}.}
 \end{figure}

In \Flexi, we use the MPI concept of non-blocking communication extensively, which allows initiating a message transfer at the
earliest possible instance but does not require immediate execution of this command. I/O of solution data is achieved by use of the
\emph{HDF5} library~\cite{hdf5}, where we additionally gather operations over a compute node to reduce file access for large scale
computations. With the help of these features, simulations with over 1 billion degrees of freedom (DOF) on 100,000 cores and perfect strong scaling
performance are possible. Fig.~\ref{fig:tbl} shows a sample computation with \Flexi: a DNS of a supersonic turbulent boundary layer,
reaching a momentum thickness based Reynolds number $Re_\theta=3878$, computed with close to 1.5 billion DOF per solution
variable~\cite{10.1007/978-3-319-20340-9_14}. 

% !!!!!!!!! Code features
\subsubsection{Code features}
\Flexi contains a number of features, most of which are focused towards stability in underresolved scenarios and analytics for
complex flows e.g. from direct aeroacoustic simulations. The most important ones can be grouped as follows:
\begin{enumerate}
	\item Hybrid DG / FV operator: The spatial discretization in \Flexi is based on an element-wise formulation of the DG
		method of arbitrary order. As discussed before, high order schemes have favorable properties in smooth regions of
		the solution, but lack stability in underresolved regions, in particular when strong gradients occur, as is typical in
		compressible flows. These solution types thus require some form of shock capturing mechanism. In \Flexi, this is
		achieved by a reinterpretation of the DG polynomial in troubled cells as a finite volume solution on a structured
		subcell grid, i.e. a DG element is conceptually switched to a local, block-structured FV grid. On this grid, $1st$
		or $2nd$ order FV total variation diminishing (TVD) schemes with slope limitation are then solved to provide element-local, accurate shock
		capturing. The inverse functionality is also included to revert back to the DG representation, once sufficient
		smoothness of the solution can be detected. Different, element-local indicators are available to judge the local
		smoothness and determine the operator choice. Due to the shared data structure of DG and FV and careful
		implementation, the hybrid method suffers almost no additional computational cost compared to pure DG. More
		details on this feature can be found in Sec.~\ref{sec:shockcapturing}. Fig.~\ref{fig:riemann2d_11-12} shows the
		application of this approach to 2D Riemann problems.
	\begin{figure}
		\centering
		% ----------- END OF HEADER -------------

\begin{tikzpicture}[x=6cm,y=6cm]
      \coordimg[scale=6,xpos=0,ypos=0,xmin=-0.5,xmax=0.5,ymin=-0.5,ymax=0.5,xtics=0.25,ytics=0.25,xmtics=0.025,ymtics=0.025]{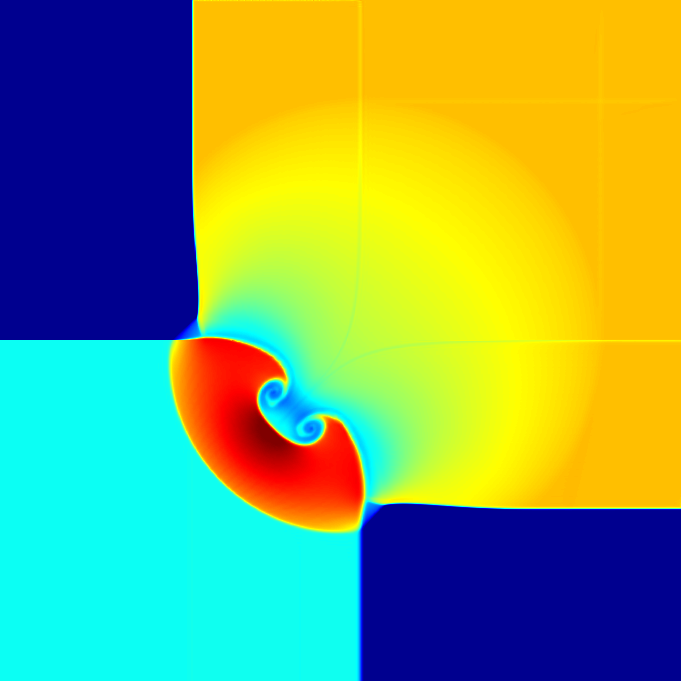};
      \coordlegend[scale=6, xpos=1.1, ypos=0,height=1.0,min=0.55,max=1.21,tics=0.25,mtics=0.025,label=$\rho$]{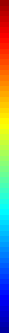}

      \coordimg[scale=6,xpos=1.4,ypos=0,xmin=-0.5,xmax=0.5,ymin=-0.5,ymax=0.5,xtics=0.25,ytics=0.25,xmtics=0.025,ymtics=0.025,yside=right]{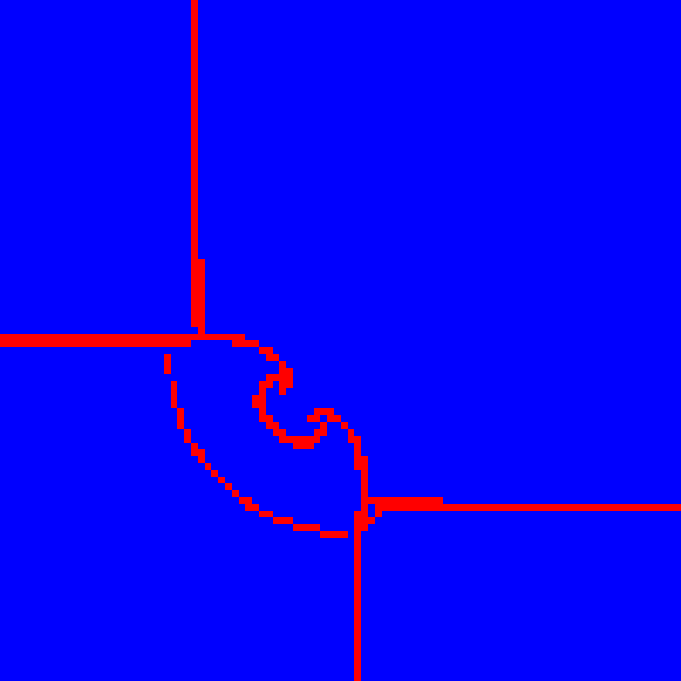};

      \def\yshift{-1.2}
      \coordimg[scale=6,xpos=0,ypos=\yshift,xmin=-0.5,xmax=0.5,ymin=-0.5,ymax=0.5,xtics=0.25,ytics=0.25,xmtics=0.025,ymtics=0.025]{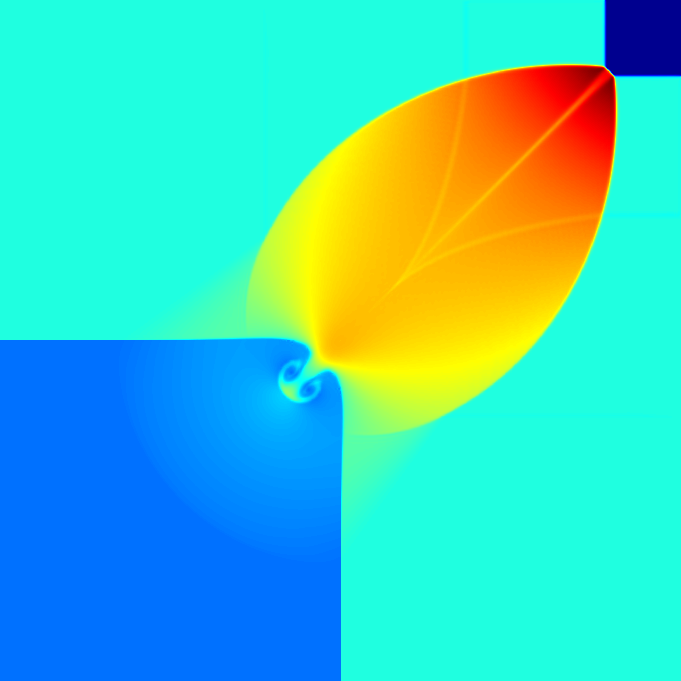};
      \coordlegend[scale=6, xpos=1.1, ypos=\yshift,height=1.0,min=0.54,max=1.7,tics=0.5,mtics=0.05,label=$\rho$]{./figures/fv/density_legend.png}

      \coordimg[scale=6,xpos=1.4,ypos=\yshift,xmin=-0.5,xmax=0.5,ymin=-0.5,ymax=0.5,xtics=0.25,ytics=0.25,xmtics=0.025,ymtics=0.025,yside=right]{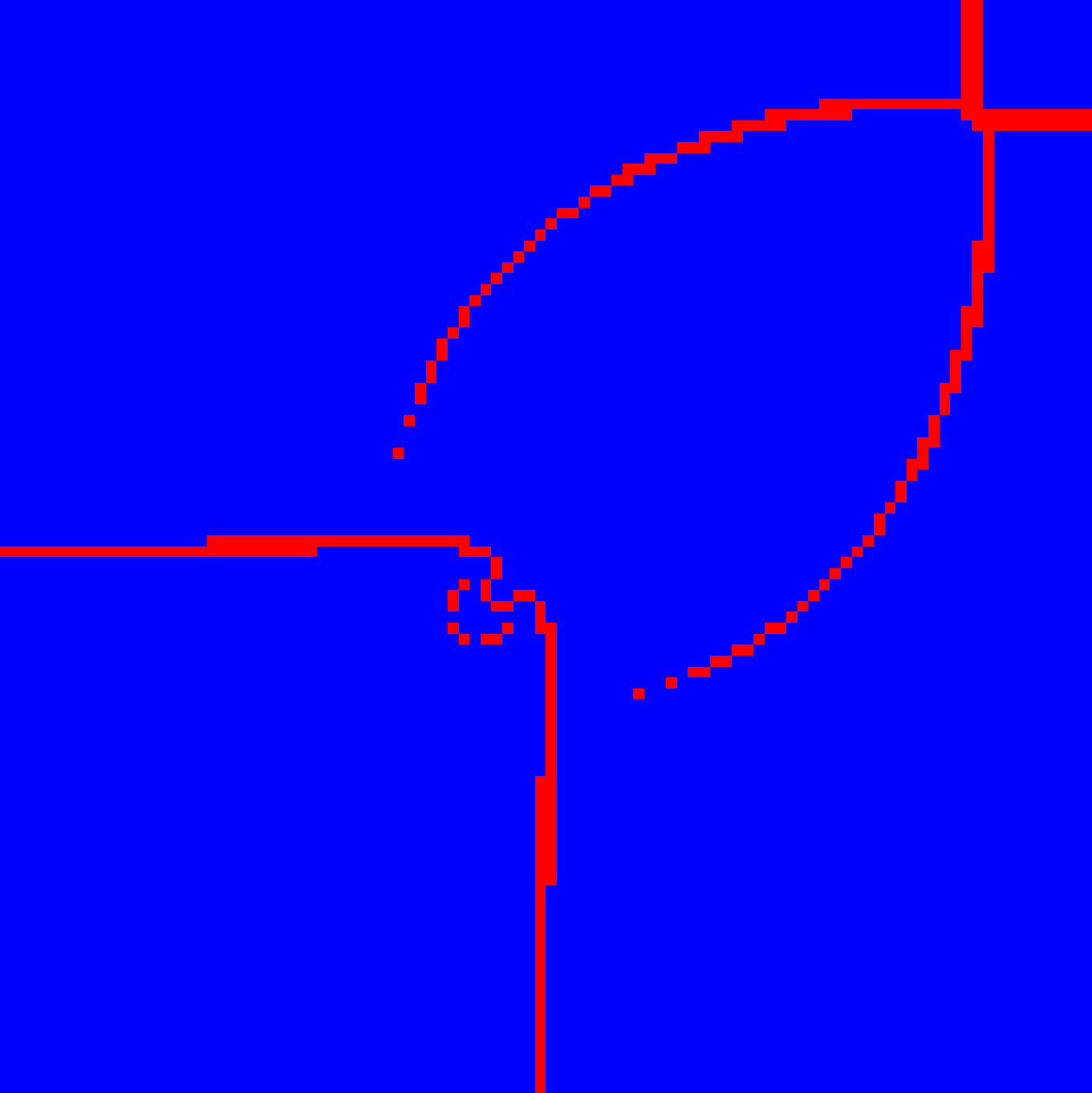};

      %\def\yshift{-2.4}
      %\coordimg[scale=6,xpos=0,ypos=\yshift,xmin=-0.5,xmax=0.5,ymin=-0.5,ymax=0.5,xtics=0.25,ytics=0.25,xmtics=0.025,ymtics=0.025]{numerics/riemann2d/cfg13/density.png};
      %\coordlegend[scale=6, xpos=1.1, ypos=\yshift,height=1.0,min=0.53,max=2.4,tics=0.5,mtics=0.05,label=$\rho$]{numerics/riemann2d/density_legend.png}

      %\coordimg[scale=6,xpos=1.4,ypos=\yshift,xmin=-0.5,xmax=0.5,ymin=-0.5,ymax=0.5,xtics=0.25,ytics=0.25,xmtics=0.025,ymtics=0.025,yside=right]{numerics/riemann2d/cfg13/fvelems.png};
\end{tikzpicture}

% ----------- FOOTER -------------
		% 11 = E, 12 = F
		\caption{Two dimensional Riemann problem, configurations E and F from~\cite{schulz1993classification}. Left: Density
		at final time. Right: FV (red) and DG (blue) elements. }
		\label{fig:riemann2d_11-12}
	\end{figure}
	
	\item Non-linear stability: For under-resolved simulations of
		non-linear problems, stability becomes an important aspect of the numerical method. Specifically high order methods
		tend to show oscillatory behaviour, ultimately leading to the crash of the simulation. Classical stabilization
		approaches are mostly based on the addition of artificial viscosity - either explicitly or implicitly, e.g. by using
		upwind operators. While the surface contribution in the DG method employs upwind Riemann solvers, the volume
		integral is completely free of artificial viscosity, such that the amount of added viscosity might not be enough to
		stabilize the simulation for even moderately under-resolved simulations. Adding viscosity to the volume might be a
		solution for the stability problem, but possibly also restricts accuracy. Thus, a method to stabilize the simulation
		without the need for excessive artificial viscosity is desirable. A possible way to achieve this is by using the
		so-called split form of the advection operator, in which aliasing errors can be avoided to guarantee preservation of
		kinetic energy (or entropy) at a discrete level~\cite{gassner2016split}. Within \Flexi, several split variants are
		available. As an alternative stabilization methods, consistent integration of the non-linear terms can also be
		selected~\cite{Beck2016}.
	
	\item LES formulations: As common in CFD, LES in \Flexi is based on the implicitly filtered form of
		the governing equations, i.e. the scale separation filter is determined by the cut-off behaviour of the
		discretization. For well-resolved LES, implicit closure through the discretization operator, in particular the
		choice of the numerical flux function, is available~\cite{Beck2016,flad2017use,beck2014high}. A number of flux
		formulations for both the advective fluxes (e.g. Roe's approximate Riemann solver, Rusanov flux, HLL-type
		formulations~\cite{toro2013riemann}) as well as the viscous fluxes (i.e. two formulations
		according to Bassi and Rebay)~\cite{bassi1997high,bassi1997turbo} can be used.  For typical LES resolutions that require
		additional modelling, the Smagorinsky, the $\sigma$ and Vreman's eddy viscosity models can be
		selected~\cite{smagorinsky1963general,nicoud2011using,vreman2004eddy}. Common filtering operations like volume or  surface
		filtering with sharp or smooth filter kernels are also available and can thus be used for filter-based LES
		or advanced multiscale models like Variational Multiscale (VMS) approaches or other structural
		models~\cite{flad2016simulation}. In order to reduce the resolution requirements for wall-bounded flows at high
		Reynolds numbers, \Flexi uses a wall-stress model according to Larsson~\cite{larsson2016large}. Initial solutions
		for the LES can be obtained by a Reynolds-averaged Navier–Stokes (RANS) formulation with a Spalart-Allmaras model~\cite{spalart1992one} using
		pseudo-time stepping. 
	\item Boundary Conditions: Boundary conditions in DG methods are typically enforced weakly through the choice of the
		corresponding interface flux. This choice leads to a more stable and accurate solution at the
		boundaries~\cite{collis2002discontinuous}, and makes implementation of a new boundary type rather easy. Standard
		boundary conditions for compressible flows are available in \Flexi, among them Dirichlet conditions, super- and
		subsonic inlet and outlet conditions, solutions to a Blasius boundary layer profile and the possibility to read in a 
		given state from file, e.g. when using information from a precursor RANS computation as far field boundaries. Generating synthetic inflow turbulence for
		boundary layer computations is achieved by the anisotropic linear forcing (ALF)
		formulation~\cite{de2015anisotropic}.  In order to avoid reflection at the outflow faces in particular for
		aeroacoustic simulations, an absorbing layer variant called a sponge layer is available. In this method, a retarding
		volume forcing term is introduced in the vicinity of the outflow, which dampens the solution towards a baseflow. A
		flexible and general method to determine a suitable base flow is to generate it from a moving time-average of the
		solution which can be computed by an exponential temporal filter~\cite{pruett2003temporally}.
	\item Flow analytics: \Flexi features a number of flow analysis options. The most basic one is that the solution (or any
		derived quantity) can be written to a solution file at a specified interval. Also, error norms with respect to a
		given exact function can be computed and stored at any time. Other information gathered and reported are the time
		history of the forces acting on immersed bodies, the velocity slip at solid walls, the bulk quantities and overall
		balances. For more advanced analysis of turbulent flows, \Flexi employs an on-the-fly time-averaging procedure for
		arbitrary linear and non-linear quantities. This information is stored at the same instance as the solution files,
		and can be reassembled in post-processing to produce volume information about the fluctuating quantities, e.g. the
		Reynolds stresses. For time-dependent solutions, \Flexi also supports data probes or record points, where the time accurate history of the full solution vector is stored for later analysis. 
\end{enumerate}

% !!!!!!!!! Posti
\subsection{POSTI: Post-processing suite}\label{subsec:posti}

Since the main focus of \Flexi is on unsteady scale-resolving computations, the output of a simulation potentially consists of a
large amount of files each containing the solution variables on tens or hundreds of million degrees of freedom for a single time
step. Besides demanding large resources for storage, post-processing such amounts of data becomes a real challenge.
The classical approach to visualization includes a conversion step, where the data from the file format native to the solver is
transformed to an alternative format suitable for the visualization tool of choice. This not only introduces the need for additional
storage space for the converted files, but also neither the conversion process nor the visualization tool itself might be
suitable for parallel execution, which could lead to a prohibitively large cost of visualizing simulation results. Another important
aspect to consider is that the large majority of available tools can not efficiently or accurately handle the
high order data provided by \Flexi.

Thus, the visualization process within the \Flexi framework follows a custom approach. We use the open-source software
ParaView~\cite{ahrens2005paraview} as our primary visualization tool. One of the main features of ParaView is the capability to use
remote computing resources for visualization in a server-client environment. Thus, a user can connect from a local workstation to
e.g. a server on the cluster system used to run a large-scale simulation and directly use the processing power available there.
Instead of converting the native HDF5 file format of \Flexi to a format suitable for ParaView, a custom read-in plugin has been created,
which allows to directly open the simulation results without any prior conversion. This reader effectively provides an interface to
a fully parallel visualization software, specifically designed for our framework. Since it is incorporated in the software package,
it can make use of routines originally designed for the solver itself, e.g. domain decomposition algorithms or the calculation of
gradients. It is also fully aware of the high order nature of the solution, which is used to super sample the
solution and visualize that version on ParaView. Additionally, the equation system used for the calculation is known to the software,
and it provides routines to calculate derived quantities from the solution variables, e.g. temperature or vorticity for the
Navier-Stokes equations. The results will then be returned directly to ParaView without first writing them to the hard drive.

As an alternative, the visualization tool itself can be used to follow the classical approach and convert the \Flexi file format to
the native ParaView file format. Next to the visualization software, the post-processing suite \Posti consists of several other
programs.  Those include tools to perform temporal or spatial averaging, analysis based on Fourier transformation, direct mode
decomposition and several other, smaller programs. As a side project, we created a pipeline~\cite{blenderPipeline} which allows to
render results from \Flexi in the open-source 3D computer graphics toolset Blender~\cite{blender}. It can be used to present 3D
simulation results in a pleasing visual style, and some examples can be found in this article.

%%%%%%%%%%%%%%%%%%%%%%%%%%%%%% ------ NUMERICAL METHODS ----- %%%%%%%%%%%%%%%%%%%%%%%%%%%%%%
\section{Numerical methods}\label{sec:numericalmethods}

\subsection{Discontinuous Galerkin spectral element method}\label{subsec:dgsem}

In the following section, the DGSEM as it is implemented in \Flexi will be derived for a general system of hyperbolic-parabolic
conservation laws for the conserved variables $\vec{U}$  defined on $\Omega \subseteq \R^3$:
\begin{equation}
	\label{eqn:3DconsLaw}
	\pderivative{\vec{U} \left( \vec{x},t \right) }{t}+ \gradient \cdot \left( \vec{F}^c \left( \vec{U} \right)-\vec{F}^v
	\left( \vec{U},\gradient \vec{U} \right) \right) = 0,
\end{equation}
with convective  and viscous fluxes $\vec{F}^c$  and  $\vec{F}^v$ respectively. We combine them into a single
flux $\vec{F}$ to keep the notation compact.

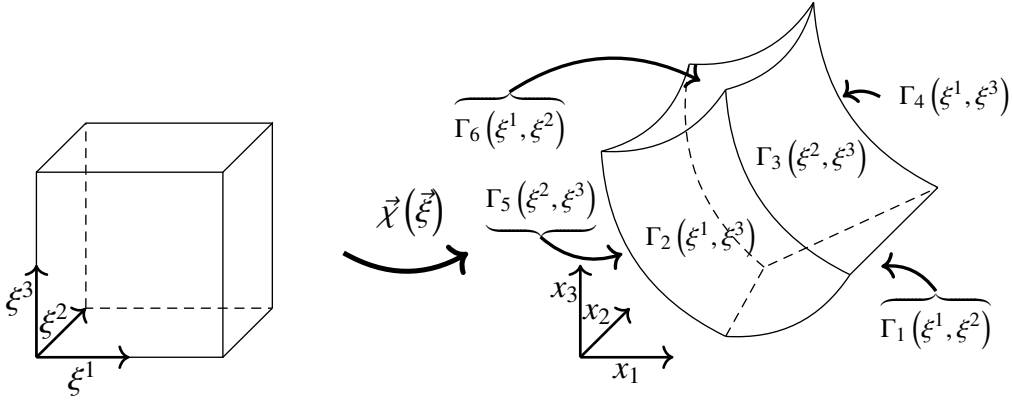
\begin{figure}
\begin{center}
	\begin{tikzpicture}[%
x={(1cm,0cm)},
y={(0cm,1cm)},
z={({0.5*cos(45)},{0.5*sin(45)})},scale=0.65]

% Cell physical domain
\coordinate (A1) at (3,1,1); 
\coordinate (B1) at (5,2,1) ;
\coordinate (C1) at (3,5,1); 
\coordinate (D1) at (1,4,1); 
\coordinate (E1) at (1.5,0,7); 
\coordinate (F1) at (5,2,5); 
\coordinate (G1) at (3,5,5); 
\coordinate (H1) at (1,4,5);

% Corners cell physical domain
\node at (3,0.75,1) {};
\node at (5.2,1.8,1) {};
\node at (2.7,5,1.5) {};
\node at (0.8,4,1) {};
\node at (1.7,-0.1,7) {};
\node at (5.2,2,5) {};
\node at (3,5.1,5) {};
\node at (1,4.2,5) {};

% Faces cell physical domain: 
% Gamma 1 
\node at (2.6,2.6,1)  {\footnotesize{$\Gamma_{2}\left(\xi^{1},\xi^{3}\right)$}};
% Gamma 2
\node at (5.8,4,3.5)  {\footnotesize{$\Gamma_{4}\left(\xi^{1},\xi^{3}\right)$}}; 
\draw[line width=1pt,bend right,->] (4.6,4,3.5) to (4,4,3.5);
% Gamma 3
\node at (5.5,0.4,3.5)  {\footnotesize{$\overbrace{\Gamma_{1}\left(\xi^{1},\xi^{2}\right)}$}};
\draw[line width=1pt,bend right,->] (5.5,0.85,3.5) to (4.62,1.25,3.5);
% Gamma 5
\node at (-1,4,2.5)  {\footnotesize{$\overbrace{\Gamma_{6}\left(\xi^{1},\xi^{2}\right)}$}};
\draw[line width=1pt,bend left,->] (-0.69,4.75,1.6) to (2.15,4.7,2.25);
% Gamma 4
\node at (3.5,3,3.5)  {\footnotesize{$\Gamma_{3}\left(\xi^{2},\xi^{3}\right)$}};
% Gamma 6
\node at (-0.5,2.5,2.5)  {\footnotesize{$\underbrace{\Gamma_{5}\left(\xi^{2},\xi^{3}\right)}$}};
\draw[line width=1pt,bend right,->] (-0.19,2.4,1.6) to (1.25,2.25,1.25);

% Corners
\draw[bend right,-] (A1) to   (B1);
\draw[bend left,-] (B1) to (C1); 
\draw[bend left,-]  (C1) to  (D1); 
\draw[bend right,-] (D1) to (A1); 
\draw[-] (B1) to (F1);
\draw[bend left,-] (F1) to (G1); 
\draw[bend left,-] (G1) to (C1);
\draw[bend left,-] (G1) to (H1); 
\draw[bend left,-] (H1) to (D1);
\draw[densely dashed] (A1) -- (E1) -- (F1);
\draw[bend left,densely dashed] (E1) to (H1);

% Physical domain coordinates 
\draw[thick,->] (0.75,0.75,0.75) -- (2.25,0.75,0.75); 
\node at (1.5,0.5,0.75)  {$x_{1}$};
\draw[thick,->] (0.75,0.75,0.75) -- (0.75,0.75,3); 
\node at (0.52,0.95,2.15)  {$x_{2}$};
\draw[thick,->] (0.75,0.75,0.75) -- (0.75,2.25,0.75);
\node at (0.5,1.8,0.75)  {$x_{3}$};

% Reference element 
\coordinate (A) at (-8,0.75,0.75); 
\coordinate (B) at (-5,0.75,0.75) ;
\coordinate (C) at (-5,3.75,0.75); 
\coordinate (D) at (-8,3.75,0.75); 
\coordinate (E) at (-8,0.75,3); 
\coordinate (F) at (-5,0.75,3); 
\coordinate (G) at (-5,3.75,3); 
\coordinate (H) at (-8,3.75,3);

% Corners
\draw[] (A) -- (B)  -- (C)  -- (D)  -- (A);
\draw[] (B) to (F);
\draw[] (F) -- (G) -- (C);
\draw[] (G) -- (H) -- (D);
\draw[densely dashed] (A) -- (E) -- (F);
\draw[densely dashed] (E) to (H);

% Reference domain coordinates 
\draw[thick,->] (-8,0.75,0.75) -- (-6.5,0.75,0.75); 
\node at (-7.25,0.4,0.75)  {$\xi^{1}$};
\draw[thick,->] (-8,0.75,0.75) -- (-8,0.75,3); 
\node at (-8.2,0.84,2.15)  {$\xi^{2}$};
\draw[thick,->] (-8,0.75,0.75) -- (-8,2.25,0.75);
\node at (-8.25,1.7,0.75)  {$\xi^{3}$};

% Mapping 
\draw[line width=1.5pt,bend right,->] (-3.5, 2,2) to (-1.5, 2,2);
\node at (-2.4,2.65,2)  {$\vec{\chi}\left(\vec{\xi}\right)$};
\end{tikzpicture}
\caption{\label{fig:ReferencePhysicalCell} Left the reference element $E=[-1,1]^3$ and on the right a general hexahedral element
$e_{\kappa}$ with the curved faces $\vec{\Gamma}_{i}$.
The mapping $\vec{\chi}\left(\vec{\xi}\right)$ connects $E$ and $e_{\kappa}$.}
\end{center}
\end{figure}

The domain $\Omega$ is subdivided into $K \in \N$ non-overlapping and (for the moment) conforming hexahedral elements $e_K$, each
bounded by six curved faces $\vec{\Gamma}_i,\;i=1,2,3,4,5,6$, as sketched in Fig.~\ref{fig:ReferencePhysicalCell}.  The geometry of the
faces $\vec{\Gamma}_i(r,s),(r,s) \in [-1,1]^2$ is approximated as tensor products of one-dimensional Lagrange interpolating
polynomials up to degree $\Ngeo$ :
\begin{equation}
\vec{\Gamma}_{i}(r,s):=\sum_{j,k=0}^{\Ngeo}\vec{\Gamma}_{i}\left(r_{j},s_{k}\right)\ell_{j}\left(r\right)\ell_{k}\left(s\right).
\end{equation}
The usual definition of the Lagrange polynomials is
\begin{equation}
	\ell_{j}\left(r\right) = \prod_{i=0,i \neq j}^{\Ngeo} \frac{r-r_i}{r_j-r_i}.
\end{equation}
The nodal points $r_i$ are taken from either Legendre-Gauss (LG) or Legendre-Gauss-Lobatto (LGL) quadrature.

The DGSEM will be constructed on a reference cube, such that all operators only need to be defined once, regardless of the actual
shape of the considered element. Thus, a transfinite map $\vec{x}=\vec{\chi}\left(\vec{\xi},t\right)$ from
the reference cube $E=[-1,1]^3$ in \textit{computational} space $\vec{\xi}=[\xi^1,\xi^2,\xi^3]^T$ onto the element in
\textit{physical} space $\vec{x}=[x_1,x_2,x_3]^T$ is introduced.
We then define the Jacobian of the mapping $\J$ as the determinant of the Jacobian matrix $\gradientXI \vec{\chi}$. The operator
$\gradientXI$ represents the nabla operator evaluated in the computational coordinates. For the required transformation between
physical and reference space, the contravariant basis vectors $\J \vec{a}^i$ (also called metric terms) are introduced, calculated in 
a guaranteed discretly divergence free (so called curl) form~\cite{kopriva2006metric}:
\begin{equation}
	\label{eqn:contravariantBasisVectorsCurl}
	\J \vec{a}^i_n= -\hat{x}_i \cdot \gradientXI \times \left( x_l \gradientXI x_m  \right), \quad i=1,2,3, \quad n=1,2,3, \quad
	(n,m,l)\; cyclic,
\end{equation}
where $\hat{x}_i$ is the unit vector in the $i$th Cartesian direction. This formulation ensures that the so-called metric identities
are fulfilled, which is necessary to achieve free-stream preservation.

As is detailed in~\cite{hindenlang2015mesh}, the mapping is constructed by blending the curved boundary faces together. This ensures 
that the map $\vec{\chi}$ itself is a polynomial of degree $\Ngeo$ or less, and that required derivatives can be computed by
derivation of that polynomial.

The conservation law~\eqref{eqn:3DconsLaw} can now be formulated in the reference cube. Inserting the
transformation rule for the divergence 
\begin{equation}
	\label{eqn:3DDivergenceTransform}
	\gradient \cdot \vec{F} = \frac{1}{\J} \sum_{i=1}^3 \pderivative{\J \vec{a}^i \cdot \vec{F}}{\xi^i},
\end{equation}
yields
\begin{equation}
	\label{eqn:3DconsLawTransformed}
	\J \pderivative{\vec{U} \left( \vec{x},t \right) }{t} + \gradientXI \cdot  \vec{\F} = 0,
\end{equation}
with the so-called contravariant fluxes $\vec{\F}$, which are defined as
\begin{equation}
    \vec{\F^i} = \J \vec{a}^i \cdot \vec{F}, \quad i=1,...,3.
\end{equation}
We can obtain the weak formulation of the conservation law in reference coordinates by
multiplying~\eqref{eqn:3DconsLawTransformed} with an arbitrary test function $\phi=\phi(\vec{\xi})$ and by integration over the reference element
\begin{equation}
	\label{eqn:weakFormulation_1}
	\iiint_E \left(\J \pderivative{\vec{{U}}}{t} + \gradientXI \cdot \vec{\F} \right) \phi \; \mathrm{d}\vec{\xi} =
	\left<\J\pderivative{\vec{{U}}}{t} + \gradientXI \cdot \vec{\F},\phi
	\right> = 0.
\end{equation}
Here, $\left<\cdot , \cdot  \right>$ defines the inner product between two functions.
The divergence can be split into interior and boundary contributions by applying Green's identity:
\begin{equation}
	\label{eqn:weakFormulation}
	\left< \J  \pderivative{\vec{{U}}}{t} ,\phi \right> + \iint_{\partial E} \phi \left( \vec{\F} \cdot \noutRef \right)^* \;\mathrm{d}S -
	\left<\vec{\F},\gradientXI \phi \right> = 0,
\end{equation}
where $\noutRef$ is the outward pointing normal vector. Since no continuity is required between elements, numerical flux functions
are used to approximate a unique normal flux $\left( \vec{\F} \cdot \noutRef \right)^*$ from the double-valued solution on the
boundary.

Both the solution $\vec{{U}}$ and the components of the contravariant fluxes $\vec{\F}^l,\; l=1,2,3$ are approximated by tensor products
of Lagrange polynomials of degree $N$ on the reference element, e.g.
\begin{equation}
	\label{eqn:approxSol}
	\vec{{U}}\left( \xi^1,\xi^2,\xi^3,t \right) \approx \vec{{\mathbf{U}}} \left( \xi^1,\xi^2,\xi^3,t \right) = \sum_{m,n,o=0}^N
	\vec{{\mathbf{U}}}_{mno}(t) \ell_{m}(\xi^1) \ell_{n}(\xi^2) \ell_{o}(\xi^3),
\end{equation}
where the bold font indicates the polynomial interpolation.  The notation $\vec{{\mathbf{U}}}_{mno}(t)$ stands for the degree of
freedom associated with the basis function $\psi_{mno}=\ell_{m}(\xi^1) \ell_{n}(\xi^2) \ell_{o}(\xi^3)$. For the chosen nodal
basis, the degrees of freedom are also identical to the value of the interpolation at the position $(\xi^1_m,\xi^2_n,\xi^3_o)$.  The nodes are again
chosen from either LG or LGL quadrature.
To determine the nodal values of the contravariant flux components $\vec{\mathbfcal{F}}^{l}_{mno}$, the physical
flux is evaluated using the interpolated solution and multiplied by the weighted contravariant basis vectors, transforming it to
computational space:
\begin{equation}
	\vec{\mathbfcal{F}}^{l}_{mno}(t) = \sum_{d=1}^3 \J a_d^l \left( \xi^1,\xi^2,\xi^3,t \right) \vec{F}^d \left(
	\vec{{\mathbf{U}}} \left( \xi^1,\xi^2,\xi^3,t \right) \right).
\end{equation}

The polynomial approximations can now be inserted into the weak form~\eqref{eqn:weakFormulation}. Integrals will be evaluated using
Gauss-Legendre quadrature with collocation of the interpolation and quadrature nodes. The basis functions are also used as test
functions, following the Galerkin ansatz.

\subsubsection*{Time derivative integral}

Substituting the polynomial approximation into the weak form~\eqref{eqn:weakFormulation} and replacing the integrals by
Gauss-Legendre quadrature yields for the first term, containing the time derivative:
\begin{align}
	&\left<\J \pderivative{\vec{{U}}}{t} ,\phi \right> = \nonumber\\
	&\pderivative{}{t} \sum_{\alpha,\beta,\gamma}^N \J \left( \vec{\xi}_{\alpha \beta \gamma} \right) \sum_{m,n,o}^N  \left(
	\vec{\mathbf{U}}_{mno}(t)
	\ell_{m}(\xi^{1}_{\alpha}) \ell_{n}(\xi^{2}_{\beta}) \ell_{o}(\xi^{3}_{\gamma}) \right) \psi_{ijk} ({\xi}_{\alpha \beta
	\gamma}) \omega_{\alpha} \omega_{\beta} \omega_{\gamma} \quad \forall\, i,j,k.
\end{align}
Here, $\{\omega_i\}_{i=0}^N$ are the weights associated with the quadrature rule and $\{\xi_i\}_{i=0}^N$ the position of the nodes.
Since the Lagrange polynomials satisfy the property
\begin{equation}
	\ell_i(\xi_j) = \delta_{ij} = \begin{cases}
1 & i=j \\
0 & i \neq j 
\end{cases} \quad \forall i,j=0,...,N ,
\end{equation}
the above equation simplifies to 
\begin{align}
	\left<\J \pderivative{\vec{{U}}}{t} ,\phi \right> = \pderivative{}{t} \sum_{m,n,o}^N
	\J \left( \vec{\xi}_{mno} \right) \vec{\mathbf{U}}_{mno} \psi_{ijk} ({\xi}_{m n o}) \omega_{m} \omega_{n} \omega_{o} \quad \forall\, i,j,k.
\end{align}
Taking once more advantage of the Lagrange property when substituting the basis functions leads to
\begin{align}
	\label{eqn:discTemp}
	&= \pderivative{}{t} \sum_{m,n,o}^N \J \left( \vec{\xi}_{mno} \right) \vec{\mathbf{U}}_{mno}
	\underbrace{\ell_{i}(\xi^{1}_{m})}_{\delta_{im}} \underbrace{\ell_{j}(\xi^{2}_{n})}_{\delta_{jn}}
	\underbrace{\ell_{k}(\xi^{3}_{o})}_{\delta_{ko}}  \omega_{m} \omega_{n} \omega_{o} \nonumber \\
	&=  \J \left( \vec{\xi}_{ijk} \right) \pderivative{\vec{\mathbf{U}}_{ijk}}{t}  \omega_{i} \omega_{j} \omega_{k}  \quad \forall\, i,j,k.
\end{align}

Our choice of basis functions and the collocation approach lead to significant simplifications of the scheme. We obtain a diagonal
mass matrix $M_{ij}=\left<\ell_i,\ell_j\right>$, a manifestation of the discrete orthogonality of the basis functions.  It should be
noted that (for a constant Jacobian) the mass matrix is exactly computed for LG nodes, since the quadrature in this case is exact up
to a degree $2N+1$, and the product of the two functions is of degree $2N$. Quadrature on LGL nodes is only exact up to degree
$2N-1$, and the analytical mass matrix for LGL nodes would thus not be diagonal. The discrete version is, which is referred to as
\textit{mass lumping}, a property that has an influence on dispersion and dissipation characteristics and which was investigated
in~\cite{gassner2011comparison}.

\subsubsection*{Volume Integral}

The rightmost term in the weak formulation~\eqref{eqn:weakFormulation} is the scalar product of the contravariant fluxes
with the gradient of the test function, compactly written as
\begin{equation}
	\label{eqn:volInt}
	\left<\vec{\F},\gradientXI \phi \right> = \sum_{d=1}^3 \left<  \vec{\F}^d ,\pderivative{\phi}{\xi^d} \right>.
\end{equation}
Only the first entry in the sum ($d=1$) will be considered in the following, since they are all of the same structure.
Applying Gauss quadrature, substituting the polynomial approximation and using the Lagrange property leads to
\begin{align}
	&\left<  \vec{\F}^1 ,\pderivative{\phi}{\xi^1} \right>  \nonumber \\
	&= \sum_{\alpha,\beta,\gamma=0}^N \sum_{m,n,o=0}^N \left(
	\vec{\mathbfcal{F}}^1_{mno} \underbrace{\ell_{m}(\xi^{1}_{\alpha})}_{\delta_{m\alpha}}
	\underbrace{\ell_{n}(\xi^{2}_{\beta})}_{\delta_{n\beta}}\underbrace{\ell_{o}(\xi^{3}_{\gamma})}_{\delta_{o\gamma}} \right)
	\pderivative{\psi_{ijk}}{\xi^1}  \omega_{\alpha} \omega_{\beta} \omega_{\gamma} \nonumber  \\
	&= \sum_{m,n,o=0}^N  \vec{\mathbfcal{F}}^1_{mno} \left. \pderivative{\ell_i \left( \xi^1 \right)}{\xi^1}
	\right|_{\xi=\xi^1_m} \underbrace{\ell_j \left( \xi^2_n \right)}_{\delta_{jn}} \underbrace{\ell_k \left( \xi^3_o
	\right)}_{\delta_{ko}} \omega_{m} \omega_{n} \omega_{o}\nonumber  \\
	&=  \omega_{j} \omega_{k} \sum_{m=0}^N  \vec{\mathbfcal{F}}^1_{mjk} \left. \pderivative{\ell_i \left( \xi^1 \right)}{\xi^1}
	\right|_{\xi=\xi^1_m} \omega_{m} \quad \forall\, i,j,k.
\end{align}
We can define   the differentiation matrix
\begin{equation}
	D_{rs} = \left. \pderivative{\ell_s \left( \xi \right)}{\xi}\right|_{\xi=\xi_r}, \quad r,s=0,...,N,
\end{equation}
containing the derivatives of the Lagrange polynomials at the interpolation nodes.
This allows us to write the sum over the derivative of the Lagrange polynomials as a matrix vector product, which can be efficiently
implemented:
\begin{equation}
	\sum_{m=0}^N   \left. \pderivative{\ell_i \left( \xi^1 \right)}{\xi^1} \right|_{\xi=\xi^1_m} \omega_{m} = 
	\sum_{m=0}^N  D_{mi}  \omega_{m}.
\end{equation}
Using this notation, the volume integral can be written as
\begin{multline}
	\label{eqn:discVolInt}
	\left<\vec{\F},\gradientXI \phi \right> = \\ \omega_{j} \omega_{k} \sum_{m=0}^N  \vec{\mathbfcal{F}}^1_{mjk} D_{mi} \omega_{m}
	+ \omega_{i} \omega_{k} \sum_{n=0}^N  \vec{\mathbfcal{F}}^2_{ink} D_{nj}  \omega_{n}+  \omega_{i} \omega_{j} \sum_{o=0}^N  \vec{\mathbfcal{F}}^3_{ijo} D_{ok} \omega_{o}  \quad \forall\, i,j,k.
\end{multline}

\subsubsection*{Surface integral}

The second term of the weak form~\eqref{eqn:weakFormulation} consists of the surface integral, which can be evaluated on all six sides
of the reference cube independently and summed up subsequently:
\begin{align}
	\label{eqn:SurfInt_1}
	&\iint_{\partial E} \phi \left( \vec{\F} \cdot \noutRef \right)^* \;\mathrm{d}S = \nonumber \\
	& \left. \int_{-1}^1 \int_{-1}^1 \phi \left( \vec{\F} \cdot \noutRef \right)^* \, \mathrm{d}\xi_2 \, \mathrm{d}\xi_3
	\right|_{\xi_1 = 1} + \left. \int_{-1}^1 \int_{-1}^1 \phi \left( \vec{\F} \cdot \noutRef \right)^* \, \mathrm{d}\xi_2 \,
	\mathrm{d}\xi_3 \right|_{\xi_1 = -1} \nonumber \\ +
	& \left. \int_{-1}^1 \int_{-1}^1 \phi \left( \vec{\F} \cdot \noutRef \right)^* \, \mathrm{d}\xi_1 \, \mathrm{d}\xi_3
	\right|_{\xi_2 = 1} + \left. \int_{-1}^1 \int_{-1}^1 \phi \left( \vec{\F} \cdot \noutRef \right)^* \, \mathrm{d}\xi_1 \,
	\mathrm{d}\xi_3 \right|_{\xi_2 = -1} \nonumber \\+ 
	& \left. \int_{-1}^1 \int_{-1}^1 \phi \left( \vec{\F} \cdot \noutRef \right)^* \, \mathrm{d}\xi_1 \, \mathrm{d}\xi_2
	\right|_{\xi_3 = 1} + \left. \int_{-1}^1 \int_{-1}^1 \phi \left( \vec{\F} \cdot \noutRef \right)^* \, \mathrm{d}\xi_1 \,
	\mathrm{d}\xi_2 \right|_{\xi_3 = -1}.
\end{align}
The Riemann solvers used for the numerical fluxes are in general implemented to compute the flux in the normal direction in
physical space on the surface $(\vec{F} \cdot \hat{n})^*$, so we need to relate those quantities to the computational system.
Realizing that a non-normalized physical normal vector $\tilde{n}$ can be computed from the metric terms at the surface
\begin{equation}
	\tilde{n}_d = \sum_{i=1}^3 \J a_d^i  \noutRef^i, \quad d=1,2,3,
\end{equation}
this relationship can be used for the following transformation:
\begin{align}
	(\vec{F} \cdot \hat{n}) &= \sum_{d=1}^3 \vec{F}_d \cdot \hat{n}_d = \frac{1}{|\tilde{n}|} \sum_{d=1}^3 \vec{F}_d \cdot \tilde{n}_d \nonumber\\
	& = \frac{1}{|\tilde{n}|} \sum_{d=1}^3 \vec{F}_d \cdot 	\sum_{i=1}^3 \J a_d^i  \noutRef^i = \frac{1}{|\tilde{n}|} \sum_{i=1}^3 \sum_{d=1}^3 \J a_d^i \vec{F}_d \cdot \noutRef^i.
\end{align}
Thus, we can state that
\begin{equation}
	\left( \vec{\F} \cdot \noutRef \right)^* = \left( \vec{F} \cdot \nout \right)^* \hat{s} = \vec{f}^* \left(
	\vec{{U}}_L, \vec{{U}}_R  , \nout \right) \hat{s},
\end{equation}
with the surface element (the norm of the non-normalized physical unit vector)
\begin{equation}
	\hat{s} = \sqrt{\sum_{d=1}^3  \left( \J a_d^1 \left( 1,\xi_2,\xi_3 \right) \right)^2}.
\end{equation}
By $\vec{f}^*$ we denote the Riemann solver that takes the double-valued solution at the left and the right side
($\vec{{U}}_L$ and $\vec{{U}}_R$, respectively) at an interface (along with the physical normal unit vector) as arguments and
returns an unique flux in the direction of that normal vector. The left state is obtained from the polynomial interpolation of the
solution in the mesh element currently considered, while the right state is obtained from the solution in the adjacent element in
the considered direction. Riemann solvers only handle the convective part of the fluxes, the treatment of the viscous part is
described later.

\begin{figure}
\begin{center}
	\includegraphics[width=\textwidth]{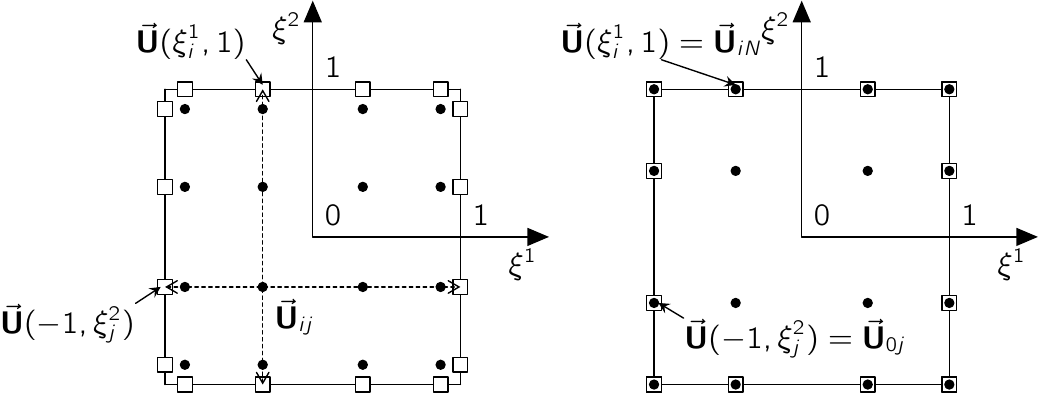}
	\caption{\label{fig:refelement}  Two-dimensional sketch of the location of the interpolation nodes in the volume \protect\tikz[x=2cm,y=2cm,baseline=-0.5ex]\protect\draw[fill=black] (0,0) circle (0.03); and on the
	surface \protect\tikz[x=2.0cm,y=2.0cm,baseline=-0.5ex]\protect\draw[] (0,0)  +(-0.03,-0.03) rectangle +(0.03,0.03);, both
	for LG nodes (left) and LGL nodes (right) with $N=3$.}
\end{center}
\end{figure}
The numerical flux is again approximated using Lagrange interpolation, e.g. on the side $\xi^1=1$
\begin{equation}
	\label{eqn:numFlux}
	\left( \vec{\F} \cdot \noutRef \right)^* = \sum_{m,n=0}^N \left[ \vec{f}^*  \left( \vec{{U}}_L, \vec{{U}}_R  ,
	\vec{\meshVel}, \nout \right) \hat{s} \right]_{m,n}^{+\xi^1} \ell_m \left( \xi^2 \right) \ell_n \left( \xi^3 \right),
\end{equation}
Here, the superscript $+\xi^1$ indicates the evaluation at the face where $\xi^1=1$ and the subscript $m,n$ identifies the $m,n-th$
Gauss node on the cell boundary. For LGL nodes, since the boundaries are included in the set of nodes, the interface nodal points
coincide with nodes from the volume, and the solution can be directly obtained from the degrees of freedom of the Lagrange
interpolation. For LG nodes , the solution at the face must be obtained by evaluating the interpolation at $\xi^1=1$. Exploiting the
tensor product nature of the interpolation, this reduces to a one-dimensional operation:
\begin{equation}
	\vec{\mathbf{U}}_{m,n}^{+\xi^1} = \sum_{o=0}^N  \vec{\mathbf{U}}_{mno} \ell_o (1)
\end{equation}
A sketch of the situation for both LG and LGL nodes is given in Fig.~\ref{fig:refelement}, in two dimensions.

We now insert the numerical flux~\eqref{eqn:numFlux} into the corresponding part of the surface integral~\eqref{eqn:SurfInt_1},
again apply Gauss quadrature, use the Galerkin ansatz and make use of the Lagrange property:
\begin{align}
	\int_{-1}^1 \int_{-1}^1 &\phi \left. \left( \vec{\F} \cdot \noutRef \right)^* \, \mathrm{d}\xi_2 \, \mathrm{d}\xi_3
	\right|_{\xi_1 = 1}\nonumber  \\
	&=  \sum_{\alpha,\beta=0}^N \left(   \sum_{m,n=0}^N \left[ \vec{f}^*  \hat{s} \right]_{m,n}^{+\xi^1} \underbrace{  \ell_m \left(
	\xi^2_{\alpha} \right)}_{\delta_{m\alpha}} \underbrace{\ell_n \left( \xi^3_{\beta} \right)}_{\delta_{n\beta}}     \right)
	\ell_i(1) \underbrace{  \ell_j(\xi^2_{\alpha})}_{\delta_{j\alpha}} \underbrace{  \ell_k(\xi^3_{\beta})}_{\delta_{k\beta}}
	\omega_{\alpha} \omega_{\beta}\nonumber  \\
	&=  \left[ \vec{f}^*  \hat{s} \right]_{j,k}^{+\xi^1} \ell_i(1) \omega_{j} \omega_{k}  \quad \forall\, i,j,k.
\end{align}
We omit the arguments of the Riemann solver to shorten the notation. Repeating the same procedure for the remaining five
sides yields:
\begin{align}
	\label{eqn:discSurfInt}
	\iint_{\partial E} \phi \left( \vec{\F} \cdot \noutRef \right)^* \;\mathrm{d}S &= 
	\left(  \left[ \vec{f}^*  \hat{s} \right]_{j,k}^{+\xi^1} \ell_i (1) + \left[ \vec{f}^*  \hat{s} \right]_{j,k}^{-\xi^{1}}
	\ell_i (-1)  \right)  \omega_{j} \omega_{k}  \nonumber   \\
	&+ \left(  \left[ \vec{f}^*  \hat{s} \right]_{i,k}^{+\xi^2} \ell_j (1) + \left[ \vec{f}^*  \hat{s} \right]_{i,k}^{-\xi^{2}}
	\ell_j (-1)  \right)  \omega_{i} \omega_{k}  \nonumber   \\
	&+ \left(  \left[ \vec{f}^*  \hat{s} \right]_{i,j}^{+\xi^3} \ell_k (1) + \left[ \vec{f}^*  \hat{s} \right]_{i,j}^{-\xi^{3}} \ell_k (-1)  \right)  \omega_{i} \omega_{j}
	\quad \forall\, i,j,k.
\end{align}

\subsubsection*{Semi-discrete formulation}

Now each part of the weak formulation~\eqref{eqn:weakFormulation} is discretized, and we can combine the results and gain
the semi-discrete version of the DGSEM, with a still continuous time:
\begin{align}
	\label{eqn:semiDiscrete}
	\pderivative{\vec{\mathbf{U}}_{ijk}}{t} = 
	- \frac{1}{\J_{ijk}}\Bigg[ &\sum_{\alpha=0}^N \vec{\mathbfcal{F}}^1_{\alpha jk} \hat{D}_{i \alpha}  + \left(     \left[ \vec{f}^*  \hat{s}
	\right]_{j,k}^{+\xi^1} \hat{\ell}_i (1) + \left[ \vec{f}^*  \hat{s} \right]_{j,k}^{-\xi^{1}} \hat{\ell}_i (-1)    \right)
	\nonumber  \\
	+ &\sum_{\beta=0}^N \vec{\mathbfcal{F}}^2_{i \beta k} \hat{D}_{j \beta}  + \left(     \left[ \vec{f}^*  \hat{s}
	\right]_{i,k}^{+\xi^2} \hat{\ell}_j (1) + \left[ \vec{f}^*  \hat{s} \right]_{i,k}^{-\xi^{2}} \hat{\ell}_j (-1)
	\right)\nonumber  \\
	+ &\sum_{\gamma=0}^N \vec{\mathbfcal{F}}^3_{i j \gamma} \hat{D}_{k \gamma}  + \left(     \left[ \vec{f}^*  \hat{s} 
	\right]_{i,j}^{+\xi^3} \hat{\ell}_k (1) + \left[ \vec{f}^*  \hat{s} \right]_{i,j}^{-\xi^{3}} \hat{\ell}_k (-1)    \right)
	\Bigg]  \quad \forall\, i,j,k.
\end{align}
The operator can be split in a \textit{volume integral} (left half) and a \textit{surface integral} (right half). For the surface
integral, the \textit{numerical flux function} must first be computed. Those three procedures are the main working points of the DG method.
Choosing the tensor product basis directly lead to a tensor product structure of the operator itself: Each of the three lines
correspond to the application of a one-dimensional DG operator in one of the reference directions. Thus, we can make use of the
precomputed one-dimensional operators
\begin{equation}
	\hat{\ell}_i = \frac{\ell_i}{\omega_i} \quad \mathrm{and} \quad \hat{D}_{ij} = -\frac{\omega_i}{\omega_j} D_{ji},\quad
	i,j=0,...,N,
\end{equation}
for an efficient computation.

\subsubsection*{Strong formulation}

If Green's identity is applied once more to the weak form~\eqref{eqn:weakFormulation}, and only using the interior contribution
$\vec{\F}^{int}$ in the newly emerging surface integral, this leads to
\begin{equation}
	\label{eqn:strongFormulation}
	\left<\J \pderivative{\vec{{U}}}{t} ,\phi \right> + \iint_{\partial E} \phi \left( \left( \vec{\F} \cdot \noutRef \right)^*-
	\vec{\F}^{int} \cdot \noutRef \right) \;\mathrm{d}S +	\left<\gradientXI \vec{\F}, \phi \right> = 0.
\end{equation}
In this form, the volume flux still needs to be differentiable, and it is thus referred to as \textit{strong form}.  It consists of
the local residual of the equation (the last term) and a surface penalty term, that vanishes when the solution is continuous at the
boundary.

Repeating the discretizations steps for this version of the equations leads to the semi-discrete formulation
\begin{align}
	\label{eqn:semiDiscreteStrong}
	\pderivative{\vec{\mathbf{U}}_{ijk}}{t} &= \nonumber  \\
	- \frac{1}{\J_{ijk}} \Bigg[ &\sum_{\alpha=0}^N \vec{\mathbfcal{F}}^1_{\alpha jk} {D}_{i \alpha}  + \left(     \left[ \vec{f}^*  \hat{s} - \vec{\mathbfcal{F}}^1
	\right]_{j,k}^{+\xi^1} \hat{\ell}_i (1) + \left[ \vec{f}^*  \hat{s}  - \vec{\mathbfcal{F}}^1\right]_{j,k}^{-\xi^{1}}
	\hat{\ell}_i (-1)    \right) \nonumber  \\
	+ &\sum_{\beta=0}^N \vec{\mathbfcal{F}}^2_{i \beta k} {D}_{j \beta}  + \left(     \left[ \vec{f}^*  \hat{s} - \vec{\mathbfcal{F}}^2
	\right]_{i,k}^{+\xi^2} \hat{\ell}_j (1) + \left[ \vec{f}^*  \hat{s}  - \vec{\mathbfcal{F}}^2\right]_{i,k}^{-\xi^{2}}
	\hat{\ell}_j (-1)    \right)\nonumber  \\
	+ &\sum_{\gamma=0}^N \vec{\mathbfcal{F}}^3_{i j \gamma} {D}_{k \gamma}  + \left(     \left[ \vec{f}^*  \hat{s} - \vec{\mathbfcal{F}}^3
	\right]_{i,j}^{+\xi^3} \hat{\ell}_k (1) + \left[ \vec{f}^*  \hat{s}  - \vec{\mathbfcal{F}}^3\right]_{i,j}^{-\xi^{3}} \hat{\ell}_k (-1)    \right)
	\Bigg] \nonumber \\ & \quad \quad \quad \quad \quad \quad \quad \quad \quad \quad \quad \quad \quad \quad \quad \quad \quad
	\quad \quad \quad \quad \quad\forall\, i,j,k.
\end{align}
For the volume integral, the difference between weak and strong form is simply that the strong form uses directly the $D$ matrix
instead of $\hat{D}$, meaning the same algorithm can be used for both of them. The interior surface contribution must be computed by
interpolating the volume flux to the boundary, introducing additional operations. Analytically and also discretely, the weak and the
strong formulation are identical for DGSEM for LGL quadrature, and for LG quadrature with special care for the internal fluxes
~\cite{kopriva2010quadrature}. The main use for the strong form is that it can be used as a starting point for the construction of
non-linear stable schemes, as will be described in Sec.~\ref{sec:splitDG}.

\subsubsection*{Approximation for second order equations}

To handle viscous fluxes, that also depend on the gradients of the solution, we follow a method by Bassi and
Rebay~\cite{bassi1997high}, now known as BR1. While \Flexi also supports the BR2 method~\cite{bassi1997turbo}, we restrict ourselves here
to the details of BR1.  For this procedure, known as \textit{lifting}, we are going to introduce an additional gradient quantity
\begin{equation}
	\vec{g}^d=\pderivative{\vec{U}}{x_d}, \quad d=1,2,3.
\end{equation}
Those are used to express second-order equations as a system of first-order equations:
\begin{align}
	\label{eqn:firstOrderSystem}
	\pderivative{\vec{U} \left( \vec{x},t \right) }{t} + \gradient \cdot \left( \vec{F}^c(\vec{U})-\vec{F}^v(\vec{U},
	\vec{g}) \right) &=0, \\
	\vec{g}^d-\pderivative{\vec{U}}{x_d} &=0, \quad d=1,2,3.
\end{align}
An additional $3m$ (in the case of three spatial dimensions) gradient equations must now be solved for $m$ variables that need to
be lifted. We note that depending on the viscous fluxes, not all of the conservative variables might be lifted, or alternative sets
of variables can be used.

We apply the same discretization steps to the gradient equations, starting with the transformation to the computational space:
\begin{equation}
	\label{eqn:gradientEquationTransformed}
	\vec{g}^d - \frac{1}{\J} \gradientXI \cdot \vec{\mathcal{U}}^d = 0,
\end{equation}
with the contravariant solution in direction $d$
\begin{equation}
	\vec{\mathcal{U}}^d = \left( \J a_d^1,\J a_d^2,\J a_d^3   \right)^T  \vec{U}  .
\end{equation}
The weak formulation of equation~\eqref{eqn:gradientEquationTransformed} is again obtained by projection onto the space of test functions and applying Green's first identity
\begin{equation}
	\label{eqn:weakFormulationGradient}
	\left<\J \vec{g}^d ,\phi \right> - \iint_{\partial E} \phi \left( \vec{\mathcal{U}}^d \cdot \noutRef \right)^* \;\mathrm{d}S +
	\left<\vec{\mathcal{U}}^d,\gradientXI \phi \right> = 0.
\end{equation}
The discretized version of the weak gradient equations then reads
\begin{align}
	\label{eqn:discreteGradient}
	\vec{\mathbf{g}}_{ijk}^d = 
	\frac{1}{\J_{ijk}} \Bigg[ &\sum_{\alpha=0}^N \vec{\mathbfcal{U}}^{1,d}_{\alpha jk} \hat{D}_{i \alpha}  + \left(     \left[ \vec{U}^*  \nout_d\hat{s}
	\right]_{j,k}^{+\xi^1} \hat{\ell}_i (1) + \left[ \vec{U}^*  \nout_d\hat{s} \right]_{j,k}^{-\xi^{1}} \hat{\ell}_i (-1)
	\right) \nonumber  \\
	+ &\sum_{\beta=0}^N \vec{\mathbfcal{U}}^{2,d}_{i \beta k} \hat{D}_{j \beta}  + \left(     \left[ \vec{U}^*  \nout_d\hat{s}
	\right]_{i,k}^{+\xi^2} \hat{\ell}_j (1) + \left[ \vec{U}^*  \nout_d\hat{s} \right]_{i,k}^{-\xi^{2}} \hat{\ell}_j (-1)
	\right)\nonumber  \\
	+ &\sum_{\gamma=0}^N \vec{\mathbfcal{U}}^{3,d}_{i j \gamma} \hat{D}_{k \gamma}  + \left(     \left[ \vec{U}^*  \nout_d\hat{s} 
	\right]_{i,j}^{+\xi^3} \hat{\ell}_k (1) + \left[ \vec{U}^* \nout_d  \hat{s} \right]_{i,j}^{-\xi^{3}} \hat{\ell}_k (-1)    \right)
	\Bigg]\quad \forall\, i,j,k.
\end{align}
The specific lifting procedure now must define the unique solution $\vec{U}^*$ used as the numerical surface flux, and how to
calculate the numerical viscous flux at the cell boundaries for the main equations. The BR1 scheme simply uses the arithmetic mean
as the surface flux for the lifting equations
\begin{equation}
	\vec{U}^* = \frac{1}{2} \left( \vec{U}_L +  \vec{U}_R \right),
\end{equation}
and the arithmetic mean of the viscous fluxes in the numerical surface flux for the main equation
\begin{equation}
	\vec{F}^{v*} = \frac{1}{2} \left( \vec{F}^v \left(\vec{U}_L,\vec{g}_L \right) +  \vec{F}^v \left(\vec{U}_R,\vec{g}_R
	\right)\right).
\end{equation}

\subsection{Non-conforming meshes}\label{subsec:mortar}

\begin{figure}
\begin{center}
    \def\svgwidth{0.8\textwidth}
    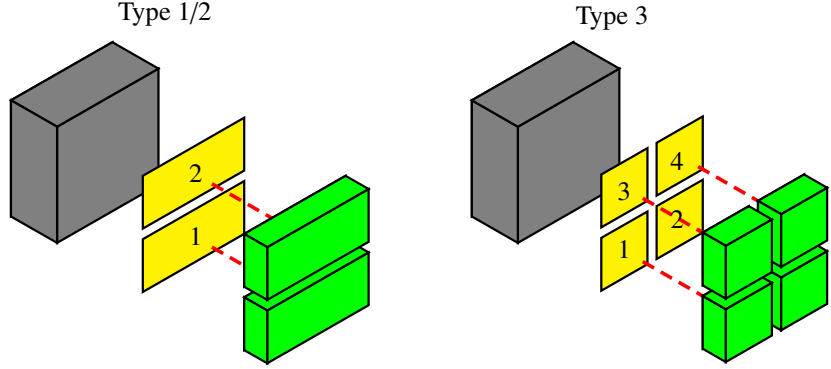
	\caption{\label{fig:mortar}Sketch of the supported mortar types. Yellow sides represent virtual small sides. Types one and
	two cut in different reference directions of the side.}
\end{center}
\end{figure}
Hexahedral mesh generation for arbitrary geometries in three dimensions is an area of active research and still a challenge. Many
algorithms used in such situations rely on unstructured and non-conforming meshes. Thus, to expand
the applicability of the solver towards more complex examples, a support for non-conforming meshes is needed. The so-called mortar
technique was introduced by Kopriva~\cite{kopriva1996conservative} to handle such interfaces, in a conservative and high order
accurate way. We restrict ourselves to three simple types of non-conforming interfaces, which are shown in Fig.~\ref{fig:mortar}.
For types one and two, a big side is halved into two smaller sides. The first type specifically cuts in the first computational
direction on the side, type two in the second direction. A combination of the first two results in the third type, where the big
side is cut into four smaller sides. It should be noted that the cuts are always made in the middle of the surface in reference
space.

The main idea of the mortar technique can be summarized as follows:
\begin{itemize}
	\item Interpolate the solution from the big side to the small sides.
	\item Calculate the numerical flux on the small sides.
	\item Project the flux onto the big side.
	\item Compute the surface integral for the small sides and the big sides.
\end{itemize}
The only additional steps introduced by the mortar technique are the interpolation to the small sides and the projection unto the
big sides, since flux calculation and surface integral are also performed for conforming sides. In the following, the interpolation
and projection operators will be derived. We only investigate the type one mortars in detail, since the other types follow in a
similar fashion. To distinguish the computational coordinates used on the surface from those in the volume, we will use the notation
$(\eta^1,\eta^2)$ for the latter.

For the type one mortar, the side is cut in half along the $\eta^1$ direction, which means that the $\eta^2$ coordinate is still
congruent. Thus, only one-dimensional operations are needed to interpolate the solution from the big side to the small sides, and
the degrees of freedom on the small sides can be computed as
\begin{equation}
	\vec{\mathbf{U}}_{ij}^L = \sum_{m=0}^N \vec{\mathbf{U}}_{mj}^B \ell_m \left( \eta_i^L \right)\quad \textrm{ and } \quad
	\vec{\mathbf{U}}_{ij}^U = \sum_{m=0}^N \vec{\mathbf{U}}_{mj}^B \ell_m \left( \eta_i^U \right) \quad \forall i,j.
\end{equation}
The superscripts are denoting the sides where the respective quantities are evaluated. The big side is marked with $B$, the two
smaller sides with $U$ and $L$ for upper and lower. Since we are interpolating from the big side, the positions of the interpolation
nodes $\eta_i^U$ and $\eta_i^L$  must be evaluated in the coordinate system of the big side. As the side is cut in half in
reference space, the positions can be expressed as
\begin{equation}
	\eta^{\mathrm{L}}_i = \frac{\eta_i-1}{2} \quad  \textrm{ and } \quad
	\eta^{\mathrm{U}}_i = \frac{\eta_i+1}{2}, \quad i=0,...,N.
\end{equation}
Here, $\eta_i$ are the positions of the interpolation nodes in the reference element $[-1,1]$. The node coordinates of the lower
element are in the range $[-1,0]$, and the upper element lies in the interval $[0,1]$. We can now define interpolation matrices for
both of the small sides
\begin{equation}
	I^L_{im} = \ell_m \left( \eta_i^L \right) \quad \textrm{ and } \quad 
	I^U_{im} = \ell_m \left( \eta_i^U \right),
\end{equation}
and use them to efficiently compute the interpolation for each position along $\eta^2$ using a matrix-vector multiplication:
\begin{equation}
	\underline{\vec{\mathbf{U}}}_{:j}^U = I^L \cdot \underline{\vec{\mathbf{U}}}_{:j}^B \quad \textrm{ and } \quad 
	\underline{\vec{\mathbf{U}}}_{:j}^L = I^U \cdot \underline{\vec{\mathbf{U}}}_{:j}^B \quad \forall j.
\end{equation}
The vector $\underline{\vec{\mathbf{U}}}_{:j}$ here contains all the degrees of freedom in $\eta^1$ direction at a specific position
$\eta^2=\eta_j$.

Once the values are interpolated to the small sides, the usual Riemann solvers can be used to compute the numerical flux functions.
No adaption to the mortar technique is necessary here. The surface integral for the small side can then be performed as usual, but
special care has to be taken for the big side. Following Kopriva~\cite{kopriva1996conservative}, we use $L^2$ projection to compute
the flux for the big side from the numerical fluxes of the two small sides. Thus, we are looking for a flux $\vec{f}^B$ on the big side
that satisfies for every test function $\phi$:
\begin{align}
	\int_{-1}^1 \int_{-1}^1 \vec{f}^B \left( \eta^{1B},\eta^2 \right) \phi \,\mathrm{d}\eta^{1B} \, \mathrm{d}\eta^{2} =
	&\int_{-1}^1 \int_{-1}^1 \vec{f}^L \left( \eta^{1L},\eta^2 \right) \phi \left( \eta^{1L},\eta^2 \right)\,\mathrm{d}\eta^{1L} \, \mathrm{d}\eta^{2} \nonumber\\
       +&\int_{-1}^1 \int_{-1}^1 \vec{f}^U \left( \eta^{1U},\eta^2 \right) \phi \left( \eta^{1U},\eta^2 \right)\,\mathrm{d}\eta^{1U} \, \mathrm{d}\eta^{2}.
\end{align}
The test function is defined on the big side, thus we need to transform the first coordinate for the integrals over the small sides.
As usual, we apply Gauss quadrature, insert the polynomial approximation of the flux and use the basis functions $\ell_i(\eta^{1B})
\ell_j(\eta^2) $ as test functions to gain
\begin{align}
	\sum_{\alpha,\beta=0}^N \sum_{m,n=0}^N &\vec{\mathbf{f}}^B_{mn} \underbrace{\ell_m(\eta_{\alpha})}_{\delta_{m\alpha}}
	\underbrace{\ell_n(\eta_{\beta})}_{\delta_{n\beta}}
	\underbrace{\ell_i(\eta_{\alpha})}_{\delta_{i\alpha}} \underbrace{\ell_j(\eta_{\beta})}_{\delta_{j\beta}} \omega_{\alpha} \omega_{\beta} \nonumber\\
	 = &
	\sum_{\alpha,\beta=0}^N \sum_{m,n=0}^N \vec{\mathbf{f}}^L_{mn} \underbrace{\ell_m(\eta_{\alpha})}_{\delta_{m\alpha}}
	\underbrace{\ell_n(\eta_{\beta})}_{\delta_{n\beta}}
	\ell_i\left(\frac{\eta_{\alpha}-1}{2}\right) \underbrace{\ell_j(\eta_{\beta})}_{\delta_{j\beta}} \omega_{\alpha} \omega_{\beta} \nonumber\\
	 + &
	\sum_{\alpha,\beta=0}^N \sum_{m,n=0}^N \vec{\mathbf{f}}^U_{mn} \underbrace{\ell_m(\eta_{\alpha})}_{\delta_{m\alpha}}
	\underbrace{\ell_n(\eta_{\beta})}_{\delta_{n\beta}}
	\ell_i\left(\frac{\eta_{\alpha}+1}{2}\right) \underbrace{\ell_j(\eta_{\beta})}_{\delta_{j\beta}} \omega_{\alpha}
	\omega_{\beta} \quad \forall i,j.
\end{align}
The Lagrange property allows us to simplify this equation to 
\begin{equation}
	\vec{\mathbf{f}}^B_{ij} = 
	\sum_{\alpha=0}^N \vec{\mathbf{f}}^L_{\alpha j} 
	\ell_i\left(\frac{\eta_{\alpha}-1}{2}\right) \frac{\omega_{\alpha}}{\omega_{i}} +
	\sum_{\alpha=0}^N \vec{\mathbf{f}}^U_{\alpha j} 
	\ell_i\left(\frac{\eta_{\alpha}+1}{2}\right) \frac{\omega_{\alpha}}{\omega_{i}} \quad \forall i,j. 
\end{equation}
We now introduce projection matrices
\begin{equation}
	P^L_{i\alpha} = \ell_i \left(\frac{\eta_{\alpha}-1}{2}\right) \frac{\omega_{\alpha}}{\omega_{i}}\quad \textrm{ and } \quad 
	P^U_{i\alpha} = \ell_i \left(\frac{\eta_{\alpha}+1}{2}\right) \frac{\omega_{\alpha}}{\omega_{i}},
\end{equation}
again allowing us to write and implement the projection as a matrix-vector multiplication for each position $j$:
\begin{equation}
	\underline{\vec{\mathbf{f}}}_{:j}^B = P^L \cdot \underline{\vec{\mathbf{f}}}_{:j}^L + P^U \cdot	\underline{\vec{\mathbf{f}}}_{:j}^U.
\end{equation}

The other types of mortars follow in a similar fashion. For type two, one simply needs to interchange the indices for the first
and second computational direction. The third type can then be constructed by concatenating the two operators defined for types one
and two, again showcasing the advantages of the tensor product structure.

\subsection{Split DG}\label{sec:splitDG}

A possible way to achieve robustness without the need for excessive artificial damping is to construct a numerical method
that is able to mimic energy or entropy estimates, that are valid in the continuous setting, on a discrete level. Recently, Gassner
et al.~\cite{gassner2016split} have shown how to design such schemes based on the strong form DGSEM on LGL nodes. This specific
version of DG possesses the diagonal norm summation-by-parts (SBP) property, a discrete analogon to integration by parts.  As Fisher
and Carpenter~\cite{fisher2013high} have shown, the application of a differentiation operator that fulfills the SBP property can be
re-written to make use of numerical volume flux functions. This allows us to express the volume integral in the
strong form on LGL nodes (discarding the influence of the Jacobian for simplicity, and only for the first direction) as:
\begin{equation}
	\sum_{\alpha=0}^N \vec{\mathbf{F}}^1_{\alpha jk} {D}_{i \alpha} \approx 2 \sum_{\alpha=0}^N
	\vec{F}^{\#,1}(\vec{\mathbf{U}}_{ijk},\vec{\mathbf{U}}_{\alpha jk}) {D}_{i \alpha},
\end{equation}
with the two-point flux function $\vec{F}^{\#,1}(\vec{\mathbf{U}}_{ijk},\vec{\mathbf{U}}_{\alpha jk})$. While every
two-point flux function that is consistent and symmetric can be used for a valid numerical scheme, additional properties might be
achieved by choosing specific flux functions. These properties include the conservation of entropy, following the definition of
Tadmor~\cite{tadmor1987numerical}, or preservation of kinetic energy as defined by Jameson~\cite{jameson2008formulation}. Gassner et
al.~\cite{gassner2016split} also showed how to relate the two-point fluxes to the discretization of split forms of the advection
terms of non-linear equations, which have been used extensively in the finite difference community to stabilize simulations (see
e.g.~\cite{blaisdell1996effect,kennedy2008reduced}). Thus, a framework was created that allowed for the direct translation of many
available split-formulations into the DG scheme, leading to the so-called \textit{split DG} method.

With the introduced discretization of the differentiation operator based on volume flux functions, the semi-discrete version of the
split DGSEM formulation can be written as:
\begin{align}
	\label{eqn:semiDiscreteSplit}
	& \pderivative{\vec{\tilde{\mathbf{U}}}_{ijk}}{t} = \nonumber\\
	- \Bigg[ &2 \sum_{\alpha=0}^N \vec{\mathbfcal{F}}^{\#,1} (\vec{\mathbf{U}}_{ijk},\vec{\mathbf{U}}_{\alpha jk}) {D}_{i \alpha}  + \left(     \left[ \vec{f}^*  \hat{s} - \vec{\mathbfcal{F}}^1
	\right]_{j,k}^{+\xi^1} \hat{\ell}_i (1) + \left[ \vec{f}^*  \hat{s}  - \vec{\mathbfcal{F}}^1\right]_{j,k}^{-\xi^{1}} \hat{\ell}_i (-1)    \right)  \nonumber\\
	+ &2 \sum_{\beta=0}^N \vec{\mathbfcal{F}}^{\#,2} (\vec{\mathbf{U}}_{ijk},\vec{\mathbf{U}}_{i \beta k}) {D}_{j \beta}  + \left(     \left[ \vec{f}^*  \hat{s} - \vec{\mathbfcal{F}}^2
	\right]_{i,k}^{+\xi^2} \hat{\ell}_j (1) + \left[ \vec{f}^*  \hat{s}  - \vec{\mathbfcal{F}}^2\right]_{i,k}^{-\xi^{2}} \hat{\ell}_j (-1)    \right) \nonumber\\
	+ &2 \sum_{\gamma=0}^N \vec{\mathbfcal{F}}^{\#,3} (\vec{\mathbf{U}}_{ijk},\vec{\mathbf{U}}_{i j \gamma}) {D}_{k \gamma}  + \left(     \left[ \vec{f}^*  \hat{s} - \vec{\mathbfcal{F}}^3
	\right]_{i,j}^{+\xi^3} \hat{\ell}_k (1) + \left[ \vec{f}^*  \hat{s}  - \vec{\mathbfcal{F}}^3\right]_{i,j}^{-\xi^{3}} \hat{\ell}_k (-1)    \right)
	\Bigg] \nonumber\\ & \quad \quad \quad \quad \quad \quad \quad \quad \quad \quad \quad \quad \quad \quad \quad \quad \quad \quad \quad \quad \quad \quad\forall\, i,j,k.
\end{align}
Here, the contravariant two-point fluxes are defined in the following way, for e.g. the first computational direction:
\begin{align}
	\vec{\mathbfcal{F}}^{\#,1} (\vec{\mathbf{U}}_{ijk},\vec{\mathbf{U}}_{\alpha jk}) =  \Big[ 
		 &\vec{F}^{\#,1} (\vec{\mathbf{U}}_{ijk},\vec{\mathbf{U}}_{\alpha jk}) \avg{ \J a^1_1}_{(i,\alpha)jk} \nonumber\\
		+& \vec{F}^{\#,2} (\vec{\mathbf{U}}_{ijk},\vec{\mathbf{U}}_{\alpha jk}) \avg{ \J a^1_2}_{(i,\alpha)jk}\nonumber\\
		+& \vec{F}^{\#,3} (\vec{\mathbf{U}}_{ijk},\vec{\mathbf{U}}_{\alpha jk}) \avg{ \J a^1_3}_{(i,\alpha)jk} \Big],
\end{align}
where the arithmetic mean is denoted by $\avg{\cdot}$.  

For the surface fluxes, the same two-point fluxes as in the volume can be used. If those fluxes would be e.g. conservative in an
entropy, then the (semi-discrete) scheme would also be entropy-conserving. Since the goal is mainly to construct a scheme that is
guaranteed dissipative in the respective quantity, additional dissipation terms are added for the surface fluxes, based e.g. on Roe-type
dissipation.

\section{Shock Capturing}\label{sec:shockcapturing}
High order methods suffer from well-known stability problems in non-smooth regions of the solution, for example across the shocks
and discontinuities occurring in compressible gas dynamics. Adopting high order methods for these situations despite this inherent
mismatch usually follows along two lines of thought: The non-smooth solution is regularized to bring it into the approximation space
of the discretization method. This is typically achieved by the introduction of a local dissipative operator that acts on the high
gradients of the solution and essentially smears them out. A classical example of this is the method of artificial viscosity,
originally introduced for finite difference schemes by von Neumann and Richtmyer~\cite{vonneumann1950method}. An adaption to high
order DG schemes was described by Persson and Peraire~\cite{persson2006sub}.  The second general approach to shock capturing is to
combine or replace the high order discretizations by more robust operators, for example WENO reconstruction~\cite{qiu2004hermite}.
Another option is to reduce the local order of the approximation, e.g. revert to first order, often in combination with
$h$-refinement~\cite{huerta2012simple}. The method implemented in \Flexi for shock capturing follows a variant of this method. The
general idea is to replace the DG operator in troubled cells by a finite volume scheme and then take advantage of the excellent
shock capturing properties of this formulation. Since each DG cell contains $(N+1)^d$ degrees of freedom, the finite volume
formulation can take advantage of this information richness by transferring the high order polynomial into $(N+1)^d$ first order
finite volume representations. This can be understood as a reinterpretation of the DG solution on a structured finite volume mesh.
This approach not only moderates the loss in effective resolution when switching from DG to FV, but also makes the implementation
easy and efficient through the reliance on the same data structure. In this sense, for shock capturing, we append the DG scheme by
the option of locally switching to a block structured FV scheme. Details on the method are given in the following sections, the
implications for the implementation are discussed in Sec.~\ref{subsec:implementation:fv}.

\subsection{Finite Volume Scheme on Subcells}
In Sec.~\ref{subsec:dgsem}, we have derived the DGSEM formulation in arbitrary dimensions. In order to keep the formulation simple
and the equations manageable, we will restrict the discussion of the FV operator in the following sections to 2D. However, it will
become apparent that the extension to 3D is naturally given by the tensor product structure of both the DG and FV discretizations.
The discussion herein follows the more extensive descriptions in~\cite{sonntag2017efficient,sonntag2014shock}. We thus start the
discussion by recalling that in each single DG element, we approximate the solution by a tensor product of 1D Lagrange polynomials of
degree $N$, which are associated with $N+1$ Legendre-Gauss(-Lobatto) nodes, leading to $(N+1)^2$ number of degrees of freedom
$nDOF_{DG}$ in each DG element $E$. In theory, the number of finite volume subcells and thus $nDOF_{FV}$ per element is
completely independent from $nDOF_{DG}$. However a strong reason exists to choose $nDOF_{FV}=nDOF_{DG}$: In this case, the
\emph{same} data structures for storing both solution representations can be used, which saves memory and avoids data copying
operations. It is thus a natural expression of the notion outlined above that FV and DG are a different interpretation of
the same underlying solution. In addition, this choice for the number of finite volume subcells also makes sense from a
load-balancing point of view: For an optimal load distribution, both DG and FV elements should incur identical computational costs.
As we will show later, our choice of $nDOF_{FV}=nDOF_{DG}$ leads to nearly even computational effort for both operators. The next
discretization choice concerns the distribution, size and location of the FV subcells within a DG element. A natural choice is an
isotropic, equispaced distribution, leading to a regular FV grid, depicted in Fig.~\ref{fig:fv_subcells}. It should be noted that
with this distribution, the FV subcells inside a DG element can be interpreted as a local ``block'' grid. Adjacent blocks are then
connected via the surface fluxes, and thus may be interpreted as a block-structured FV discretization.

\begin{figure}
   \centering

      % ----------- END OF HEADER -------------

\begin{tikzpicture}[line cap=round,line join=round,>=triangle 45,x=1.5cm,y=1.5cm]

   \foreach \a in {-1,1} {
       \draw (\a, -1) -- (\a, 1); 
       \draw (-1, \a) -- (1, \a); 
   }
   \foreach \a in \innerborderseq {
       \draw [dash pattern=on 2 off 2] (\a, -1) -- (\a, 1); 
       \draw [dash pattern=on 2 off 2] (-1, \a) -- (1, \a); 
   }
   \foreach \x in \xGP {
       \foreach \y in \xGP {
           \draw [fill=black] (\x,\y) circle (0.02);
       }
   }
   \xdef\lasta{-1}
   \edef\tmp{\innerborderseq, 1}
   \foreach \a in \tmp {
      \draw [>=stealth,<->] (\lasta, -1.2) -- node[anchor=north] {$w\vphantom{\omega_3}$} (\a,-1.2); 
      \draw [>=stealth,<->] (-1.2, \lasta) -- node[anchor=east] {$w$} (-1.2,\a); 
      \xdef\lasta{\a}
   }

   \foreach \x in {-1,1} {
      \foreach \y in \xGPeq {
         \draw[fill=white] (\x,\y) +(-0.03,-0.03) rectangle +(0.03,0.03);
         \draw[fill=white] (\y,\x) +(-0.03,-0.03) rectangle +(0.03,0.03);
      }
   }
   \foreach \x in \innerborderseq {
      \foreach \y in \xGPeq {
         \draw[fill=gray] (\x,\y) +(-0.03,-0.03) rectangle +(0.03,0.03);
         \draw[fill=gray] (\y,\x) +(-0.03,-0.03) rectangle +(0.03,0.03);
      }
   }

\end{tikzpicture}

% ----------- FOOTER -------------
   \caption{DG reference element split into FV subcells \protect\tikz[x=2cm,y=2cm,baseline=-0.5ex]\protect\draw[dash pattern=on 2 off 2] (0,0) -- (0.18,0); 
      with Gauss points~\protect\tikz[x=2cm,y=2cm,baseline=-0.5ex]\protect\draw[fill=black] (0,0) circle (0.02); 
      of the original DG reference element,
      locations of the inner~\protect\tikz[x=2.0cm,y=2.0cm,baseline=-0.5ex]\protect\draw[fill=gray] (0,0)  +(-0.03,-0.03) rectangle +(0.03,0.03); 
      and the interface~\protect\tikz[x=2.0cm,y=2.0cm,baseline=-0.5ex]\protect\draw[] (0,0)  +(-0.03,-0.03) rectangle +(0.03,0.03); 
      boundary fluxes as well as the sizes $w$ of the subcells.
   }
   \label{fig:fv_subcells}
\end{figure}
Since the distribution is constructed in reference space, each FV subcell has an area of$w \cdot w$, with $w=\frac{2}{N+1}$.
Each of the subcells now contains a DG solution point $ij$, so we label the FV cells as $\kappa_{ij}$ and formulate the FV scheme of
the transformed equation Eq.~\eqref{eqn:3DconsLawTransformed} on each of them as
\begin{equation}
	\int_{\kappa_{ij}} \J \pderivative{\vec{U} }{t} + \gradientXI \cdot  \vec{\F} \,\,\mathrm{d}\vec{\xi} = 0, \quad \forall{\kappa_{ij} \in E}.
\end{equation}
Applying the divergence theorem yields the classical FV formulation
\begin{equation}\label{eq:fv:continuous}
\int_{\kappa_{ij}} \J \pderivative{\vec{U}^{FV} }{t} \,\,\mathrm{d}\vec{\xi}+ \int_{\partial \kappa_{ij}}\vec{\F}(\vec{U}^{FV}) \cdot \noutRef \,\,\mathrm{d}S_{\kappa_{ij}} = 0, \quad \forall{\kappa_{ij} \in E},
\end{equation}
where we have also introduced the superscript $FV$ to distinguish the DG and FV solution. We will now first derive the operators
that allow consistent transformation between the DG and FV solutions, and then discretize Eq.~\eqref{eq:fv:continuous}. To ensure
conservation, we require that
\begin{equation}\label{eq:fv:cons}
\int_E \vec{U} \,\,\mathrm{d}\vec{\xi} = \int_E \vec{U}^{FV} \,\,\mathrm{d}\vec{\xi} = \sum_{i,j=0}^N \int_{\kappa_{ij}} \vec{U}^{FV} \,\,\mathrm{d}\vec{\xi} 
\end{equation}
holds.
Taking again advantage of the tensor product structure of both the FV and DG operators, we can consider Eq.~\eqref{eq:fv:cons} for the one-dimensional case as
\begin{equation}\label{eq:fv:cons1d}
\int_E \vec{U} \,\,\mathrm{d}\vec{\xi} = \int_{-1}^{+1} \vec{U} \,\,\mathrm{d}\vec{\xi}  = \sum_{k=0}^N \int_{-1+k\cdot	w}^{-1+(k+1)\cdot w} \vec{U}^{FV} \,\,\mathrm{d}\vec{\xi}, 
\end{equation}
where we have defined the FV subcells on the reference line as 
\begin{equation}
[-1 + k\cdot w, -1+(k+1)\cdot w]  \quad \forall k =0, \ldots, N.
\end{equation}
Replacing the second integral in Eq.~\eqref{eq:fv:cons1d} by the sum over the FV subcells, inserting the DG solution according to
Eq.~\eqref{eqn:approxSol} and noting that the FV solution $\vec{U}^{FV}_{ij}$ is constant in every subcell $\kappa_{ij} $ yields
\begin{equation}\label{eq:fv:cons_dof}
\sum_{k=0}^N \int_{-1+k\cdot w}^{-1+(k+1)\cdot w}  \sum_{i=0}^{N}\vec{U}_i\ell_i(\xi) \,\,\mathrm{d}\vec{\xi}  = \sum_{k=0}^N \vec{U}^{FV}_k w.
\end{equation}
This formulation ensures a conservative definition of the mean values $\vec{U}^{FV}$ in each subcell. In order to solve
Eq.~\eqref{eq:fv:cons_dof}, the integration of the DG polynomial is achieved by Gauss quadrature in each subcell, leading to 
\begin{equation}\label{eq:fv:vdm}
\frac{w}{2}\sum_{k=0}^{N}\sum_{\lambda=0}^{N}\sum_{i=0}^{N} \vec{U}_i\ell_i(\xi^k_\lambda) \omega_\lambda = \sum_{k=0}^N \vec{U}^{FV}_k w.
\end{equation}
Here, the scaling factor $\frac{w}{2}$ relates the reference line $ [-1,1 ] $
to the length $w$ of the subintervalls, and $\xi^k_\lambda$ are the integration
nodes in the $k$-th subcell. Also note that
$\ell_i(\xi^k_\lambda)\neq\delta_{ik}$, since the Lagrange basis functions are
constructed w.r.t. the integration nodes in the
reference interval and not the local subintervall. With the help of the Vandermonde matrix 
\begin{equation}
\label{eq:FVvdm}
V_{DG\Rightarrow FV} := \left\lbrace \frac{1}{2} \sum_{\lambda=0}^N \omega_\lambda \ell_i(\xi^k_\lambda) \right\rbrace_{k,i=0}^N.
\end{equation}
Equation~\eqref{eq:fv:vdm} can be rewritten as
\begin{equation}\label{eq:dg2fv}
\vec{U}^{FV} = V_{DG\Rightarrow FV} \vec{U},
\end{equation}
which can be implemented as a simple matrix-vector multiplication in one spatial dimension. The backward transformation
$FV\Rightarrow DG$ is facilitated by $V_{FV\Rightarrow DG} = V_{DG\Rightarrow FV}^{-1} $. Turning now to the discretization of
Eq.~\eqref{eq:fv:continuous}, the usual steps for an FV scheme are followed, i.e. the introduction of a numerical flux function at the
cell interfaces and the approximation of the integrals by the midpoint rule. With $J_{ij}$ being the integral value of the original
Jacobian determinant inside subcell $\kappa_{ij}$ and $f_{i,j+\frac{1}{2}}$ for example being the flux at the top edge of the
$ij$-th subcell, the FV scheme on the block-structured grid reads
\begin{equation}
\label{eq:fv_dof_time_derivative}
\begin{aligned}
w J_{ij} \fracp{\vec{U}^{FV}_{ij}}{t} = 
&-  \left[ f_{i-\frac{1}{2},j}\left({\vec{U}^{FV-}},{\vec{U}}^{FV+},\noutRef\right) + f_{i+\frac{1}{2},j}\left(\vec{U}^{FV-},\vec{U}^{FV-},\noutRef\right)\right] \\
&-  \left[ f_{i,j-\frac{1}{2}}\left(\vec{U}^{FV-},\vec{U}^{FV+},\noutRef\right) + f_{i,j+\frac{1}{2}}\left(\vec{U}^{FV-},\vec{U}^{FV+},\noutRef\right)\right]. \\
\end{aligned}
\end{equation}

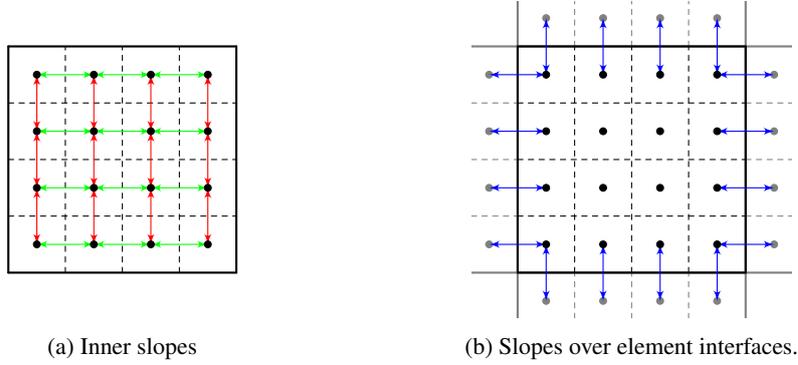
\begin{figure}[htb] 
   \centering
   \begin{subfigure}[t]{0.49\textwidth}
      \centering
      % ----------- END OF HEADER -------------

\begin{tikzpicture}[line cap=round,line join=round,>=triangle 45,x=1.5cm,y=1.5cm,axis/.style={->}]
   \clip(-1.4,-1.4) rectangle (1.4,1.4);

   \foreach \a in {-1,1} {
      \draw[thick] (\a, -1) -- (\a, 1); 
      \draw[thick] (-1, \a) -- (1, \a); 
   }
   \foreach \a in \innerborderseq {
      \draw [dash pattern=on 2 off 2] (\a, -1) -- (\a, 1); 
      \draw [dash pattern=on 2 off 2] (-1, \a) -- (1, \a); 
   }

   \foreach \y in \xGPeq {
      \foreach \x[count=\xi from 1,remember=\x as \lastx] in \xGPeq {
         \ifnum \xi > 1
         \draw[green,<->,>=latex'] (\lastx,\y) -- (\x,\y);
         \fi
      }
   }

   \foreach \y in \xGPeq {
      \foreach \x[count=\xi from 1,remember=\x as \lastx] in \xGPeq {
         \ifnum \xi > 1
         \draw[red,<->,>=latex'] (\y,\lastx) -- (\y,\x);
         \fi
      }
   }

   \foreach \x in \xGPeq {
      \foreach \y in \xGPeq {
         \draw [fill=black] (\x,\y) circle (0.03);
      }
   }
\end{tikzpicture}

% ----------- FOOTER -------------
      \caption{Inner slopes}
      \label{fig:slope_inner}
   \end{subfigure}
   \begin{subfigure}[t]{0.49\textwidth}
      \centering
      % ----------- END OF HEADER -------------

\begin{tikzpicture}[line cap=round,line join=round,>=triangle 45,x=1.5cm,y=1.5cm,axis/.style={->}]

   \clip(-1.4,-1.4) rectangle (1.4,1.4);

   \foreach \x/\y in {-2/0,2/0,0/-2,0/2} {
      \begin{scope}[shift={(\x,\y)}]
         \foreach \a in {-1,1} {
            \draw[gray,thick] (\a, -1) -- (\a, 1); 
            \draw[gray,thick] (-1, \a) -- (1, \a); 
         }
         \foreach \a in \innerborderseq {
            \draw [gray,dash pattern=on 2 off 2] (\a, -1) -- (\a, 1); 
            \draw [gray,dash pattern=on 2 off 2] (-1, \a) -- (1, \a); 
         }

         \foreach \x in \xGPeq {
            \foreach \y in \xGPeq {
               \draw [gray,fill=gray] (\x,\y) circle (0.03);
            }
         }
      \end{scope}
   }
   
   \foreach \a in {-1,1} {
      \draw[thick] (\a, -1) -- (\a, 1); 
      \draw[thick] (-1, \a) -- (1, \a); 
   }
   \foreach \a in \innerborderseq {
      \draw [dash pattern=on 2 off 2] (\a, -1) -- (\a, 1); 
      \draw [dash pattern=on 2 off 2] (-1, \a) -- (1, \a); 
   }

   \foreach \y in \xGPeq {
      \draw[blue,<->,>=latex'] ({\xGPeq}[0],\y) -- ($(-2-{\xGPeq}[0],\y)$);
      \draw[blue,<->,>=latex'] ({\xGPeq}[3],\y) -- ($( 2-{\xGPeq}[3],\y)$);
      
      \draw[blue,<->,>=latex'] (\y,{\xGPeq}[0]) -- ($(\y,-2-{\xGPeq}[0])$);
      \draw[blue,<->,>=latex'] (\y,{\xGPeq}[3]) -- ($(\y, 2-{\xGPeq}[3])$);
   }

   \foreach \x in \xGPeq {
      \foreach \y in \xGPeq {
         \draw [fill=black] (\x,\y) circle (0.03);
      }
   }
\end{tikzpicture}

% ----------- FOOTER -------------
      \caption{Slopes over element interfaces.}
      \label{fig:slope_outer}
   \end{subfigure}
   \caption{Computation of slopes for second order reconstruction (green: $\xi^1$-direction, red: $\xi^2$-direction, blue: interface slopes).}
\end{figure}
The first order FV method described up to now is overly dissipative away from the shocks. To ameliorate the properties of the scheme
in smooth regions adjacent to the shock, we extend it by a second order reconstruction procedure. Slope limiters are introduced to
avoid an associated loss of the TVD (total variation diminishing, see e.g ~\cite{harten1983high,van1979towards}) property; we apply
the reconstruction and limiting procedure to primitive variables to prevent violation of thermodynamic relations.  In a first step,
the inner slopes between the centroids of the FV subcells inside a DG element are computed along $\xi^1$ and $\xi^2$ lines (see
Fig.~\ref{fig:slope_inner}). These slopes can then be limited with standard approaches, e.g. those described
in~\cite{Roe1986,sweby1984,vanLeer1974}. In a second step, the slopes that require neighboring information are evaluated, see
Fig.~\ref{fig:slope_outer}). 
%If this requires crossing an MPI interface in a parallel context, additional bilateral communication is
%introduced. 

\subsection{FV/DG Interfaces}
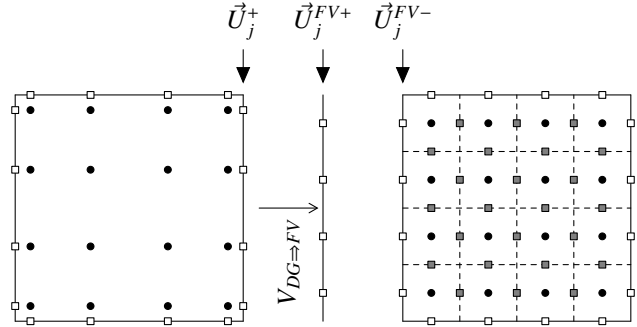
\begin{figure}
	\centering
	% ----------- END OF HEADER -------------

\begin{tikzpicture}[line cap=round,line join=round,>=triangle 45,x=1.5cm,y=1.5cm,axis/.style={->}]
   \foreach \a in {-1,1} {
      \draw (\a, -1) -- (\a, 1); 
      \draw (-1, \a) -- (1, \a); 
   }
      \foreach \x in \xGP {
         \foreach \y in \xGP {
         \draw [fill=black] (\x,\y) circle (0.03);
      }
   }
   \foreach \x in {-1,1} {
      \foreach \y in \xGP {
         \draw[fill=white] (\x,\y) +(-0.03,-0.03) rectangle +(0.03,0.03);
         \draw[fill=white] (\y,\x) +(-0.03,-0.03) rectangle +(0.03,0.03);
      }
   }
   \draw[->] (1,1.4) node[anchor=south]{$\vec{U}^+_j$} -- (1,1.1);

   \begin{scope}[shift={(1.7,0)}]
      \draw[->] (0,1.4) node[anchor=south]{$\vec{U}^{FV+}_j$} -- (0,1.1);
      \draw[->,>=angle 60] (-0.56, 0) -- node[anchor=east,rotate=90]{$V_{DG\Rightarrow FV}$} (-0.04, 0);
      \draw (0, -1) -- (0, 1);
      \foreach \y in \xGPeq {
         \draw[fill=white] (0,\y) +(-0.03,-0.03) rectangle +(0.03,0.03);
      }
   \end{scope}

   \begin{scope}[shift={(3.4,0)}]
   \draw[->] (-1,1.4) node[anchor=south]{$\vec{U}^{FV-}_j$} -- (-1,1.1);
      \foreach \a in {-1,1} {
         \draw (\a, -1) -- (\a, 1); 
         \draw (-1, \a) -- (1, \a); 
      }
      \foreach \a in \innerborderseq {
         \draw [dash pattern=on 2 off 2] (\a, -1) -- (\a, 1); 
         \draw [dash pattern=on 2 off 2] (-1, \a) -- (1, \a); 
      }
      \foreach \x in \xGPeq {
         \foreach \y in \xGPeq {
            \draw [fill=black] (\x,\y) circle (0.03);
         }
      }
      %\foreach \a [count=\i from 0] in \xGPeq {
      %\draw[color=black] (\a,-1) node [anchor=north] {$\omega_\i$};
      %}
      %\foreach \a [count=\i from 0] in \xGPeq {
      %\draw[color=black] (-1,\a) node [anchor=east] {$\omega_\i$};
      %}
      \foreach \x in {-1,1} {
         \foreach \y in \xGPeq {
            \draw[fill=white] (\x,\y) +(-0.03,-0.03) rectangle +(0.03,0.03);
            \draw[fill=white] (\y,\x) +(-0.03,-0.03) rectangle +(0.03,0.03);
         }
      }
      \foreach \x in \innerborderseq {
         \foreach \y in \xGPeq {
            \draw[fill=gray] (\x,\y) +(-0.03,-0.03) rectangle +(0.03,0.03);
            \draw[fill=gray] (\y,\x) +(-0.03,-0.03) rectangle +(0.03,0.03);
         }
      }
   \end{scope}
\end{tikzpicture}

% ----------- FOOTER -------------
	\caption{Coupling of a discontinuous Galerkin element with equidistant distributed finite volume subcells. }
	\label{fig:coupling_equidistant}
\end{figure}

It is clear from the discussion on properties of high order methods that switching to a FV representation should only occur in
regions of shocks or unresolvably strong gradients. Detecting such troubled cells is an open area of research in itself, and we will
briefly present our approach using indicators in Sec.~\ref{subsec:fv:indicators}. In any case, once a troubled DG element is
detected and switched to a FV representation, a DG/FV interface as depicted in Fig.~\ref{fig:coupling_equidistant} appears. Since
both operators are coupled to the surroundings weakly via flux functions, it is natural to use this for the DG/FV interface
definition as well. However, since the points of flux evaluation of the FV and DG schemes differ along their interfaces, a direct computation
of the pointwise flux functions is not possible. Instead, the DG solution $\vec{U}^+_j$ is first evaluated at the FV solution points
with the help of the Vandermonde matrix defined in Eq.~\eqref{eq:FVvdm}. The only difference to Eq.~\eqref{eq:dg2fv} is that here it is
applied to face data only and not to the whole volume. Once this operation yields $\vec{U}^{FV+}_j$, the pointwise numerical flux
function is then evaluated. It can directly be applied to the DOF in the FV element to update the solution in subcells. The transfer
to the DG solution points occurs by the inverse of Eq.~\eqref{eq:FVvdm}. Note that a FV/DG interface can also occur across the mortar
interfaces discussed in Sec.~\ref{subsec:mortar}. While in principle the coupling in this case uses the same concepts that are presented
here, some subtleties and special cases need to be considered. For a detailed discussion, we refer the reader
to~\cite{sonntag2017shape}.

\subsection{Troubled Cell Indicators}\label{subsec:fv:indicators}
\begin{figure}
	\centering
	\includegraphics{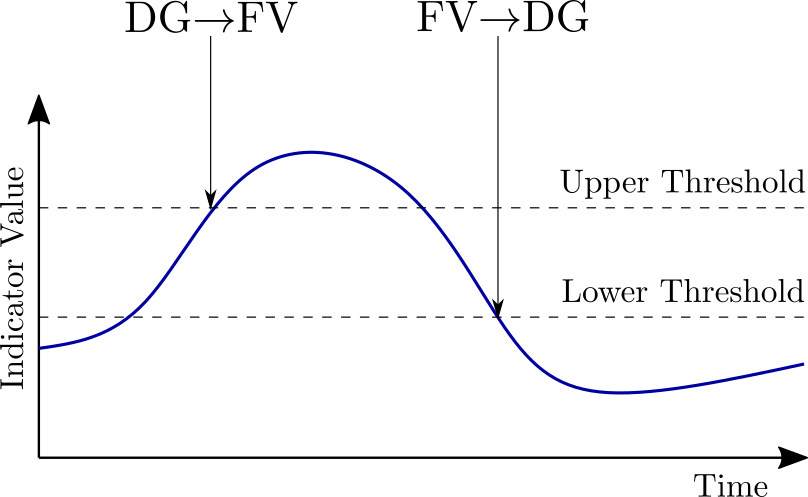}
	\caption{Indicator based switching between DG and FV sub-cell method. }
	\label{fig:fv:indicator}
\end{figure}

In the previous sections, we have presented the operators that allow a fast and efficient switching of the element type from a DG to
a FV-subcell element and vice versa. It is clear that the switching itself costs some computational effort and should occur as
infrequently as possible. Also, cells that require FV treatment to ensure stability should be detected and switched in time, while
they should be able to revert back to the DG representation if the solution permits. Detecting candidate cells for switching thus
requires the evaluation of some criterion - either in an a priori or an a posteriori approach. In the latter, less frequently used case,
all elements are computed at each time step with the DG scheme and the solution is evaluated afterwards. If it shows
signs of instability, the time step is repeated, with the marked cells being treated by a more stable
method~\cite{dumbser2014posteriori}. In the more prevalent a priori method, the solution at the beginning of each time step is
examined, and cells are then flagged accordingly and treated by the appropriate numerical scheme. In any case, evaluating the cells
is achieved by some form of indicator function. Often, these are loosely called "shock detectors", however, it is worthwhile
pointing out they are used more as troubled cell indicators. What makes a cell suffer from instabilities is not just a function of
the solution (i.e. the presence of a shock, steep gradient or not) itself, but even more so of the discretization scheme and its
properties. A situation that might lead to oscillations in one discretization scheme might be treated more robustly in another, and
thus, troubled cell indicators need to be tuned to the specific discretization at hand. A large number of indicators exist which are
based on physical reasoning, jump conditions or model solution
representation~\cite{persson2006sub,jameson1981numerical,ducros1999large}. All of these cited indicators are available in \Flexi. They are
implemented to return a large value in regions of non-smoothness and a low value otherwise. Starting from purely DG elements, if the
indicator increases over a given threshold, the element in question is switched to a FV subcell element before the next time step.
It remains in this discretization state until the indicator drops below a given value. The solution in this element is then seemed
smooth enough to warrant a DG discretization of residual without the introducing stability issues. Numerical studies have shown that
a single threshold might lead to excessive back- and forth switching and a high sensitivity to the actual, user-defined value.
Therefore, the single threshold has been replaced by an upper and a lower threshold as visualized by the range in
Fig.~\ref{fig:fv:indicator}.

\subsection{Time integration}\label{subsec:timeintegration}

While the semi-discrete systems can be, in principle, advanced in time by any suitable time integration method, both implicit and
explicit, we restrict ourselves to explicit Runge-Kutta type schemes, specifically low-storage versions. For this class of methods,
there is no need to store the result from every Runge-Kutta stage, thus reducing the memory requirements and leading to a scheme
which is suitable for HPC. In general, the schemes used are based on the two-register formalism introduced
by Williamson~\cite{williamson1980low}, and several specific schemes are available to choose from. In the hybrid DG/FV approach,
both systems are advanced in time with the same method.

\subsubsection*{Time step restrictions}

Since we use explicit Runge-Kutta time integration, the time step is limited by the famous CFL
condition~\cite{courant1928partiellen} for a given mesh size. For our DGSEM, the condition can be written as:
\begin{equation}
	\label{eqn:CFLCondition}
	\Delta t \leq \mathrm{CFL} \cdot \gamma^1_{RK}(N) \frac{\Delta x}{\lambda^c \left( 2N+1  \right)},
\end{equation}
where CFL is the CFL number, $\Delta x$ the size of an element and $\lambda^c$ the largest convective signal velocity. Usually, a
scaling factor of $\frac{1}{2N+1}$ is included to take the high order solution in the element into account~\cite{cockburn2001runge}.
In practice, this scaling is not accurate, and the user might not be able to tell the valid range of the CFL number for each
configuration. To circumvent this problem, we introduce an additional correction factor $\gamma^1_{RK}(N)$, that depends on the
polynomial degree and the set of interpolation nodes. This factor is derived from numerical experiments such that the usual
condition $\mathrm{CFL} \leq 1$ must hold to obtain a (linear) stable time step. 

The largest signal velocity $\lambda^c$ depends on the specific set of equations currently considered, where those velocities are
given by the eigenvalues of the system matrix. While this restriction stems from the hyperbolic part of the equations, the parabolic
fluxes also introduce time step limitations. Those can be expressed analogously to the hyperbolic restriction
\begin{equation}
	\label{eqn:DFLCondition}
	\Delta t \leq \mathrm{CFL}_{\mathrm{d}} \cdot \gamma^2_{RK}(N) \frac{\Delta x^2}{\lambda^v \left( 2N+1  \right)},
\end{equation}
where a diffusive CFL number $\mathrm{CFL}_{\mathrm{d}}$ was introduced. The factor $\lambda^v$ here is the largest eigenvalue of the diffusion matrix.
Since this restrictions scales as $\Delta x^2$, it is especially restrictive in regions where the mesh size reduces significantly.

Both of these conditions must be satisfied for a stable simulation, so that the minimum of the two will be used for the time step in
the computation. The conditions are evaluated for each of the directions in the reference element, thus the physical size of the
element enters through the transformations.

For cells that are discretized with the FV method, the time step restriction can be expressed as
\begin{equation}
\label{eq:time step_fv_org}
\Delta t \le CFL \cdot \gamma_{RK}^1(0) \frac{\Delta x_{FV}}{\lambda^c},
\end{equation}
where $\Delta x_{FV}=\Delta x_{DG}/(N+1)$.
Note that for a given flow field the time step given by the DG representation is smaller than for the FV subcell representation.
\section{Implementation details}\label{sec:implementation}

This section will describe how the DGSEM and the hybrid DG/FV scheme is implemented in the solver. Aspects that will be discussed
include the data structures, parallelization approach and general structure of the algorithm. Performance aspects like the required
count of operations and the memory consumption are investigated, before the parallel scalability is reviewed. 

\subsection{Coordinate systems for unstructured grids}\label{sec:coordSys}

\begin{figure}
    \centering
    \def\svgwidth{0.8\textwidth}
    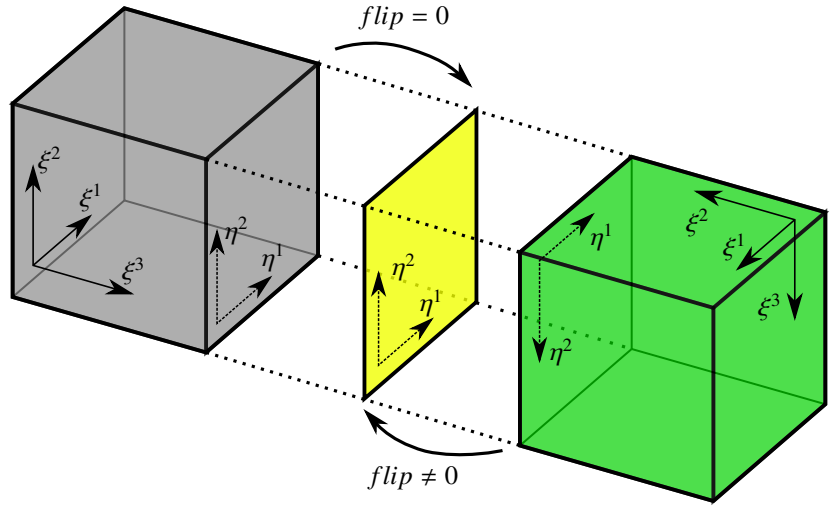
    \caption{Schematic of relationship between volume and side coordinate systems. Volume systems denoted by solid coordinate axis,
	side systems by dotted axis. The grey element is the master for the yellow side, the green element the slave.}
    \label{fig:coordSys}
\end{figure}

Operations in the DG method either act on the volume of a cell or on sides, which are shared by two adjacent elements (excluding the
boundaries). In our mesh structure, elements and sides are thus handled separately, and sides are only stored once and connected to
the respective elements through a mapping. To be able to handle unstructured grids, several different coordinate systems and their
relationship must first be introduced. Figure~\ref{fig:coordSys} shows a sketch of the used systems. Each element has a
\textit{local volume coordinate system}, denoted by $\xi^1$, $\xi^2$ and $\xi^3$, which is always a right-hand system. On each side
of the element, a \textit{local side coordinate system} ($\eta^1$ and $\eta^2$) is introduced, following the definitions given by
the CGNS (see e.g.~\cite{rumsey2012recent}) standard. But, as each side is shared by two elements, local side coordinate systems from those two elements might differ.
To obtain a definitive system that should be used for operations on that side, one of the two elements will be designated as
the \textit{master} of that side (and the other as \textit{slave}). The local coordinate system of the master element will then  be
used as the \textit{global side coordinate system} for that respective surface. To convert from local side systems to global ones, a
\textit{flip} is defined that represents how the two systems are oriented with respect two each other. A flip of $0$ is used to
indicate a master side.

The conversion between side and volume information thus requires first a mapping between the volume coordinate system and local side
systems. This only depends on the currently considered local side (e.g. $+\xi^1$) and the relation can be implemented as a
lookup-table. Second, the mapping between local and global sides is uniquely defined by the flip, and since only a limited number of
possible combinations exist, this can also be computed and stored before the actual simulation starts.

\subsection{Domain decomposition}

The natural parallelization approach for the DG method is to divide the
elements in the mesh into separate domains (each containing one or more mesh elements), and to assign each domain to an available
processing unit. We want the decomposition to be flexible, so it can handle computations on an arbitrary number of processes
without the need for expensive re-calculations. Additionally, domains should be compact and contiguous, to reduce the required amount of
communication between the domains. As has been mentioned in section~\ref{sec:meshFormat}, \Hopr provides the SFC sorting to achieve this. 
When a parallel computation is started, the SFC can simply be subdivided into as many parts as processing units are available.

\subsection{Parallelization aspects}

\Flexi uses a distributed memory approach for parallelization. Thus, across boundaries of the mesh domains, communication will be
necessary. An advantage of the DG method is that, due to the local approximation space and the absence of continuity requirements on
the solution across element interfaces, only surface data needs to be communicated to the direct neighbour elements, reducing the
amount of data compared with other schemes significantly. Thus, all communications operations are based on side data structures. To
perform the actual communication, the Message Passing Interface (MPI) protocol is used. It is widely available and optimized on HPC
architecture, and allows for non-blocking operations that can be used to hide communication time. For efficient communication
hiding, one can take advantage of the high density of local operations of the DG operator, as is explained in detail
later. 

The communication patterns are build at the start of the simulation, as each processor can determine from the SFC with which other
processors it needs to communicate. Only very few operations require all-to-all communication, notably the identification of a
global smallest time step for the Runge-Kutta time integration scheme.

At MPI boundaries, the distinction between master and slave elements of a side will also influence the amount of work that the MPI
domain needs to do, since only the domain with the master element will compute the numerical flux function for that surface. Thus, to
ensure optimal load-balancing, we try to evenly distribute slave and master sides at MPI boundaries for each domain.

\subsection{Data structures}

The duality of the volume and side operations is also reflected in the data structures employed in \Flexi. Volume data are stored on
a per-element basis, but side data are not provided for each of the six sides of an element. Instead, sides are uniquely and
independently defined, such that two elements that share a side simply access the same side data. Of course, this means the data
might need to be converted to a local coordinate system, as is explained in Sec.~\ref{sec:coordSys}, and a mapping between the side
data structures and elements must be provided. The major advantage is that operations can directly be performed on either the sides
or the volume, whichever is more appropriate.

\begin{table}[!htp]
\centering
\begin{tabularx}{0.75\linewidth}{rl|rl}
	\hline
	\multicolumn{2}{c}{Element-wise data} & \multicolumn{2}{c}{Side-wise data} \\
	\hline
	$U,U_t,R$:		&$n_{Var}(N+1)^3$ 			& $U^+,U^-$:			& $3 \cdot n_{Var}(N+1)^2$ 			\\
	$\gradient U$:		&$3 \cdot n_{VarLift}(N+1)^3$ 		& $\gradient U^+,\gradient U^-$:& $3 \cdot 3 \cdot n_{VarLift}(N+1)^2$ 		\\
	$\J a$: 		&$3\cdot 3 (N+1)^3$ 			& Flux: 			& $3 \cdot 3 \cdot n_{Var}(N+1)^2$ 		\\
	$\vec{x}$:		&$3\cdot (N+1)^3$ 		        & $\vec{n},\vec{t_1},\vec{t_2}$:& $3 \cdot 3 \cdot (N+1)^2$ 			\\
	$\J$			&$(N+1)^3$ 				& $\hat{s}$:			& $3 \cdot (N+1)^2$ 				\\
	\hline
\end{tabularx}
	\caption{Overview of the main data structures and their respective sizes per elements. Each is realized as a REAL array with
	$32 \mathrm{bits}$ per variable. The number of conservative variables is denoted with $n_{Var}$, the number of lifted
	variables as $n_{VarLift}$.}
\label{tab:dataStructures}
\end{table}

Table~\ref{tab:dataStructures} shows the main data structures that are used during the simulation and their sizes per element.
Regarding the volume data, besides the actual DOFs themselves, we need to store the current time increment ($U_t$) and the last 
register ($R$) to use in the two-register time integration scheme. Other arrays hold the metric terms required
for the transformation ($\J \vec{a}_i$), the Jacobian $J$ itself and the physical coordinates of the nodes $\vec{x}$. For
computations involving gradients, also those need to be stored (with one gradient per direction per primitive variable).

Although there are six sides per element, on average we only need the storage for three of them per element since the side arrays
are shared by adjacent elements. If double-valued solutions exist on sides, they are stored in separate arrays - e.g. the solution
from the element that is the master to a specific side is stored in $U^-$, while the solution from the slave element is stored in
$U^+$. The geometry of the side is represented by normal and tangential vectors ($\vec{n},\vec{t_1},\vec{t_2}$) used for the
transformation into the coordinate system normal to the side, and the surface element $\hat{s}$.

Volume data are always stored ordered along the SFC, while side-based data structures are combined into different types for each
MPI domain. The ordered sequence for each domain is: boundary sides, inner sides, master sides on MPI boundaries and finally
slave sides. Storing them in such an ordered approach is beneficial since we can run operations only required for one type of sides
on a contiguous chunk of memory.

\subsection{Algorithm sequence}

\begin{figure}
    \centering
    % ----------- END OF HEADER -------------

\begin{tikzpicture}[x = 3.8cm, y = 1.8cm, >=triangle 45, scale=0.8]
   \def\mpisize{1.8}
   \def\dist{-0.3}
   \draw(0,0) node (un) {$U$};
   % ----------- Prolongate -----------------
\draw (un.south)++(0,\dist)  node (prolongateDG) [function,anchor=north] {Extrapolate $U \to U^-,U^+$ $\begin{array}{l}1.\, MPI \\ 2.\, other\end{array}$};
   %\draw (prolongateDG.west) node[anchor=east] {1.};
   \draw[->] (un) -- (prolongateDG);

   \draw [rounded,dashed,->] (prolongateDG.east) node[mpi]{$Send_{S \to M}\colon U^+$} -- ++(\mpisize,0);

   % ----------- Lifting -----------------
   \draw(prolongateDG.south)++(0,\dist) node[function, align=left,anchor=north, fill=gray!30] (lifting) [function] {
      Lifting Flux $\begin{array}{l}1.\, MPI \\ 2.\, other\end{array}$ \\[0.1cm]
      Lifting Volume Operator  \\[0.1cm]
      Lifting Surface Integral $\begin{array}{l}1.\, other \\ 2.\, MPI\end{array}$ \\[0.1cm]
      Extrapolate  $\gradient U \to \gradient U^-,\gradient U^+$ $\begin{array}{l}1.\, MPI \\ 2.\, other\end{array}$
   };
   %\draw (lifting.west) node[anchor=east] {2.};
   \draw [rounded,dashed,<-] (lifting.north east) node[mpi]{$Receive_{M \leftarrow S}\colon U^+$} -- +(\mpisize,0);
   \draw [rounded,->] (prolongateDG) -- (lifting);

   \draw [rounded,dashed,->] (lifting.east)++(0,0.9) node[mpi]{$Send_{M \to S}\colon \text{Lifting Flux}$} -- +(\mpisize,0);
   \draw [rounded,dashed,<-] (lifting.east)++(0,-0.1) node[mpi]{$Receive_{S \leftarrow M}\colon \text{Lifting Flux}$} -- +(\mpisize,0);
   \draw [rounded,dashed,->] (lifting.east)++(0,-1.0)node[mpi]{$Send_{S \to M}\colon \gradient U^+$} -- ++(\mpisize,0);

   % ----------- Volume -----------------
   \draw(lifting.south)++(0,\dist) node (volume) [function,anchor=north] {Volume Operator DG };
   %\draw (volume.west) node[anchor=east] {6.};
   \draw[rounded,->] (lifting) -- (volume);

   %% ----------- Fluxes -----------------
   \draw (volume.south)++(0,\dist) node (flux) [function,anchor=north] {Fluxes $\begin{array}{l}1.\, MPI \\ 2.\, other\end{array}$ };
   %\draw (flux.west) node[anchor=east] {7.};
   \draw [rounded,dashed,<-] (flux.north east) node[mpi]{$Receive_{M \leftarrow S}\colon \gradient U^+$} -- +(\mpisize,0);
   \draw [rounded,dashed,->] (flux.east)++(0,0) node[mpi]{$Send_{M \to S}\colon \text{Flux}$} -- ++(\mpisize,0);

   \draw[->] (volume) -- (flux);

   % ----------- Surface integral -----------------
   \draw (flux.south)++(0,\dist) node (surfint) [function,anchor=north] {Surface integral $\begin{array}{l}1.\, other \\ 2.\, MPI\end{array}$};
   %\draw (surfint.west) node[anchor=east] {8.};
   \draw[->] (flux) -- (surfint);
   \draw [rounded,dashed,<-] (surfint.east) node[anchor=north west]{$Receive_{S \leftarrow M} \colon \text{Flux}$} -- +(\mpisize,0);
   \draw (surfint.south)++(0,\dist) node (ut) [anchor=north] {$\pderivative{U}{t}$};
   \draw[->] (surfint) -- (ut);
    
   \draw[rounded,->] (ut) -- ++(-1.6,0) -- node[rotate=90,anchor=south]{explicit Runge Kutta time integration} ($(un) -(1.6,0)$)-- (un);
   % frame
   %\path (current bounding box.south west)++(-0.1,-0.1) coordinate (a);
   %\path (current bounding box.north east)++(0.1,0.1) coordinate (c);
   %\path (current bounding box.north west) coordinate (b);
   %\path (current bounding box.south east) coordinate (d);
   %\draw [rounded,double] (a) -- ([yshift=1.75em] b) -- ++(3cm,0) -- ++(0,-1.75em) -- (c) -- (d) -- cycle;
   %\draw (b) node[anchor=south west,minimum width=3cm] {\itshape DG operator};
\end{tikzpicture}

% ----------- FOOTER -------------
    \caption{Flowchart of the discontinuous Galerkin operator. Grey box indicates the lifting procedure.}
    \label{fig:dg_algorithm}
\end{figure}
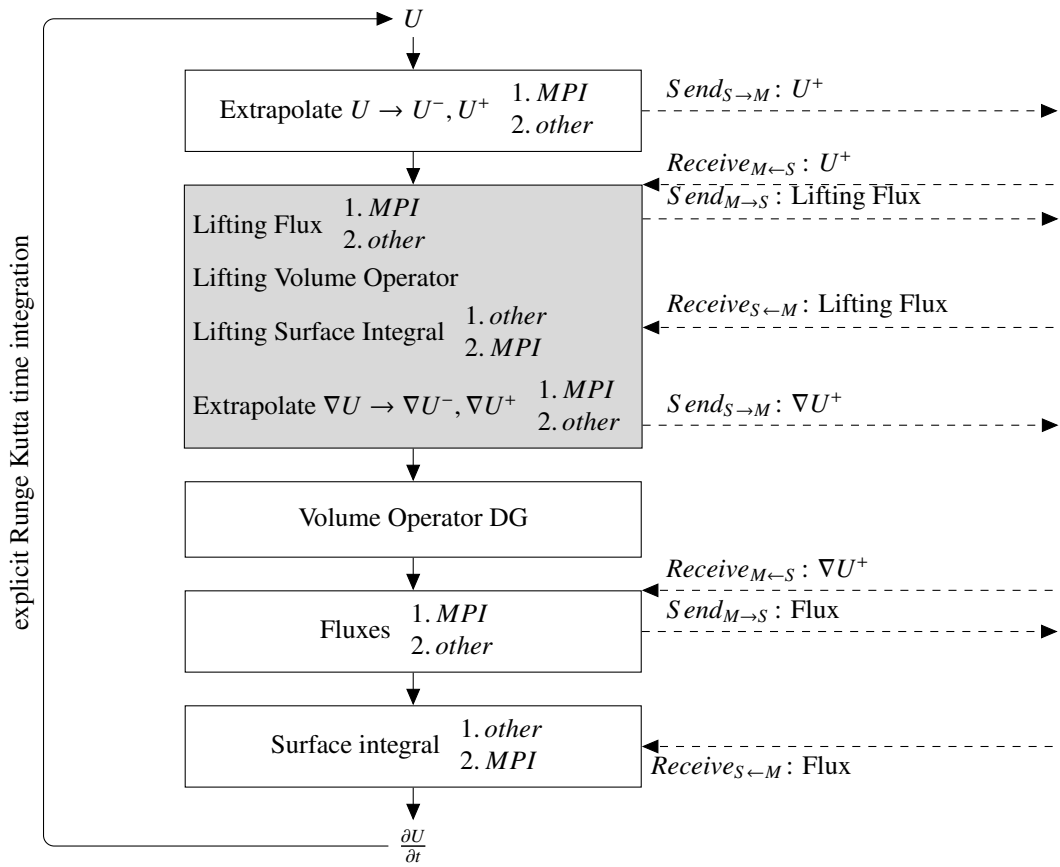

Figure~\ref{fig:dg_algorithm} shows an overview of the algorithmic implementation of the DG operator. We start by considering the
DOFs in the volume $U$ obtained from initialization or during the last time step. Since the computation of the numerical fluxes
requires the solution on the surface of the elements, as a first step we need to extrapolate the volume solution to the sides of
each element. Master and slave elements of the respective side will use different arrays to store both values of the double-valued
solution. If LGL nodes are used, the extrapolation reduces to a simple memory copy, since the surface nodes are included in this set
of nodes.

As the numerical fluxes are unique on a side, they only need to be computed once. We arbitrarily decide that the domain that owns
the master element of that side will compute the flux function.  In a parallel setting, this means that now the MPI threads owning
slave sides on the MPI boundaries must communicate the extrapolated solution to the threads owning the respective master sides. We
employ non-blocking communication, thus the communication is started, but other computations can be commenced while the actual data
transfer happens. Ignoring the lifting procedure (grey box in Fig.~\ref{fig:dg_algorithm}) for the moment, the volume integral of
the DG operator can now be computed. This is a purely cell-local operation, and can thus be done while the communication of the
surface data is still ongoing. The large amount of cell-local work enables effective hiding of the communication in the DG operator.

Once the volume integral is finished, the communication must be finished. As a next step, the master sides calculate the numerical
flux from the double-valued surface states. This process can be split in two parts, where first only the surfaces that belong to MPI
boundaries are processed. This allows us to immediately start the communication of the fluxes back to the slave cells, and then use
the computation of the fluxes at the inner cell boundaries to again hide that communication. One step further, the actual
computation of the surface integral over the numerical fluxes starts with the inner sides. Only when all inner sides are finished,
the communication of the fluxes must be closed, and the surface integral over the remaining sides can be computed. 

Once these steps are complete, the calculation of the semi-discrete DG operator is finished. The result then enters the low-storage
time integration scheme, and the operator evaluation starts again at the next stage.

If gradients are required for the considered equation system, the lifting procedure basically introduces another layer of a
similarly functioning operator. The lifting is again split into the volume and the surface contribution, and communication hiding is
performed in the same manner. 

\subsection{Operator count and memory requirements}

In this section we estimate the number of required operations for the main parts of the operator as well as the memory
requirements.

\begin{table}[!htp]
\centering
\begin{tabularx}{0.78\linewidth}{llc}
	\hline
	Part of operator				& Operation count ($\mathcal{O}$)	& Communication required 			\\
	\hline
	Volume fluxes					&$3(N+1)^3$ 				& No 						\\
	Flux transformation				&$9(N+1)^3$ 				& No 						\\
	Volume integral					&$3(N+1)^4$ 				& No 						\\
	Surface fluxes					&$3(N+1)^2$ 				& Yes 						\\
	Surface integral LG				&$6(N+1)^3$ 				& No 						\\
	Surface integral LGL				&$6(N+1)^2$ 				& No 						\\
	Prolongation LG 				&$6(N+1)^3$ 				& Yes 						\\
	Prolongation LGL				&$6(N+1)^2$ (copy)			& Yes 						\\
	\hline
\end{tabularx}
	\caption{Operation count estimates of a single cell for important parts of the DG operator, for three-dimensional
	computations, on a per variable (of the considered equation system) basis.}
\label{tab:operatorCount}
\end{table}
Table~\ref{tab:operatorCount} gives the estimated operation count for several important parts of the DG operator. Each of them is
given on a per variable (of the considered equation system) basis. These numbers can only be estimates, since the actual number of
operations depends strongly on implementation, compiler optimization and the considered equation system (e.g. the complexity
of the flux functions). For the surface fluxes, only half of the six sides are considered, since they are computed once per 
interface. The difference in LG and LGL operators is due to the fact that LGL nodes include the boundary, thus e.g. the surface
integral only acts on the DOFs at the surface and not in the volume (as it is the case with LG nodes).

It can be seen that the volume integral makes up for a large part of the operations due to its scaling with $(N+1)^4$. If the
polynomial degree increases, the volume contribution thus get's even larger compared to the surface contributions. This is an
important reason for the excellent scaling qualities of the DGSEM especially at higher polynomial degrees, since the ratio of purely
local volume operations to surface operations that require communication is increased compared to low-order stencils. If we
consider the volume integral with its $(N+1)^4$ scaling to be the defining factor of the operation count, and recalling that the
number of DOFs per cell is $(N+1)^3$, we can conclude that the number of operations per DOF scales as $\mathcal{O}(N)$.

\setlength{\tabcolsep}{3.7pt}
\begin{table}[!htp]
\centering
\small
\begin{tabularx}{0.9\linewidth}{c|cccccccccccc}
	\hline
	N &$1$&$2$&$3$&$4$&$5$&$6$&$7$&$8$&$9$&$10$&$11$&$12$ \\
	\hline
	KiB &$1.47$&$1.08$&$0.906$&$0.805$&$0.74$&$0.693$&$0.659$&$0.632$&$0.611$&$0.594$&$0.58$&$0.567$ \\
	\hline
\end{tabularx}
	\caption{Measured memory requirements per DOF for the Navier-Stokes equation system.}
\label{tab:memoryReq}
\end{table}
Regarding the memory consumption, estimates could be based on the required storage for e.g. the solution, metric terms, gradients
and so on, as given in Tbl.~\ref{tab:dataStructures}. Since for production runs these estimates might not be accurate due to
overhead for additionally stored quantities, e.g. for analyze functionalities, we measure the memory consumption in a
real-world situation. Table~\ref{tab:memoryReq} shows the measured memory requirements for a single DOF for the Navier-Stokes
equation system. It becomes obvious that the memory consumption decreases with increasing polynomial degree. This is again due to
the fact that the ratio between volume and surface DOFs increases, leading to a more memory-conserving scheme at higher order. Also,
the overall low memory consumption is evident: At $N=5$, we only need around $740 \, MB$ to store $1$  Mio. DOFs. This
low-consumption property is of course shared with many other explicit numerical schemes. 

If we consider a typical system used for HPC, e.g. the Cray XC40 system \textit{Hazel Hen} used at the High Performance Computing
Center (HLRS) in Stuttgart, where a compute node with $24$ processor is equipped with $128 \, GB$ of memory, this would allow us to
store over $7$ Mio. DOFs per processor, several orders of magnitude more than typically used in production runs. While there is
plenty of memory available, CPU cache storage is much more scarce, and higher-order computations with their reduced memory footprint
help us to keep a large amount of the required data in the cache. Since the cache has not only a higher bandwidth than the RAM, but
also ultra-low latency, cache-hits are an important factor for any implementation. Thus, with increasing polynomial degree two
advantages show: The scheme becomes more cache friendly and the increased percentage of volume operations helps to localize the data
even further. Both factors contribute to the scaling performance, which 
is illustrated in Sec.~\ref{sec:scaling}.

\subsection{FV shock capturing}\label{subsec:implementation:fv}

In this section, we present the implementation of the hybrid DG/FV method. A main design principle was to keep the general structure
of the DG operator intact, and to find equivalents for each of the components in the FV operator. In combination with the fact that we
keep the same data structure for the solution in the hybrid scheme, this allows for an implementation that is only minimally
intrusive for the DG operator.

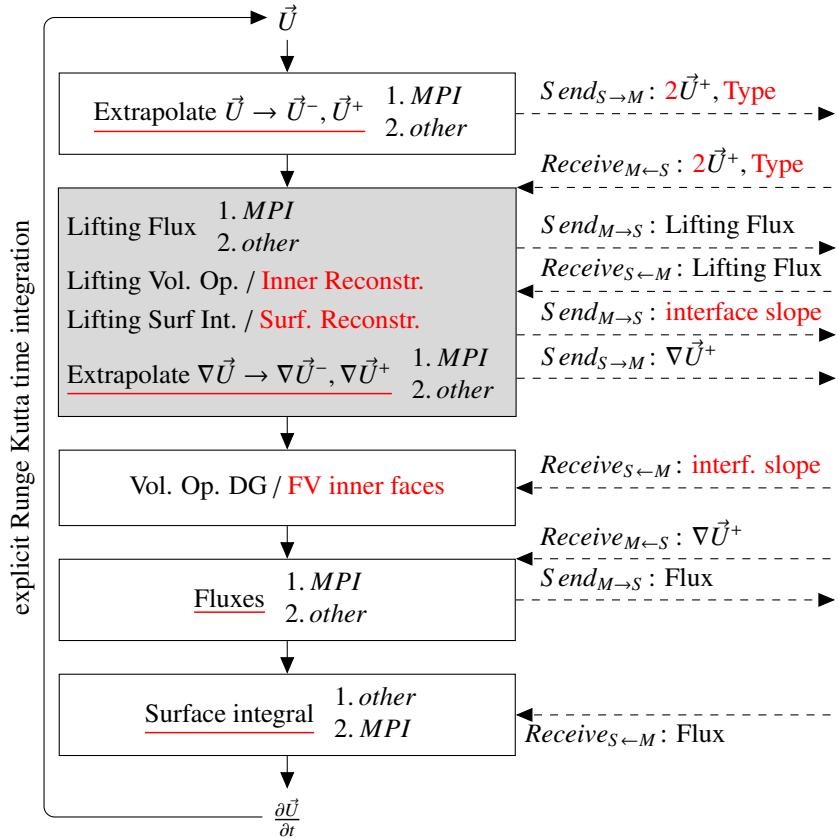
\begin{figure}
	\centering
	% ----------- END OF HEADER -------------

\begin{tikzpicture}[x = 2.5cm, y = 1.8cm, >=triangle 45, scale=0.8]
   \def\mpisize{2.1}
   \def\dist{-0.3}
   \draw(0,0) node (un) {$\vec{U}$};
   % ----------- Prolongate -----------------
   \draw (un.south)++(0,\dist)  node (prolongateDG) [function,anchor=north] {\cunderline{red}{Extrapolate $\vec{U} \to \vec{U}^-,\vec{U}^+$} $\begin{array}{l}1.\, MPI \\ 2.\, other\end{array}$};
   %\draw (prolongateDG.west) node[anchor=east] {1.};
   \draw[->] (un) -- (prolongateDG);

   \draw [rounded,dashed,->] (prolongateDG.east) node[mpi]{$Send_{S \to M}\colon {\color{red}2} \vec{U}^+, \color{red}{\text{Type}}$} -- ++(\mpisize,0);

   % ----------- Lifting -----------------
   \draw(prolongateDG.south)++(0,\dist) node[align=left,anchor=north, fill=gray!30] (lifting) [function] {
      Lifting Flux $\begin{array}{l}1.\, MPI \\ 2.\, other\end{array}$ \\[0.1cm]
      Lifting Vol. Op.  / \color{red}{Inner Reconstr.} \\[0.1cm]
      Lifting Surf Int. / \color{red}{Surf. Reconstr.} \\[0.1cm]
      \cunderline{red}{Extrapolate  $\gradient \vec{U} \to \gradient \vec{U}^-, \gradient \vec{U}^+$} $\begin{array}{l}1.\, MPI \\ 2.\, other\end{array}$
   };
   %\draw (lifting.west) node[anchor=east] {2.};
      \draw [rounded,dashed,<-] (lifting.north east) node[mpi]{$Receive_{M \leftarrow S}\colon {\color{red}2}  \vec{U}^+, \color{red}{\text{Type}}$} -- +(\mpisize,0);
   \draw[rounded,->] (prolongateDG) -- (lifting);

   \draw [rounded,dashed,->] (lifting.east)++(0,0.5) node[mpi]{$Send_{M \to S}\colon \text{Lifting Flux}$} -- +(\mpisize,0);
   \draw [rounded,dashed,<-] (lifting.east)++(0,0.1) node[mpi]{$Receive_{S \leftarrow M}\colon \text{Lifting Flux}$} -- +(\mpisize,0);
   \draw [rounded,dashed,->] (lifting.east)++(0,-0.3)node[mpi]{$Send_{M \to S}\colon {\color{red} \text{interface slope}}$} -- ++(\mpisize,0);
   \draw [rounded,dashed,->] (lifting.east)++(0,-0.7)node[mpi]{$Send_{S \to M}\colon \gradient \vec{U}^+$} -- ++(\mpisize,0);

   % ----------- Volume -----------------
   \draw(lifting.south)++(0,\dist) node (volume) [function,anchor=north] {Vol. Op. DG / \color{red}{FV inner faces} };
   \draw [rounded,dashed,<-] (volume.east) node[mpi]{$Receive_{S \leftarrow M}\colon {\color{red} \text{interf. slope}}$} -- +(\mpisize,0);
   %\draw (volume.west) node[anchor=east] {6.};
   \draw[rounded,->] (lifting) -- (volume);

   %% ----------- Fluxes -----------------
   \draw (volume.south)++(0,\dist) node (flux) [function,anchor=north] {\cunderline{red}{Fluxes} $\begin{array}{l}1.\, MPI \\ 2.\, other\end{array}$ };
   %\draw (flux.west) node[anchor=east] {7.};
   \draw [rounded,dashed,<-] (flux.north east) node[mpi]{$Receive_{M \leftarrow S}\colon \gradient \vec{U}^+$} -- +(\mpisize,0);
   \draw [rounded,dashed,->] (flux.east) node[mpi]{$Send_{M \to S}\colon \text{Flux}$} -- ++(\mpisize,0);

   \draw[->] (volume) -- (flux);

   % ----------- Surface integral -----------------
   \draw (flux.south)++(0,\dist) node (surfint) [function,anchor=north] {\cunderline{red}{Surface integral} $\begin{array}{l}1.\, other \\ 2.\, MPI\end{array}$};
   %\draw (surfint.west) node[anchor=east] {8.};
   \draw[->] (flux) -- (surfint);
   \draw [rounded,dashed,<-] (surfint.east) node[anchor=north west]{$Receive_{S \leftarrow M} \colon \text{Flux}$} -- +(\mpisize,0);
   \draw (surfint.south)++(0,\dist) node (ut) [anchor=north] {$\fracp{\vec{U}}{t}$};
   \draw[->] (surfint) -- (ut);
    
   \draw[rounded,->] (ut) -- ++(-1.6,0) -- node[rotate=90,anchor=south]{explicit Runge Kutta time integration} ($(un) -(1.6,0)$)-- (un);
   % frame
   %\path (current bounding box.south west)++(-0.1,-0.1) coordinate (a);
   %\path (current bounding box.north east)++(0.1,0.1) coordinate (c);
   %\path (current bounding box.north west) coordinate (b);
   %\path (current bounding box.south east) coordinate (d);
   %\draw [rounded,double] (a) -- ([yshift=1.75em] b) -- ++(3cm,0) -- ++(0,-1.75em) -- (c) -- (d) -- cycle;
   %\draw (b) node[anchor=south west,minimum width=3cm] {\itshape DG operator};
\end{tikzpicture}

% ----------- FOOTER -------------
 \caption{Flow chart of the hybrid discontinuous Galerkin/finite volume subcells
	operator: Procedures underlined in red are modified to perform their specific task either for DG or FV elements. At the same
	time the lifting of the DG elements is computed, the 2nd order reconstruction of the FV subcells is built. The
	counterpart to the DG volume integral are the fluxes over inner FV subcell interfaces. Additional communication is
	required. Besides the information of which type (DG or FV) an element at a MPI interfaces is, for the FV reconstruction a
	second array of face data has to be transmitted, which is indicated by the red numbers~2. }
	\label{fig:fv_algorithm}
\end{figure}
Figure~\ref{fig:fv_algorithm} presents a flowchart of the combined DG/FV scheme, which we will use to discuss the differences to the
pure DG operator (Fig.~\ref{fig:dg_algorithm}). For FV cells, the extrapolation of the solution to the boundaries changes to a
simple copy of the cell mean value at the element interface to the respective arrays, such that this information can be used to
compute a slope, and is communicated to the master element in a parallel setting.
Additionally, to prepare for later reconstruction, the slope between the first and the second subcell away from the interface is
computed and also communicated. We need to do this since, as for the pure DG case, only the master side will compute the flux
functions. For this, it needs to know the limited slope from the slave element, which is a function of the slope across the
interface itself and the one between the next two subcells (the one we communicate). If a DG element is considered, instead we
communicate the solution at the first Gauss point. This means we need to send and receive twice the amount of data as in the DG
case, both the solution and the values needed for the reconstruction. We also communicate a single integer indicating the type of
the cell (DG or FV), such that the algorithm can decide what type of interface it needs to handle.

The equivalent for the lifting procedure in DG is the computation (and limitation) of the slopes between the subcells for the FV
elements.  While there is no real volume work for the FV scheme, we perform the slope computation across the inner subcell
interfaces at the same time as the volume integral is computed for the DG scheme. The reason is that both of those require no
communication between MPI domains, and can thus be used to hide the communication latency. While the surface integral is done for
the DG method, the remaining slopes across the cell-interfaces can be computed, on the master sides of the elements. Here, the
limitation requires the additional data that was communicated beforehand. The limited slopes at the cell interface need to be send
back from the master element to the slave sides, since they are needed to limit the slopes for those elements.

With the lifting/computation of the slopes complete, the volume integral of the DG operator is performed. For FV elements, here the
fluxes at the inner subcell interfaces are computed and directly applied, since again those require no communication. The
reconstruction of the values at the subcell interfaces from the limited slopes is also done here. In the next step the fluxes at the
cell interfaces are computed. In the case of a mixed interface the DG solution must first be transformed into the FV representation.
The last step is the surface integral, where for DG cells at mixed interfaces the flux is projected back to the DG representation,
and the computed fluxes are simply applied to the FV subcells at element boundaries.

While the arrays containing the solution are shared between the two schemes, other additional quantities must be stored for the
hybrid scheme. This includes the metric terms in the FV representation of a cell, geometric information (normal and tangential
vectors, surface elements) for inner faces as well as the physical distances between subcells in each direction to perform the
reconstruction. This additional memory consumption has some impact on the performance of the scheme, and will be discussed in the
next section.

\subsection{Parallel performance}\label{sec:scaling}

In this section, we will investigate the strong scaling behaviour of both the pure DGSEM and the hybrid DG/FV method developed for
shock capturing. All the subsequent test were performed on the Cray XC40 system \textit{Hazel Hen} at the HLRS, which is
equipped with compute nodes that consists of two Intel Xeon E5-2680 CPUs (12 cores each) and 128 GB of memory. The communication
between nodes is realized using the Cray Aries high-speed interconnect. We ran a simple simulation of a free-stream on cubic
Cartesian domain, with a varying number of mesh elements. The smallest mesh consisted of $[6 \times 6 \times 6]$ elements, and
bigger meshes were created by doubling the number of elements individually for each direction. So the next mesh would have $[12
\times 6 \times 6]$ elements, then $[12 \times 12 \times 6]$ and so on until $[96 \times 96 \times 48]$ elements were reached. The
simulations were all first executed on a single node, except for the largest mesh, where two nodes were needed to provide enough
memory for that case. The number of nodes was then repeatedly doubled, until for each case only $9$ elements per core remained. 
Thus, the minimum number of cores used was $24$, and the largest cases were run on up to $49.152$ cores. All
simulations were advanced for 100 times steps, and each configuration was repeated five times to gain information about variance in
the results. To reduce the possible amount of combinations that could be checked, we restrict ourselves to the Navier-Stokes
equation system, $N=6$ and choose the LG nodes. For the mixed DG/FV scheme, one half of the computational domain was always set to
be only DG, the other to be pure FV. This is not the worst-case scenario since the amount of mixed DG/FV interfaces is limited, but
a realistic one considering that regions with shocks should be compact and not spread out.

\begin{figure}
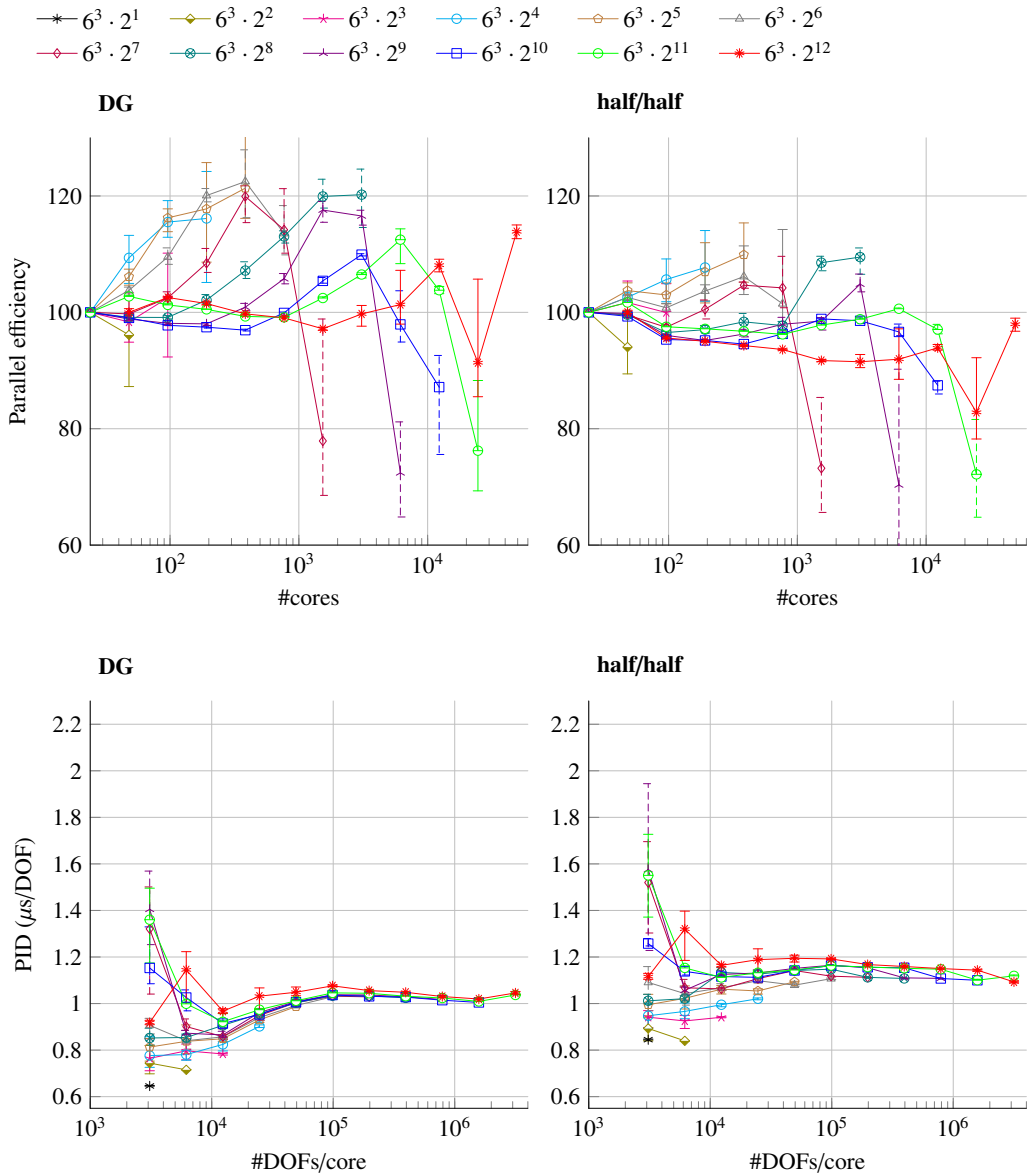

   \centering
   \includestandalone[width=\linewidth]{figures/scaling/efficiency}
   \caption{Parallel efficiency and performance for the strong scaling test case. The line color/marker combination refers to a
	specific mesh, the number of elements is given in the legend. Each test case was repeated five times, shown are the median
	values, error bars indicate the smallest and biggest values observed.}
   \label{fig:scaling}
\end{figure}
The main metric analyzed is the performance index PID, which is the time required to advance a single DOF from one stage of the
Runge-Kutta time integration scheme to the next. It is computed as
\begin{equation}
   \text{PID} = \frac{\text{wall-clock-time} \cdot \#\text{cores}}{\#\text{DOF} \cdot \#\text{time steps} \cdot \#\text{RK-stages}}.
\end{equation}
Figure~\ref{fig:scaling} shows the results from the strong scaling tests. In the top row, we compare the parallel efficiency, which
is defined as the ratio of the PID using a certain amount of nodes to the median of the PID (calculated from the repeated runs of
the same configuration) using the smallest amount of nodes. Focussing on the results obtained from the DG calculations, we observe
that all cases exhibit super-linear scaling, where the efficiency is larger than $100\%$ in certain areas. This can be explained by
the high memory consumption for calculations on a small number of nodes. When the number of compute nodes increases, the required
memory per core decreases, and more of the used data can be stored in the fast cache of the CPU, increasing performance. This also
explains why for the larger cases the super-linear behaviour emerges at a larger number of cores. If the number of nodes is
increased further, a drop in efficiency can be observed for nearly all cases. In this region, the remaining core-local work load
might no longer be enough to hide the communication effort, which ultimately outweighs the increased cache-efficiency. The
performance in this region is thus bound by latency of the underlying network. Remarkably, efficiency even for the largest cases
does not drop below $70\%$, which - in combination with the super-linear areas - highlights the already expected excellent scaling
capabilities of the DG method.

For the hybrid DG/FV approach, a similar picture emerges, but the super-linear behaviour is not as pronounced or not present at all for
the large cases. This is consistent with the fact that the FV operator increases the memory consumption of the overall method. Thus
the described cache-effects are shifted towards an increased number of compute nodes, and might not appear at all since the latency
is already the dominant factor in the cache-friendly regions. While the parallel efficiency might not reach the same level as the
pure DG operator, still the overall behaviour is deemed as sufficient.

The bottom row of Fig.~\ref{fig:scaling} presents in principle the same data, but shows absolute values for the PID over the load on
each core. Again focussing on the pure DG operator first, it is clear that starting at about $100,000$ DOFs/core the PID is nearly
constant all the way up to the largest loads, independent of the actual size of the mesh. Reducing the load below that value will
see an increase in performance, attributed again to the effects of the lower load on the cache-friendliness. An optimum value seems
to be found at $10,000$ DOFs/core. If the load is decreased further, the latency will start to become dominant and performance might
decrease. The smaller cases are more resilient against such performance losses, since they only require communication between a
smaller number of nodes. A striking result is the large spread of the measured performance for the largest meshes and the smaller
load per core. In those cases, performance is depending the most on latency, and the specific physical distribution of the
simulation on the manycore system as well as workload of the network play a huge role in realizable performance.

The results for the hybrid DG/FV approach show a very similar behaviour, but lack the clear optimum for the load per core that was
observed for the pure DG operator. The reason for that was already given in the previous paragraphs. It is important to compare the
absolute values for the performance at not just relative scaling, as we do here. This reveals that the performance for the hybrid
method is about $15-20\%$ worse than for the pure DG method, which is the price we pay for the hybrid operator structure.

\subsection{Implementation of post-processing}\label{subsec:implementation:posti}

As has been mentioned in the introduction, we implement a custom visualization toolchain, which is fully parallel and takes advantage
of the high order nature of our solution. This toolchain combines our specifically build \Posti visu tool, which handles the actual
computations, and the open-source software ParaView.

\begin{figure}
   \centering
	\includegraphics[width=0.8\textwidth]{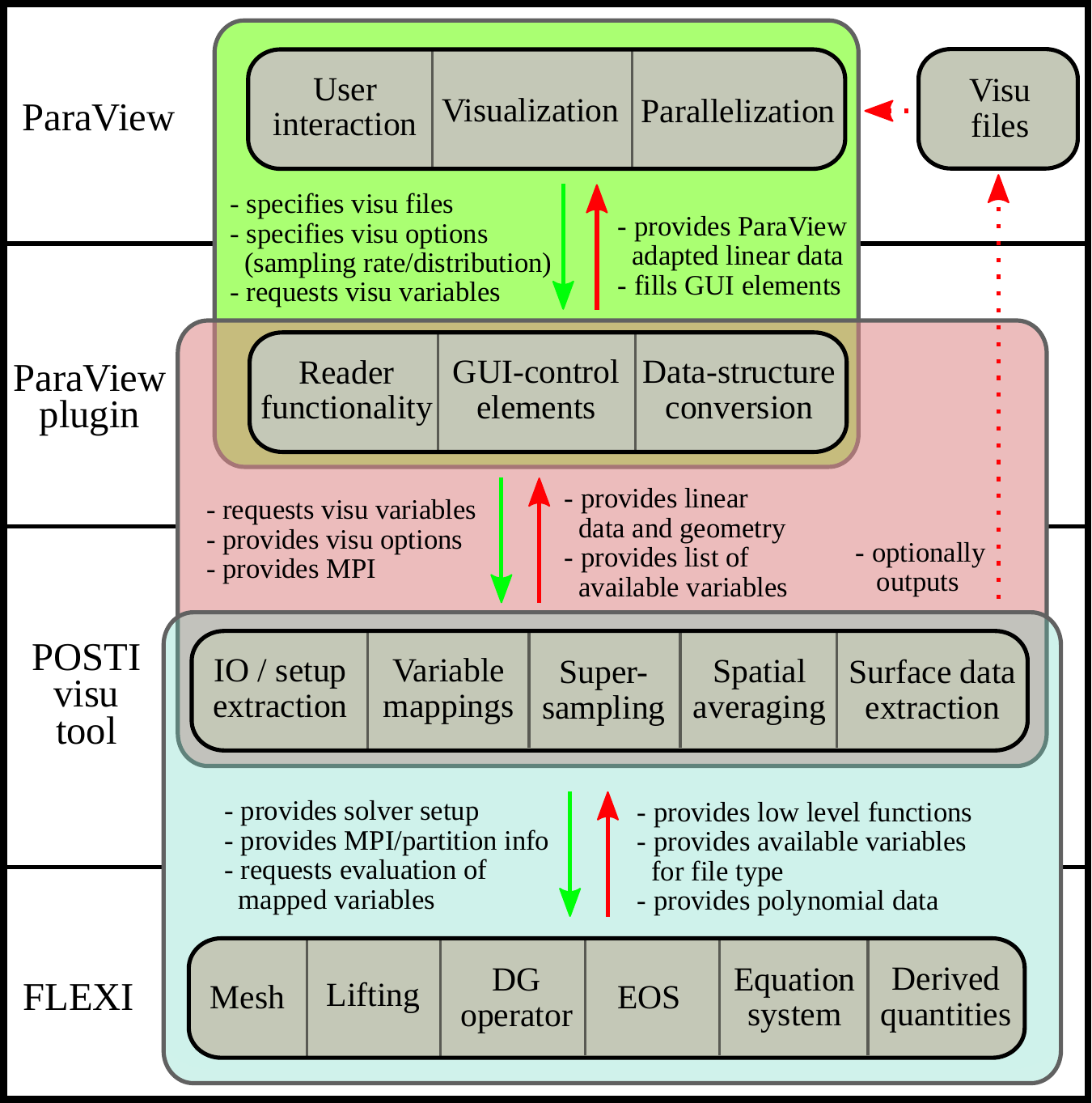}
   \caption{Schematic of the implementation of the visualization tool.}
   \label{fig:postiSchematic}
\end{figure}
Figure~\ref{fig:postiSchematic} presents the different parts of the toolchain and how they interact with each other. On the top
level, ParaView provides the user interface as well as access to a multitude of visualization options and filters. An important
aspect is also that ParaView is fully parallelized, allowing us to run the post-processing on a number of cores that is adequate for
the problem at hand. The second layer of the toolchain is a plugin that provides a reader (essentially an import functionality) for
the files written by \Flexi. A reader defines GUI elements in ParaView, that can be used to specify options which can then be
passed on to the \Posti program. 

The plugin acts as an interface between ParaView (written in C/C++) and the custom \Posti tool (written in Fortran). It will pass
the MPI communicator created by ParaView down to that tool, which means both of the visualization and the data processing part of
the toolchain use the same parallelization. The options specified by the user in the GUI are also passed down, as a simple text file
which can be parsed by the \Posti tool. Depending on the options selected by the user, the tool can now perform calculations. To
this end, it can directly use routines originally implemented for the solver itself. This includes e.g. I/O routines, initialization
of the mesh, partitioning in MPI regions, calculation of gradients using the lifting operator or the DG operator itself. Depending
on the used equation system, derived quantities can be computed from the conservative variables stored in the state files. Those can
range from simple calculations of primitive quantities like the temperature (taking the current equation of state into account) to
complex vortex identification criterion like $\lambda_2$, requiring knowledge of the derivatives and the calculation of eigenvalues
of a matrix. The advantages of doing these calculations directly at the lowest level instead of in the visualization software are a)
the direct accessibility of all necessary data and b) an performance advantage, since only the derived quantity needs to be passed
to the visualization software, not all variables needed to calculate it. Additional features are also provided by the \Posti tool,
for example the ability to average three-dimensional data along a regular direction, or the interpolation of data to the surfaces.

Once the variables specified by the user have been calculated, the \Posti tool converts them from the high order polynomial
representation used internally to the linear format needed by ParaView. To reduce the errors introduced by the linear interpolation,
usually a super-sampling of the high order polynomial is used. The data is passed back to the plugin, which provides it in the way
needed by ParaView, which is then finally able to display the requested data. All the computation and conversion steps are
hidden from the users, they simply interact with the ParaView GUI.

As will be discussed in more detail in Sec.~\ref{sec:userblock}, the state files contain next to the solution variables also all
parameters that were used to run the simulation. Besides the aspect of reproducibility, this also means we can use that information
in the post-processing. This frees the user from having to specify the details of the simulation that are necessary for the
visualization process, e.g. the polynomial degree, node type or equation system used in the simulation.

%%%%%%%%%%%%%%%%%%%%%%%%%%%%%% ------ USABILITY ----- %%%%%%%%%%%%%%%%%%%%%%%%%%%%%%
\section{Aspects of Usability and Reproducibility}

While a lot of emphasis is typically placed on stringent description of numerical algorithms and mathematically precise analysis of
the schemes encapsulated in open-source CFD codes, ease of access and use as well as reproducibility are often treated less
thoroughly. However, conveying the expert knowledge that goes into the conceptual design of the scheme and code to the final user is
important to ensure correct and reproducible results. In this section, we present some of the concepts and methods used in the
\Flexi ecosystem to help make the code more user-friendly and thus error-resistant and to support good reproducibility practices.
It is obvious that only a framework that ensures \emph{reproducible} results should have its place in research and development in
the long run. Open-source software supports reproducibility on a conceptual level, as it lays the basic algorithms open. However,
this is just a single link in a complex, interwoven chain that drives possible sources of uncertainty in the simulation stack and
thus decreases reproducibility. Other, equally important aspects that drive systemic opacity are for example the user-defined choice
of discretization parameters, the hardware and associated operating system, the dependence on third party libraries and the build or
compile options for all the code pieces involved~\cite{Resch2019}. According to Peng~\cite{peng2011reproducible}, the gold standard
in reproducibility entails not only providing all this information, but also the binaries and even the produced simulation data
itself. Taking the heterogeneous and varied hardware landscape (and again, the associated first level software like drivers and
operating systems) into account, perfect reproducibility would in the limit entail even hardware specifications. Clearly, already
the previous level demanding compiled and linked binaries as well as full simulation data is not practical, in particular in a HPC
context, where the data sets are voluminous and the user has neither full knowledge nor control over the hardware and available
libraries. Despite all these hindrances, striving for a maximum amount of reproducibility increases the trustworthiness of the
results, stability of the code during its lifetime and the usability of the CFD software. In the \Flexi framework, a number of
measures supporting these notions have been included, two of which will be represented here briefly. 

\subsection{Regression testing}

Regression tests are mainly used to ensure code stability during the development process and to uncover introduced bugs.
Additionally, they provide provide a benchmark for code efficiency, i.e. new features should not slow down execution speed
noticeably. The tests themselves are implemented in various forms. \textit{Unit tests} are run each time the code base is compiled,
and they test several core functionalities of the DG operator, e.g. computing a surface integral for a flux vector.
The success of those tests is determined by comparing them with pre-calculated results. They make sure that the very basic
building blocks of the framework continue to operate as expected. 

Another form of tests is performed using the continuous integration capabilities of the gitlab code management framework. Each time
a new commit is published, several \textit{check-in tests} are executed. Those are computationally cheap tests, that check some
basic functionalities at a higher level than the unit tests: Can the code be compiled with the most common options? Can the code be
executed and run minimalistic examples like a constant state solution on Cartesian grids? More thorough test are performed as
\textit{daily} and \textit{weekly regression tests}. In those, most possible combinations of compile options are build and tested in
various examples, aimed at testing features such as mortar interfaces, boundary conditions, Riemann solvers and time integration
methods. The custom python tool \textit{REGGIE}\footnote{https://github.com/reggie-framework/reggie2.0/} is used to compile, run and analyze those
tests, which provides advanced features such as the possibility to directly calculate convergence rates from subsequent runs and
compare those to expected values. The weekly tests consist of real-world applications, such as the flow around an airfoil or a
turbulent channel flow, and check the complex interaction of multiple features at the same time. Overall, this feature thus ensures
reproducibility on a programming level. 

\subsection{The userblock concept}\label{sec:userblock}

One level above the specific programming or implementation details lies the management of code and simulation meta-data. These
meta-data contain information about the specific code variants, the compile options and the simulation settings. While open-source
software generally is seen as a step towards reducing systemic opacity, it comes with the additional challenge of managing (even more)
meta-data. \Flexi and its supporting software can be distributed freely, modified at will and compiled on a wide range of platforms,
leading to a growing variety of variants.  Hence, identifying and obtaining the underlying code version for reproducing results can
be at least time consuming if not impossible. In order to assist in replicating a simulation setup at a later date, \Flexi
implements a robust mechanism to re-obtain source code, compiled binaries and simulation parameters from each solution file written
by \Flexi with minimal storage overhead.

\begin{figure}[t!]
	\centering
	\includegraphics[scale=0.85]{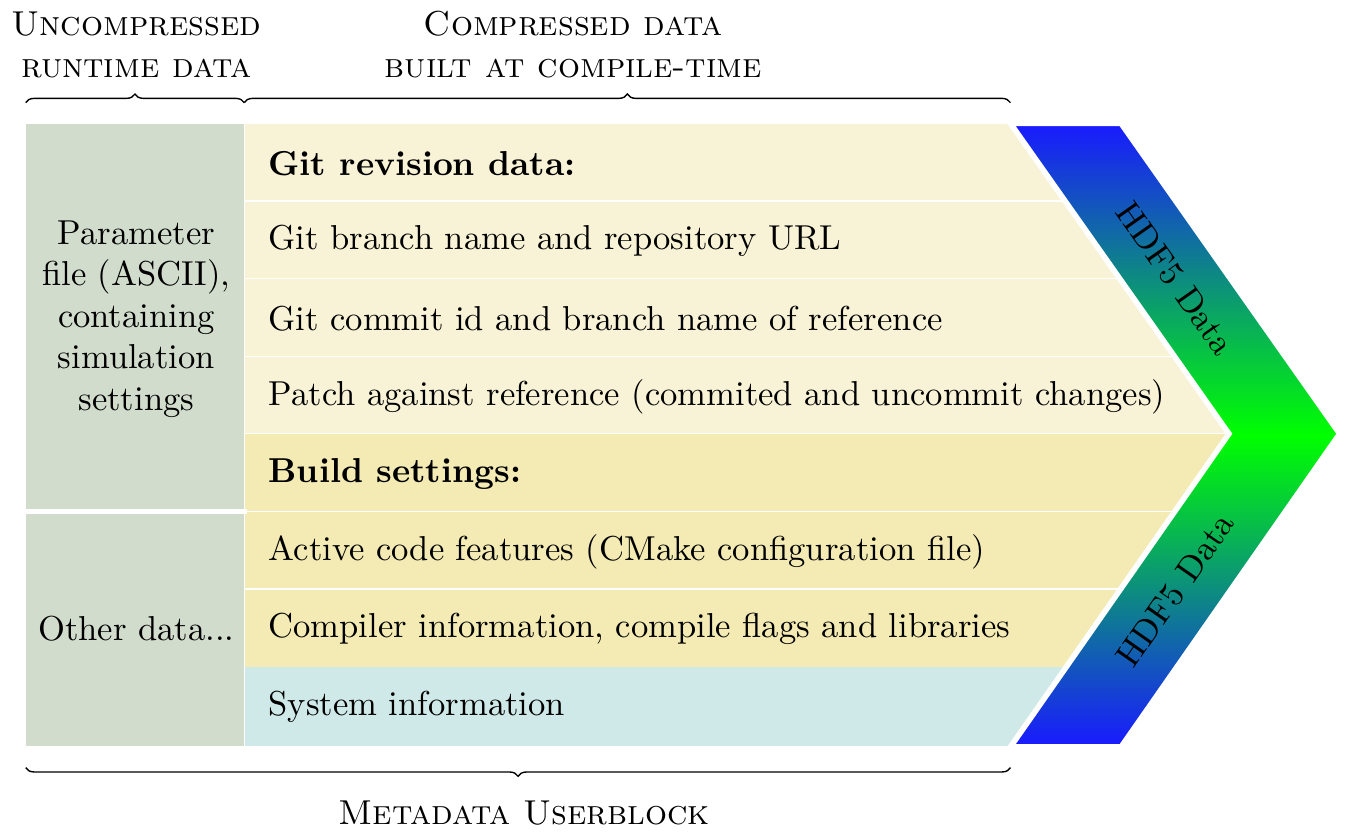}
	\caption{Format description of meta-data in the user-block section of the solution files.}
	\label{fig:userblock}
\end{figure}

The flow solution at a given time $t$ is stored in a HDF5 format. To each of these so
called "state" files, a block of custom meta-data is appended, subsequently called the "user-block". This user-block contains all
the available information at run and compile time to ensure that a) a state file can be associated with a specific simulation run in
case this information becomes inadvertently lost and b)  the simulation can be recreated with the same code variant and user
settings. The user-block format features a simple syntax with sub-groups. It can contain both text and binary data, as well as
compressed data. A schematic of the data in the user-block is depicted in Fig.~\ref{fig:userblock}. The first block provides an
exact copy of the settings file with the simulation parameters. A second compressed block contains data from an object file, which
is directly generated at compile-time and linked statically into the solver executable. It contains sufficient information to
rebuild the executable. Code version control is based on Git~\cite{chacon2014pro}, so we store the repository that was used to clone
the current code base, the git commit the user was on and any local modifications to the code which have not been uploaded to
the repository in the form of file patches, which can be applied to recover the local code variant. Compile-time flags
and options in the form of preprocessor flags that are generated and interpreted by the CMake build system are also stored. 
Access to the information contained in the user-block is facilitated by scripts that are part of the \Flexi repository. Directly
rebuilding the \Flexi variant and restoring the configuration can also be initiated. This process consists of downloading the code
from the specified repository, checking out the correct revision, applying the patches and rebuilding the binaries using the stored
CMake configuration.  Besides serving as a quick and robust mechanism to help with reproducibility, the user-block data can also be
queried by other parts of the \Flexi ecosystem. For example, the post-processing tool chain presented in Sec.~\ref{subsec:posti}
and~\ref{subsec:implementation:posti} can recover the information stored therein and incorporate it consistently into the
post-processing of the simulation data. 

\subsection{Documentation and Tutorials}

The documentation for \Flexi is available from the project website. Besides a description of the building process and the features
of \Flexi, it also contains a number of tutorials that introduce the basic functionalities and give a starter for code
modifications. For code developers, a code documentation based on Doxygen is also available from the project website. It is rebuild
automatically to reflect updates to the master branch.

%%%%%%%%%%%%%%%%%%%%%%%%%%%%%% ------ APPLICATIONS ----- %%%%%%%%%%%%%%%%%%%%%%%%%%%%%%
\section{Applications}
In this section, we briefly summarize three challenging simulations conducted with \Flexi, which highlight some of the features
presented in the previous chapters. As a first example, we discuss a wall-resolved LES of a NACA 64418 airfoil at high Reynolds
number. This simulation combines some of the challenges addressed before, as it requires $\mathcal{O}(10^8)$ DOF per solution
variable, a surface- and volume-curved mesh, a stable scheme for implicitly modelled LES and an pre- and post-processing framework
capable of handling the large-scale data seamlessly. The second example, a shock-vortex interaction, demonstrates the ability of the
hybrid DG/FV method to capture shocks and retain the advantageous high order properties in regions of smooth flow. Our last example
highlights a HO zonal LES approach for a highly sensitive acoustic feedback mechanisms. It features a wall-model as well as a
synthetic turbulent inflow, which both do not generate noticeable acoustic disturbances. A comparison against an experimental
measurement campaign shows that \Flexi is capable of capturing the intricate interactions between hydrodynamics and acoustics
accurately. Besides these highlighted examples here, \Flexi has been both extensively validated against canonical test cases as well
as applied to a variety of problems, see
e.g.~\cite{beck2014high,Beck2016,flad2016simulation,gassner2013accuracy,hindenlang2012explicit,flad2014discontinuous,10.1007/978-3-319-20340-9_14,atak2016high,frank2016direct}.

\subsection{LES of airfoil at $Re=10^6$}

The first example is used to demonstrate the applicability of \Flexi to large-scale problems in the field of aerodynamics. We
compute the flow around a NACA 64418 airfoil at a chord Reynolds-number $Re=\frac{U_{\infty}c}{\nu}=10^6$ and a free-stream Mach
number of $Ma=\frac{U_{\infty}}{a}=0.2$. Here, $U_{\infty}$ is the free-stream velocity, $c$ the chord length, $\nu$ the kinematic
viscosity and $a$ the speed of sound. The angle of attack is set to $\alpha=4\degree$, which will lead to a region of separated flow
at the trailing edge of the airfoil. The separation process is sensitive to the details of the flow and underlying modeling. 
Thus an accurate and scale-resolving simulation methodology is necessary to reliably predict such phenomena. 

\begin{figure}
\centering
\begin{subfigure}[b]{0.49\textwidth}
  \includegraphics[width=\textwidth]{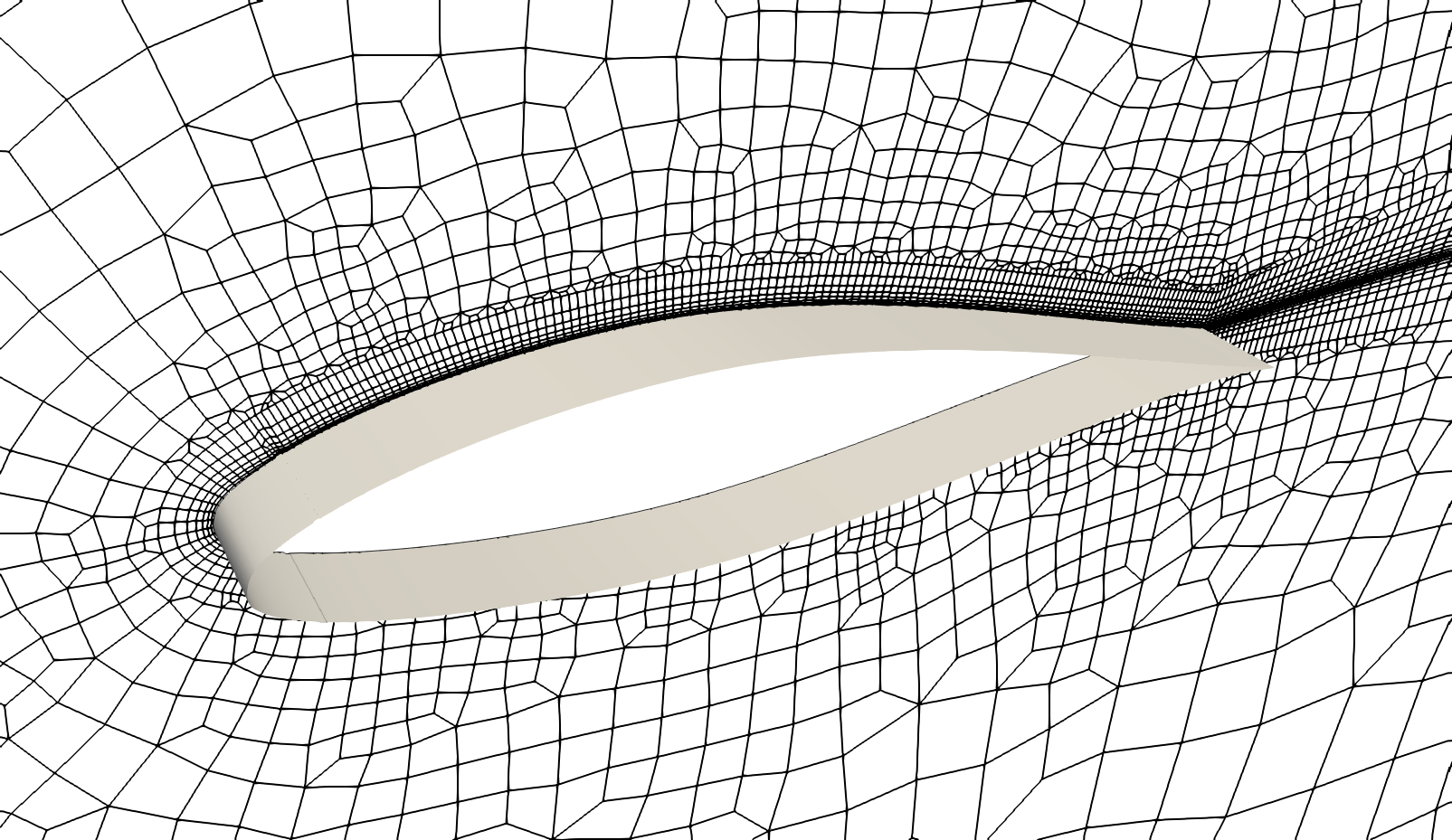}
	\vspace{1.0cm}
\end{subfigure}%
\hfill
\begin{subfigure}[b]{0.49\textwidth}
%
\end{subfigure}
	\caption{Mesh used for the NACA 64418 airfoil simulation (left) and associated non-dimensional grid spacing (right). Crosses
	denote the suction side, squares the pressure side.}
  \label{fig:NACA64418Mesh}
\end{figure}
Figure~\ref{fig:NACA64418Mesh} shows details of the used mesh. A C-grid of height $0.04c$ is employed in the close vicinity of the
airfoil, to guarantee optimal grid quality for the turbulent boundary layer. Outside of that structured layer, the grid becomes
unstructured and the mesh size is increased rapidly to minimize the total number of grid cells. A geometrical trip in the form of a
protruding step  of height $0.001c$ is incorporated in the mesh at $x=0.05c$, which reliably triggers transition to turbulence 
at this specific location. The spanwise extend is chosen as $10\%$ of the chord
length, and the outer boundary is at least $50c$ away form the airfoil surface. The mesh spacing is chosen such that
certain requirements for wall-resolved LES are met. The left plot of Fig.~\ref{fig:NACA64418Mesh} shows the grid
spacing in wall-units in the region after the turbulent trip. Since one grid cell contains several degrees of freedom per direction,
the spacing is normalized by the factor $(N+1)$. Across the airfoil, the values for $x^+$ range between $40$ and $60$, $z^+$ between
$20$ and $30$ and $y^+ \approx 3$. The mesh consists in total of  $229,620$ elements, and is represented by polynomials of fifth
order. We used \ANSA to generate the grid, and the surface elements were curved by splitting them as described in Sec.~\ref{sec:surfCurv}. The
volume mesh close to the airfoil was then curved using the RBF approach, see Sec.~\ref{sec:RBF}.

\begin{figure}
	\floatsetup{valign=t, heightadjust=all}
	\ffigbox{%
\begin{subfloatrow}
	\ffigbox{
		
%
		
		}
	{\caption{Comparison of pressure and friction coefficient ($c_p$ and $c_f$) with results from XFoil. Blue lines represent results
	from the suction side, red lines on the pressure side. (\full) $c_p$ \Flexi, (\dashed) $c_p$ XFoil, (\dotted) $c_f$ \Flexi.}
	\label{fig:NACA64418cpcf}}
	\ffigbox{\includegraphics[width=0.49\textwidth]{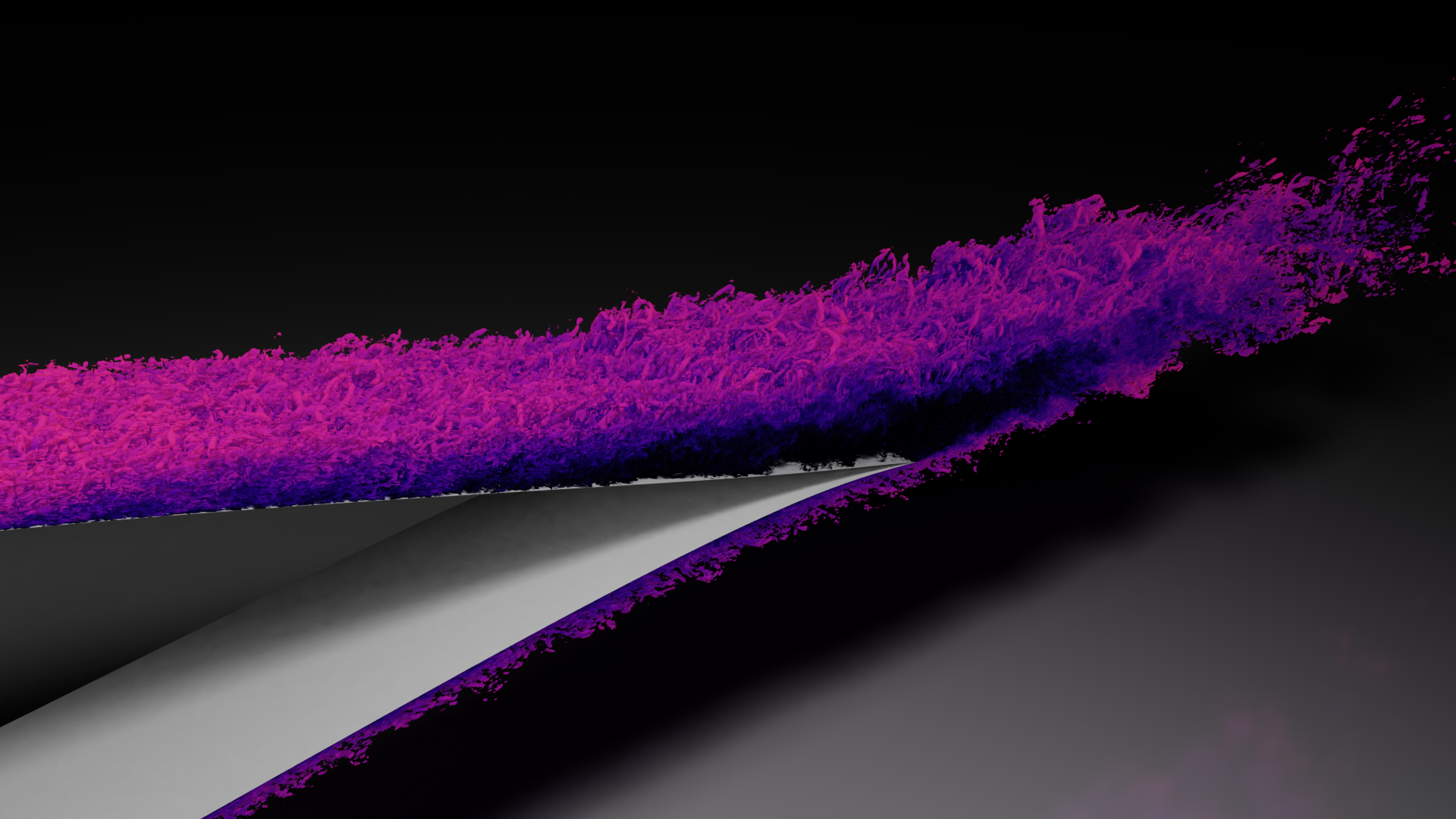}}
	{\caption{Isocontour of Q-criterion, colored with velocity magnitude. Close-up of trailing edge with separated flow region.}}
\end{subfloatrow}}%
	{\caption{Selected results from the NACA 64418 simulation.}
	\label{fig:NACA64418Results}}
\end{figure}
The simulation was run as an implicit LES, using the methodology investigated by Flad and Gassner in~\cite{flad2017use}. It is based 
on the split DG formulation with kinetic energy preserving fluxes, specifically the flux introduced by 
Pirozzoli~\cite{pirozzoli2011numerical}. The equations were discretized with an eight-order scheme, leading to approximately $117$ 
Mio. DOFs per solution variable. The simulation was advanced for $15$ non-dimensional time units $T^*=\frac{U_{\infty}}{c}$, and the 
last five time units were used to gather statistics. Selected results are shown in Figure~\ref{fig:NACA64418Results}. 
The left plot compares the pressure coefficient on the suction and pressure side with results from XFoil~\cite{XFOIL}, and they show excellent 
agreement, except for a small region at the trailing edge of the suction side. There, the flow separates, which can not be predicted by the 
low-fidelity software XFoil. The separation bubble is also evident in the negative wall friction coefficient, as seen in 
Fig.~\ref{fig:NACA64418cpcf}. In the right plot, the turbulent structures at the trailing edge are visualized in a rendering of an isocontour
of the Q-criterion, colored by the velocity magnitude. The large area of low-velocity fluid (dark color) and the sudden thickening of the boundary
layer indicate the recirculation area.

We ran the computation on $12,000$ cores of the Hazel Hen system at the HLRS, corresponding to a load of $9,800$ DOFs/core. It took 
$47,000\unit{CPUh}$ to advance the simulation by one time unit, including all analyze and I/O routines. Thus, we achieved a PID of $1.35\mu s$, 
which is larger than the values presented in Sec.~\ref{sec:scaling}. Part of this can be explained by additional analyze work (e.g. calculation 
of time-averages, time-accurate sampling of the solution at specific points), but the larger part of the increase is due to the additional work introduced
by the split formulation of the volume integral.

\subsection{Shock-vortex interaction}

This example shows how the shock-capturing mechanism based on FV subcells can be used to simulate flows with shocks, while retaining
high order accuracy in smooth regions. We simulate a shock-vortex interaction on a domain of size $[0, 2]\times[0, 1]$, where a
vertical, stationary shock of mach number $Ma=1.5$ is initialized at $x=0.5$ based on the Rankine-Hugoniot condition.
To the left of the shock, a travelling, isotropic vortex is superimposed on the flow field, see~\cite{shu1998essentially} for details
of the initial solution. We do not take viscous effects into account, so the Euler equations are considered for this simulation. The
two-dimensional mesh consists of $200 \times 100$ cells, and we run a fifth-order simulation. The left and right sides are connected
through periodic boundary conditions, while the top and bottom of the domain are treated as walls. In the DG cells, the split DG
formulation is used, and we employ entropy conserving flux formulations after Chandrashekar~\cite{chandrashekar2013kinetic}. To
detect the shock region, an indicator function origination from the Jameson-Schmidt-Turkel scheme~\cite{jameson1981numerical}
is used. It is based on pointwise comparisons of the pressure with neighbouring maximal and minimum pressure values.

\begin{figure}
\begin{center}
	\includegraphics[width=\textwidth]{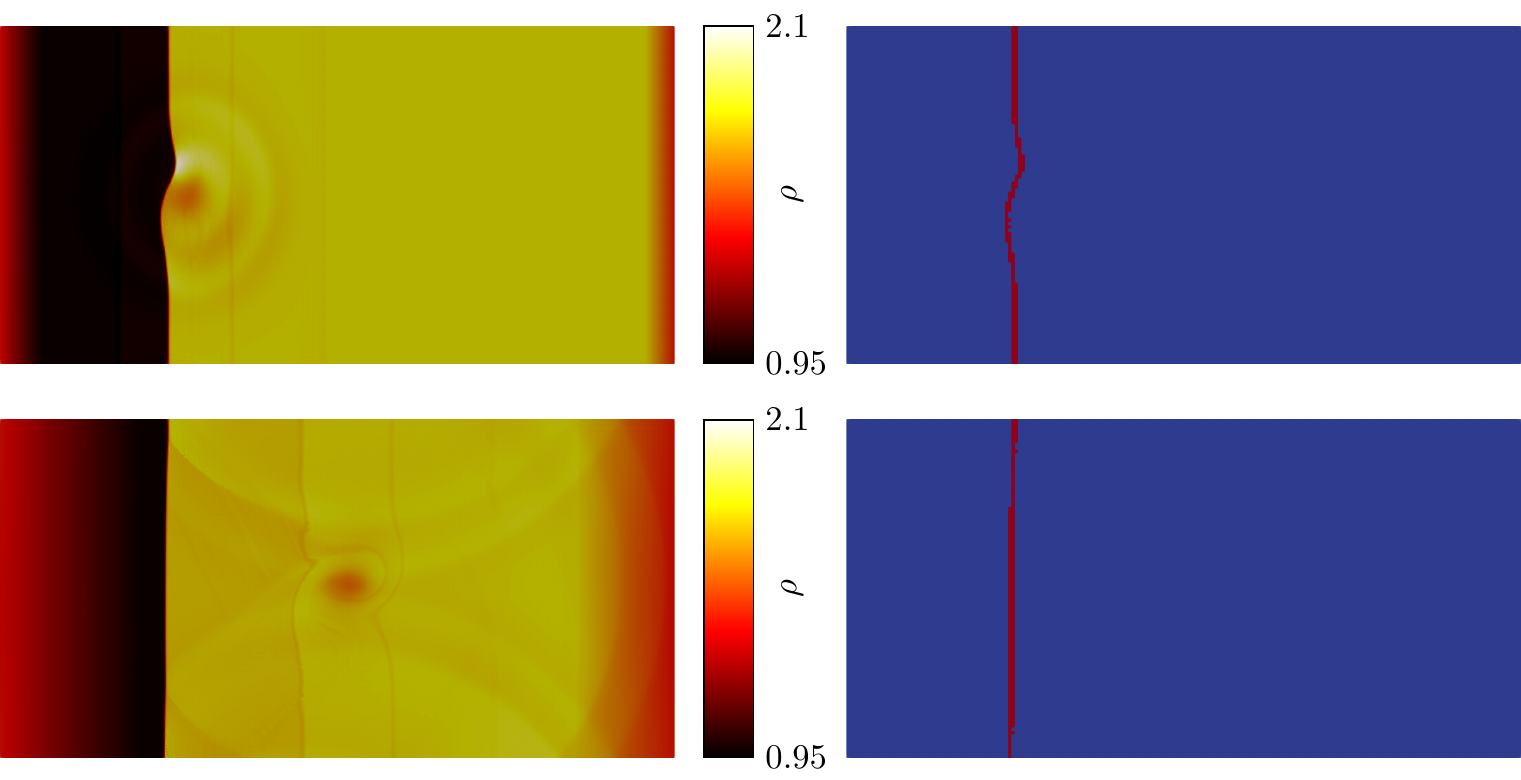}
	\caption{\label{fig:ShockVortex} Results of a shock-vortex interaction at $Ma=1.5$. Top row: $t=0.2$, bottom row: $t=0.7$.
	The left plots visualize the density $\rho$, the right plots the distribution of FV (red) and DG (blue) cells.}
\end{center}
\end{figure}
Figure~\ref{fig:ShockVortex} shows results from the simulation at two different time instances. On the left side, the density is
visualized, while the right side shows the distribution of FV and DG cells. The top row corresponds to a time where the isotropic
vortex is traveling directly through the shock, significantly distorting the formerly straight shock front. Nonetheless, the
indicator function detects the location of the front, and we are able to only thread the immediate vicinity of the shock with the FV
scheme. No artificial waves are originating from the shock front, which would be typical for the behaviour of high order schemes
when high gradients are present. At the time instance depicted in the bottom row, the vortex has traveled further to the right, and
the shock front is again nearly vertical. Due to the interaction with the shock, the vortex is severely distorted. However, it is
still clearly visible. At the chosen resolution, this is only possible because the high order DG method was used after the shock,
and it is able to convect the vortex further downstream without excessive artificial damping.

We have shown here, that the proposed hybrid DG/FV scheme can capture shocks and prevent them from introducing spurious oscillations,
while the high order method used in the smooth regions of the flow is able to leverage its low dissipation and dispersion errors.

\subsection{Aeroacoustic Cavity Noise}
The third example demonstrates the applicability of \Flexi to perform direct noise computations (DNC). As an example, we consider wall-bounded turbulent flow over cavities. 
Industrial applications can be found in the context of an automobile exterior such as sunroof buffeting as well as noise emitted from small gaps between exterior panels, e.g. the door gap. 
Here, different noise generation mechanisms have been identified such as Rossiter feedback, Helmholtz resonance and standing waves. 
All physical mechanisms have their source of noise in pressure fluctuations of the incoming turbulent boundary layer above the cavity opening. 
To capture the large bandwidth of scales between acoustics and turbulence, a DNC of those effects results in a challenging multiscale problem which requires high order numerics.
DNC is necessary to resolve the two-way interaction between hydrodynamics and acoustics. 
Therefore, both requirements, high order numerics as well as DNC, make \Flexi a promising compressible flow solver to resolve aeroacoustic noise.

To perform cavity simulations with an upstream turbulent boundary layer, a zonal LES approach is indispensable, since simulating laminar to turbulent transition in addition could exceed computational resources. 
Therefore, \Flexi has been augmented by a new turbulent inflow condition combining weak anisotropic linear forcing by de Meux \cite{de2015anisotropic} and a recycling rescaling approach similar to Lund \cite{lund1998generation}, 
in order to get a defined turbulent inflow which reveals good properties in terms of artificial noise emission. 
To reduce computational cost further, the wall-model by Kawai and Larsson \cite{kawai2012wall} is used in combination with Vreman's eddy-viscosity model \cite{vreman2004eddy} to account for small-scale dissipation.

\begin{figure}[h!]
  \centering
   \resizebox{0.45\textwidth}{!}{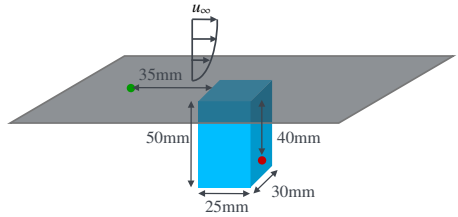}
    \caption{Schematic sketch of the open cavity configuration from \cite{kuhn2019zonal} including the acoustic sensor positions inside the cavity and in front ( \tikzcircle[red, fill=red]{1.5pt} Microphone, \tikzcircle[green, fill=green]{1.5pt} Kulite sensor). }
  \label{fig:Fuge}
\end{figure}

\begin{figure}[h!]
  \centering
   \fontsize{6.5pt}{8.5pt}\selectfont
   \def\svgwidth{3.333in}
   \resizebox{0.45\textwidth}{!}{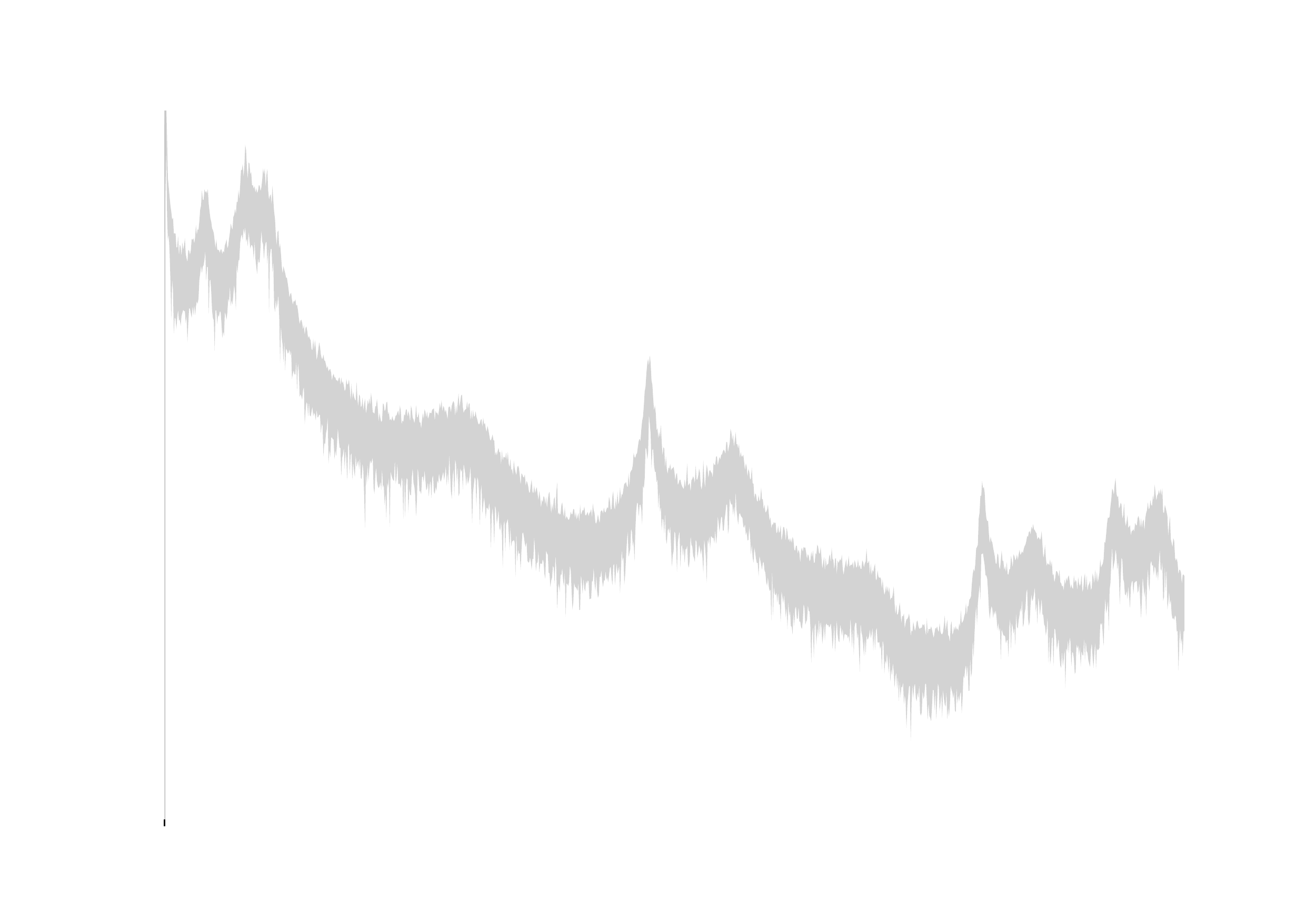}
   \fontsize{6.5pt}{8.5pt}\selectfont
   \def\svgwidth{3.333in}
   \resizebox{0.45\textwidth}{!}{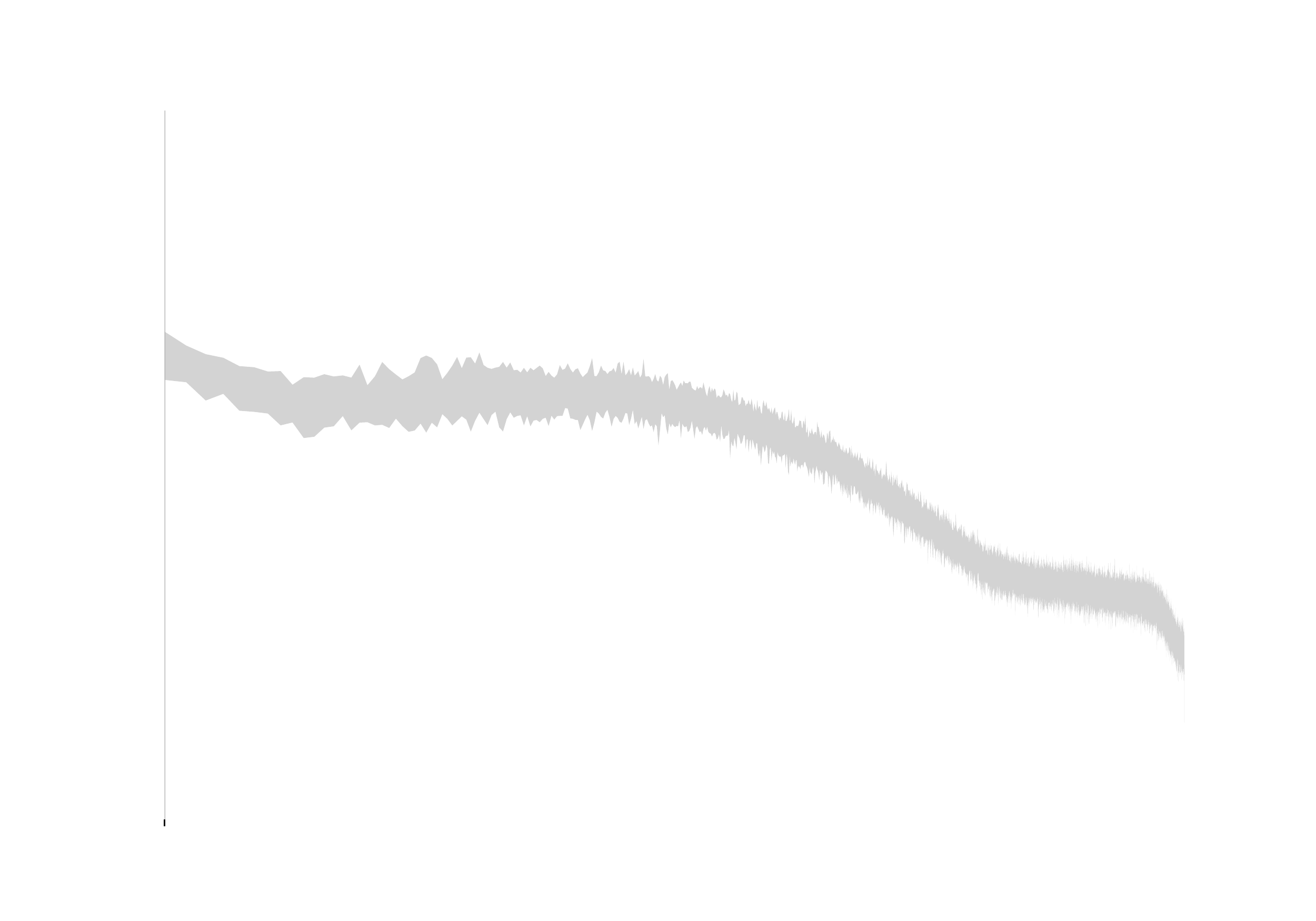}
   \caption{\label{fig:noiseSpectra} (left) Comparison of the pressure spectrum measured inside the cavity. (right) Comparison of the wall pressure spectrum measured in front of the cavity.}
\end{figure}

Based on the geometric definition of a benchmark experiment, which is described in detail in \cite{erbig2018experimental}, DNC simulations of the open rectangular cavity given in Fig.~\ref{fig:Fuge} have been performed.
This specific setup includes Rossiter-feedback as well as excited standing waves with a non-linear interaction between both. 
The inflow velocity is chosen as $U_\infty=42.5\unit{m/s}$ with a corresponding momentum thickness Reynolds number at the leading edge of $Re_\theta=6300$ at standard conditions. 
The computational setup has 30 Mio. DOFs per solution variable, a polynomial degree of N=7 is used and the equally spaced $y^+$ approximation at the wall is $y^+=15$. 
The computation was run for $\Delta t=0.2\unit{s}$ on HLRS Cray XC40 HPC system at a computational cost of $80,000\unit{CPUh}$.
To validate the accuracy of our framework, results of the wall pressure signal measured inside the cavity and in front of the cavity (see Fig.~\ref{fig:Fuge}) are compared to the results of the benchmark experiment. Fig.~\ref{fig:noiseSpectra}~(left) shows the Fourier transformed pressure signal inside the cavity. Above $f=2\unit{kHz}$ the spectrum is characterized by various peaks that all correspond to different acoustic room modes. Here, both pressure level as well as the frequency width of the peaks are in good accordance with the experimental results, especially when comparing the same time averaging intervals. Below $f=2\unit{kHz}$ the spectrum reveals the first two Rossiter feedback modes as well as the first quarter wave. The underlying interaction between hydrodynamics and acoustics plays an essential role to capture the given phenomena. This demonstrates that \Flexi is capable to depict defined multiscale physics accurately. Fig.~\ref{fig:noiseSpectra}~(right) compares the wall pressure spectrum in front of the cavity. Below $f=2kHz$ experiment and numerics agree well. Especially, Vreman's eddy-viscosity model helps to reduce artificial pressure fluctuations caused by the wall-model. Above $f=2\unit{kHz}$ the measured pressure signal drops significantly, which reveals limits of the pressure transducer used in the experiment, especially when comparing to Hu's semi-empirical wall pressure model.  
In summary, \Flexi is capable to resolve multiscale problems such as the interaction of hydrodynamics and acoustics within cavity flows, which can not be captured by state-of-the-art hybrid approaches. Further, we have demonstrated that in case of cavity noise, a great amount of turbulence modeling can be introduced to significantly reduce computational cost while still preserving accuracy of the acoustic results.

%%%%%%%%%%%%%%%%%%%%%%%%%%%%%% ------ CONCLUSION ----- %%%%%%%%%%%%%%%%%%%%%%%%%%%%%%
\section{Conclusion and Outlook}
High order schemes offer a number of potential benefits for the numerical simulation of multiscale problems. However, leveraging their advantages also requires algorithms for computation, pre- and post-processing which are specifically designed with HO in mind. Up to now, commercially available software lacks stringent and consistent HO support.
\Flexi provides an open-source toolchain for HO fluid dynamics simulations, combining the \Hopr mesh generator, the DG-based solver for the compressible Navier-Stokes equations itself and the post-processing suite \Posti into a seamless framework. We have presented \Flexi, its underlying numerical algorithms, features and capabilities as well as implementation details and aspects important to usability and reproducibility in this work here. In order to keep the discussion concise and to the point, we have focused on aspects that are a) already available as open-source and b) that are directly relevant to single-phase, compressible turbulent flows. There are however a number of ongoing extension to \Flexi towards a full multi-physics framework with a strong focus on applicability to complex engineering problems which have not been mentioned here. Among them are:
\begin{itemize}
	\item Multiphase and multicomponent capabilities, in which phase boundaries are tracked with a sharp interface approach~\cite{fechter2017sharp}
	\item Complex equations of state based on realistic models or tabulated data~\cite{hempert2017simulation}
	\item A Lagrangian particle tracking method for high order geometry, used in LES and DNS of particle laden flows~\cite{beck2019towards}
	\item Asymptotic consistent low Mach number schemes based on IMEX splittings~\cite{zeifang2018efficient}
	\item Semi-implicit and fully implicit time integration schemes
	\item Mesh deformation and mesh moving based on ALE formulations
	\item An overset / Chimera mesh module 
	\item Sliding mesh interface for stator/rotor flows
	\item A coupled particle-in-cell and direct simulation Monte Carlo solver for reactive plasma flows~\cite{copplestone2019high,ortwein2017piclas}
	\item Intrusive and non-intrusive methods for uncertainty quantification of the Navier-Stokes equation~\cite{kuhn2018uncertainty}
	\item A flexible and modular framework for the creation and management of simulation stacks and automatic, optimal scheduling on HPC systems
	\item A data-exchange interface to OpenFoam\footnote{https://openfoam.org/} and a coupling to the preCICE library~\cite{preCICE} for coupled multiphysics simulations
\end{itemize}
We plan on incorporate most of these features into the open-source version of \Flexi in order to provide a mature and feature-rich HO framework to the community.

\section*{Acknowledgement}
	The authors gratefully acknowledge the support and computing resources granted by the High Performance Computing Center
	Stuttgart (HLRS) on the national supercomputer Cray XC40 \textit{Hazel Hen} under the grants \textit{hpcdg} and \textit{SEAL}. \Flexi has been developed, tested and applied by a number of people, both from the Numerics Research Group at the University of Stuttgart and elsewhere. We acknowledge their contributions and work. A possibly incomplete list of contributors can be found in the \Flexi Doxygen:\begin{verbatim}https://www.flexi-project.org/doc/doxygen/html/\end{verbatim}
	We would  also like to thank Andrei Cimpoero from CFMS Services Ltd., for his valuable input by using and testing \Hopr, for providing the mesh of the serrated nozzle and for organizing the webinar on high order CFD technologies, where \Hopr was presented.
	Gregor Gassner has been supported by the European Research Council (ERC) under the European Union's Eights Framework Program Horizon 2020 with the research project Extreme, ERC grant agreement no. 714487.



\bibliographystyle{elsarticle-num} 
\bibliography{references}


%\begin{thebibliography}{00}


%\bibitem{}

%\end{thebibliography}
\end{document}